\documentclass{IEEEoj}
\usepackage{cite}
\usepackage{amsmath,amssymb,amsfonts}
\usepackage{algorithmic}
\usepackage{graphicx,color}
\usepackage[caption=false,font=footnotesize]{subfig}
\usepackage{booktabs}
\usepackage{tabularx}
\usepackage{array}
\usepackage{xcolor}

\usepackage{textcomp}
\def\BibTeX{{\rm B\kern-.05em{\sc i\kern-.025em b}\kern-.08em
    T\kern-.1667em\lower.7ex\hbox{E}\kern-.125emX}}
\AtBeginDocument{\definecolor{ojcolor}{cmyk}{0.93,0.59,0.15,0.02}}
\def\OJlogo{\vspace{-4pt}\hskip-4pt\includegraphics[height=18pt]{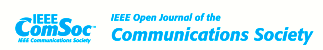}}

\begin{document}

\title{Advances in Wavelet Denoising for Communication Signals: From Parameter Selection Toward Data-Driven Optimization}

\author{Priyalakshmi Sheela\IEEEauthorrefmark{1} \IEEEmembership{(Member, IEEE)}, Indrakshi Dey\IEEEauthorrefmark{2}
\IEEEmembership{(Senior Member, IEEE)}}
\affil{Walton Institute, South East Technological University, Waterford, Ireland}
\affil{Department of Computing and Mathematics, South East Technological University, Waterford, Ireland}
\corresp{%
Corresponding author: Priyalakshmi Sheela
(e-mail: priyalakshmi.sheela@setu.ie)%
}
\authornote{%
This work was supported in part by the
EU Marie Sk{\l}odowska-Curie Actions (MSCA)
COALESCE Project under Grant Agreement
No.~101130739, the US--Ireland R\&D
Partnership Programme under Grant
RI-SFI-23/US/3924, and Taighde Eireann - Research Ireland under Grant 13/RC/2077\_P2.%
}
\markboth{Advances in Wavelet Denoising for Communication Signals: From Parameter Selection Toward Data-Driven Optimization}{Sheela \textit{et al.}}

\begin{abstract}
{Wavelet denoising suppresses nonstationary, impulsive, and interference-like disturbances in communication signals, yet efficacy hinges on coupled choices of transform, mother wavelet, decomposition level, thresholding rule, and shrinkage function. Pairing original communication-domain experimentation with a PRISMA-aligned synthesis of 2020-2025 studies across ten sources, coded by parameter-selection focus and domain (biomedical, industrial, geophysical, communication), this review traces migration from empirical defaults toward
\mbox{similarity-,} \mbox{sparsity-,} \mbox{entropy-,}
\mbox{energy-,} \mbox{sub-band-SNR-,} and task-loss-driven
selection, yet finds communication-oriented evidence scarce. This motivates benchmarking DWT, SWT, and WPT on an impulsive-noise OFDM channel under SNR gain, MSE, BER, EVM, and real-time feasibility, with Friedman and Wilcoxon testing, efficiency-index ranking, and characterising embedded DSP\slash FPGA constraints. Because these experiments show waveform-fidelity gains need not improve hard-decision performance, the resulting Unified Decision Framework for wavelet-parameter selection privileges receiver-level validation over fidelity surrogates, validated on a synthetic pilot-aided OFDM benchmark and measured IEEE~802.11n channels.
On the synthetic link the selected configuration attains the lowest channel-estimation NMSE, significantly outperforms a fixed-parameter wavelet $(p < 10^{-11})$, and is statistically indistinguishable from the strongest classical baseline in BER; on the measured channels it
additionally attains the lowest BER and EVM. The defensible claim is therefore the lowest estimation error at statistically equivalent bit-error performance, and it is conditional on the disturbance being structured or impulsive, since under AWGN alone wavelet thresholding yields no
systematic BER improvement. Within these limits the framework generalises to held-out data and adapts to the observed channel, recasting wavelet denoising into a reproducible, communication-aware design methodology extensible to deep-unfolding architectures.}
\end{abstract}


\maketitle

\section{INTRODUCTION}
\IEEEPARstart{W}{avelet} denoising has emerged as a powerful and widely adopted technique for enhancing weak signals that are otherwise difficult to analyze due to pervasive noise ~\cite{Jangjit2017,Georgieva2019, Wei2019,Bayer2019}. By improving the signal-to-noise ratio (SNR) while minimizing distortion of the underlying signal structure, wavelet-based approaches have demonstrated clear advantages over conventional filtering-based denoising methods~\cite{Georgieva2019}. Consequently, they have found application across a broad spectrum of domains, including magnetic resonance spectroscopy and imaging~\cite{Srivastava2017ESR, Zhong2018Speech, Xu2018EchoState,Peng2020Traffic}, time-series analysis~\cite{Xu2018EchoState}, traffic volume prediction~\cite{Peng2020Traffic}, wireless ultraviolet communication~\cite{Islam2021MRIReconstruction}, time-difference-of-arrival localization~\cite{Buranachai2008FLIM}, fluorescence and Raman imaging/spectroscopy~\cite{Asfour2011VoltageDye,Gan2019Microarray}, surface electromyography~\cite{Sun2020SEMG}, computed tomography~\cite{Bayer2019,Srivastava2017ESR,Zhong2018Speech}, high-resolution surface scans~\cite{Schimmack2018SurfaceScans,Mohammadi2019CT}, nanopore and ion-channel signal analysis~\cite{Shekar2019Nanopore}, electron spin resonance spectroscopy~\cite{Srivastava2017ESR,Freed2000AnnualReview}, and simulated chest wall motion~\cite{Acar2019ChestWall}. Across these domains, the overarching goal remains the same: to extract reliable, information-rich representations from noisy observations without introducing artefacts that could mislead downstream analysis or interpretation.

The selection of mother wavelet, decomposition level and thresholding criteria play a significant role in the denoising performance of wavelet denoising. The choice of the mother wavelet is critical, as it governs the concentration of most of a signal’s energy into a relatively small number of wavelet coefficients. Such concentration facilitates effective separation of signal and noise components via thresholding. In general, the closer the resemblance between the mother wavelet and the signal of interest, the more faithfully the signal can be decomposed into its constituent components in the wavelet domain. As noted in~\cite{Keng2013}, there is no universal or standardized procedure for selecting a mother wavelet; rather, the selection is inherently problem-dependent and is typically guided either by the intrinsic properties of the candidate wavelets or by their similarity to the target signal. The work in~\cite{Keng2013} classifies existing selection strategies into two broad categories: qualitative and quantitative approaches. In qualitative approaches, properties of the mother wavelet such as symmetry, regularity, number of vanishing moments, degree of shift variance, orthogonality, compact support, and availability of a closed-form expression are used as criteria for choosing the most suitable wavelet, with candidate wavelets evaluated against these attributes~\cite{Srivastava2017ESR,Fodor2003Empirical,Wink2004FMRI,Whitcher2000Atmospheric,Cazelles2008Ecology}. In other studies, adopting a qualitative perspective, the similarity between the signal and the mother wavelet is assessed through shape matching, often relying on visual inspection to identify the wavelet that best captures the salient features of the signal~\cite{Xu2018EchoState,Mohammadi2019CT,gertz2006isotropic,Katunin2015CompositeCT}. 

In quantitative approaches, the selection of the mother wavelet is guided by explicit mathematical criteria. Some researchers use the Minimum Description Length (MDL) principle to identify the most suitable mother wavelet~\cite{Brammer1998MDWavelet,Mueller2003WaveletStats}. According to this framework, the optimal wavelet is that which achieves an appropriate balance between fidelity of the reconstructed signal to the observed data and parsimony in its representation. The maximum cross-correlation coefficient between the candidate wavelet and the signal is used as an alternative selection criterion, favouring wavelets that exhibit the strongest linear similarity to the signal structure~\cite{Islam2021MRIReconstruction,Buranachai2008FLIM,Delakis2007MRIImage}. Other researchers employ the similarity between the mother wavelet and transient features within the signal quantified via the Symmetric Distance Coefficient (SDC), exploiting the observation that the wavelet coefficients of a transient closely resembling the mother wavelet exhibit a symmetric pattern.

Maximization of the Compression Ratio (CR) is proposed in~\cite{Hupfel2020RamanProc}, as another quantitative strategy, specifically for selecting the most appropriate mother wavelet for decomposition of vertical ground reaction force (VGRF) signals, under the assumption that a better-adapted wavelet yields a more compact representation. In~\cite{hu2008classification}, the Information Quality Ratio (IQR) is introduced as a novel metric for mother wavelet selection, motivated by the requirement that reconstructed signals preserve the essential informational content of the original data. A multi-criteria approach is described in~\cite{kumar2014raman}, where the mother wavelet is chosen using a weighted combination of peak signal-to-noise ratio (PSNR), mean squared error (MSE), and maximal error, with weights assigned by means of the analytic hierarchy process (AHP). In another method, remodelling factors are used for the determination of the mother wavelet~\cite{Sun2020SEMG}.

In recent years, the problem of selecting an appropriate decomposition level (DL) in wavelet-based analysis has been investigated from multiple perspectives. Yang et al.~\cite{yang2016discussion} and kim et al.~\cite{kim2004damage} reported that the choice of DL depends strongly on the properties of the signal, while Kim et al.~\cite{kim2004damage} determined the optimal DL by converting the sampling rate to the main signal frequency and its associated frequency bands. Other authors have related the maximum number of DLs to the data length ~\cite{tikkanen1997wavelet,sang2012practical, rouis2019optimal}. Chen et al.~\cite{chen2017high} further argued that the number of decomposition levels can be jointly determined by the wavelet filter size and data length to satisfy specific constraint. Despite these efforts, many studies still choose DLs heuristically, for example by selecting an integer between 2 and 5, by simply adopting the deepest level without examining the detail components, or by using ad hoc trial-and-error procedures based on wavelet characteristics~\cite{srivastava2016new}. To introduce more quantitative rigor, several criteria have been proposed to define an “optimal” DL, including linear correlation, spectrum growth index, energy-based measures, and SNR~\cite{kim2004damage,burger2007wavelet,aalborg1993vibrational,sang2010entropy,ranjeet2011retained,wang2010feature}. Wang et al.~\cite{wang2010feature} suggested selecting the DL according to the ratio of detail to approximation energy, while other studies have employed entropy-based measures~\cite{jaffery2012selection,he2015optimal,wang2009robust}; for instance, Wang et al.~\cite{wang2009robust} determined the number of DLs from the frequency bands implied by the sampling rate. Nevertheless, a generally accepted, systematic framework for DL selection has yet to be established, and the choice of DL in practice often remains problem-dependent and empirical.

\subsection{RELATED REVIEWS}
Wavelet-based denoising has become an important component of contemporary signal-processing pipelines because it offers multiresolution representation, time–frequency localization, and coefficient-domain shrinkage mechanisms that are well suited to nonstationary signals. Existing reviews demonstrate this importance across several domains. In EEG processing, prior reviews have shown that wavelet transforms are widely used for artefact removal and for preserving transient neurophysiological activity that may be degraded by Fourier-only methods~\cite{Daud2022EEGArtifacts,Grobbelaar2022EEGSurvey}. In ECG and MCG processing, review studies discuss baseline wander, electromyographic interference, power-line contamination, and hybrid denoising schemes that combine wavelet-based preprocessing with statistical or machine-learning methods~\cite{Shi2023ECGNoiseReview,Malghan2020ECGFiltering,Das2025OptimizedOrthogonal,Jia2024ECGMCGPreproc}. Similarly, reviews of single-channel EMG processing identify wavelet-based denoising as a major class of methods for reducing motion artefacts, electrical interference, and other contaminating components in neuromuscular recordings ~\cite{Boyer2023EMGReview}.

Beyond biomedical applications, wavelet and transform-domain denoising methods have also been reviewed in industrial and structural-monitoring contexts. Reviews of partial-discharge signal processing identify wavelet transforms, singular-value decomposition, and related adaptive methods as common tools for extracting weak fault-related transients from high-noise electrical environments~\cite{Huang2025PDDReview,Chen2024SwitchgearML}. Quantitative reviews of Lamb-wave-based structural health monitoring further show that transform-domain methods can improve the interpretation of dispersive guided-wave signals, although different transforms may be preferred depending on whether the priority is noise suppression, feature localization, or time–frequency representation ~\cite{Han2024LambWaveReview}. Application-oriented studies in geophysical and seismic signal analysis similarly demonstrate that wavelet configurations can be tuned to improve weak-event detection, low-frequency noise suppression, and downstream classification~\cite{Vanchak2024PulseDWT,Silkin2023SeismogramFingerprints}.

Although these reviews and studies collectively demonstrate the broad utility of wavelet-based denoising, they remain largely application-specific. Most prior reviews focus on a single signal class, such as EEG, ECG, EMG, partial-discharge measurements, or Lamb-wave inspection, and therefore do not provide a unified analysis of how wavelet parameters should be selected across domains. In particular, existing reviews rarely examine mother wavelet selection, decomposition-level selection, and thresholding-rule design as coupled decisions. This creates a methodological gap for communication and networking systems, where denoising performance is not only determined by waveform fidelity but also by receiver-level metrics such as bit error rate, error vector magnitude, synchronization accuracy, channel-estimation quality, localization error, and detection performance.
To address this gap, the present review is organized around the research questions listed in Table~\ref{tab:key_research_questions}. These questions are designed to support a systematic assessment of wavelet-denoising parameter selection, with particular attention to communication signals and receiver-oriented performance criteria.

\begin{table*}[!t]
\centering
\caption{Key research questions for wavelet-denoising parameter selection.}
\label{tab:key_research_questions}
\footnotesize
\renewcommand{\arraystretch}{1.15}
\setlength{\tabcolsep}{6pt}

\begin{tabularx}{\textwidth}{>{\centering\arraybackslash}p{0.08\textwidth}X}
\toprule
\textbf{Question} & \textbf{Research Question} \\
\midrule

Q1 &
What theoretical properties make wavelet transforms suitable for denoising nonstationary and transient-rich signals? \\

Q2 &
Which wavelet transform variants, such as DWT, SWT, WPT, and DTCWT, are used for signal denoising, and what trade-offs do they introduce? \\

Q3 &
How are mother wavelets selected, and what criteria are used to match a wavelet basis to signal morphology or task requirements? \\

Q4 &
How is the decomposition level selected, and how do sampling frequency, bandwidth, and scale separation affect this choice? \\

Q5 &
What thresholding functions and threshold-selection rules are used, and how are they adapted to Gaussian, impulsive, or nonstationary noise? \\

Q6 &
What evaluation metrics are used to assess denoising performance, including waveform-fidelity and task-level metrics? \\

Q7 &
How do wavelet-parameter selection strategies differ across biomedical, industrial, geophysical, environmental, and communication domains? \\

Q8 &
To what extent have existing studies connected wavelet-denoising parameters to communication and networking tasks such as BER, EVM, synchronization, channel estimation, spectrum sensing, localization, and interference mitigation? \\

Q9 &
How are optimization, machine learning, and data-driven methods being used to automate wavelet-parameter selection? \\

Q10 &
What limitations, open challenges, and future research directions remain for robust, transferable, and communication-aware wavelet denoising? \\

\bottomrule
\end{tabularx}
\end{table*}

Table~\ref{tab:comparison_related_reviews} compares the present review with closely related prior reviews. Because the wavelet-denoising literature is large and heterogeneous, the comparison is restricted to reviews with the strongest thematic overlap with signal denoising, biomedical preprocessing, industrial monitoring, and transform-domain analysis. The comparison shows that existing reviews provide valuable domain-specific summaries of wavelet-based denoising and related transform-domain methods, but they do not comprehensively address wavelet-parameter selection as a unified design problem. In particular, prior work has not systematically compared mother wavelet selection, decomposition-level determination, and thresholding strategy design across signal domains, nor has it translated these choices into communication-system requirements. The present review therefore extends the literature by consolidating parameter-selection methods across biomedical, industrial, geophysical, environmental, and communication-related signals, and by proposing a decision framework that links wavelet family, decomposition depth, transform type, and thresholding strategy to measurable signal characteristics, noise models, computational constraints, and receiver-level performance objectives.

\begin{table*}[!t]
\caption{Comparison of the Present Review With Closely Related Published Reviews}
\label{tab:comparison_related_reviews}
\centering
\scriptsize
\setlength{\tabcolsep}{2.8pt}
\renewcommand{\arraystretch}{1.25}

\newcolumntype{C}{>{\centering\arraybackslash}p{0.045\textwidth}}

\begin{tabular}{p{0.44\textwidth}*{10}{C}}
\toprule
\textbf{Published Review Papers} &
\textbf{Q1} & \textbf{Q2} & \textbf{Q3} & \textbf{Q4} & \textbf{Q5} &
\textbf{Q6} & \textbf{Q7} & \textbf{Q8} & \textbf{Q9} & \textbf{Q10} \\
\midrule

Wavelet-based filters for artefact elimination in EEG signals: a review~\cite{Daud2022EEGArtifacts} &
Y & P & P & P & Y & Y & N & N & P & Y \\

A survey on denoising techniques of EEG signals using wavelet transform~\cite{Grobbelaar2022EEGSurvey} &
Y & P & P & P & Y & Y & N & N & P & Y \\

A review of noise-removal techniques in ECG signals~\cite{Shi2023ECGNoiseReview} &
P & P & N & N & P & Y & N & N & P & Y \\

A review on ECG filtering techniques for rhythm analysis~\cite{Malghan2020ECGFiltering} &
P & P & N & N & P & Y & N & N & N & Y \\

Preprocessing and denoising techniques for ECG and MCG: a review~\cite{Jia2024ECGMCGPreproc} &
Y & P & P & P & Y & Y & N & N & Y & Y \\

Reducing noise, artefacts, and interference in single-channel EMG signals: a review~\cite{Boyer2023EMGReview} &
Y & P & P & N & P & Y & N & N & P & Y \\

A review on partial-discharge signal denoising methods in power grids~\cite{Huang2025PDDReview} &
Y & P & P & P & Y & Y & N & N & P & Y \\

A quantitative review of air-coupled ultrasonic Lamb-wave analysis based on signal transformations~\cite{Han2024LambWaveReview} &
Y & Y & N & N & P & Y & N & N & P & Y \\

\midrule
\textbf{Present review} &
\textbf{Y} & \textbf{Y} & \textbf{Y} & \textbf{Y} & \textbf{Y} &
\textbf{Y} & \textbf{Y} & \textbf{Y} & \textbf{Y} & \textbf{Y} \\
\bottomrule
\end{tabular}

\vspace{1mm}
\begin{minipage}{0.96\textwidth}
\footnotesize
\textit{Note:} Y indicates that the topic is addressed substantively, P indicates partial or domain-limited coverage, and N indicates that the topic is not a central focus.
\end{minipage}

\end{table*}

\subsection{CONTRIBUTIONS}

The current review offers several contributions to wavelet-based signal denoising strategies.
\begin{enumerate}
    \item  The review reframes wavelet denoising as a parameter-selection problem rather than as a generic preprocessing operation. It treats the transform family, mother wavelet, decomposition level, and thresholding strategy as coupled design variables, and organises the literature according to the specific parameter-selection problem addressed in each study. This allows the review to compare studies not only by method, but also by motivation, selection criterion, and performance objective.
    \item The study is based on a systematic search across multiple databases, using clear inclusion and exclusion criteria to identify a focused set of primary studies on wavelet-based signal denoising. The corpus spans biomedical, industrial, geophysical, environmental, and communication domains, while a dedicated telecommunications relevance tag locates the communication-bearing subset within the wider evidence base. This provides an auditable synthesis of where the literature is concentrated and where it remains sparse.
    \item The review consolidates cross-domain evidence into a unified analytical view of wavelet-parameter selection. Across diverse application areas, it identifies recurring selection criteria, including similarity, sparsity, entropy, energy concentration, subband SNR, and task-specific performance measures. Because many of these criteria are statistics of the wavelet representation rather than domain-specific descriptors, the review shows how parameter-selection principles can be transferred to communication receivers when the structural prior, disturbance model, protected spectral support, and validation endpoint are re-specified.
    \item The review translates the cross-domain evidence and communication-specific experiments into a source-grounded unified decision framework for wavelet-parameter selection. The framework maps measurable signal and system descriptors, including signal context, sampling rate, occupied bandwidth, and disturbance model, onto a reproducible denoising configuration through four explicit stages: transform-family selection, mother-wavelet selection, decomposition-level selection, and thresholding-strategy selection. It also introduces a coupled wavelet-and-level resolution step to avoid circular decision making when no prior wavelet or decomposition depth is available.
    \item  The proposed decision framework is validated at the receiver level, rather than merely introduced as a conceptual taxonomy. Validation is conducted using two complementary case studies: a synthetic pilot-aided OFDM channel-estimation benchmark with exact ground truth and a study using over-the-air measured IEEE 802.11n channels. Across BER, EVM, channel-estimation NMSE, and complex SNR gain, the selected configurations attain the lowest estimation error, outperform fixed-parameter wavelet and classical baselines, and generalise from validation to held-out test data, including measured propagation conditions.
    \item The review explicitly incorporates implementation constraints relevant to communication receivers. It relates the candidate transform families to their computational complexity classes and measured FLOP-proxy ordering, considers latency and real-time feasibility, and identifies fixed-point word length as an implementation-level parameter that can affect certified BER and EVM gains. This emphasises that wavelet-parameter selection for practical receivers must be constrained not only by signal quality but also by the target DSP or FPGA platform.
    \item This review traces the progression of wavelet parameter selection from empirical practices to optimisation-based and data-driven methods, while identifying the lack of learning-based parameter-selection studies for communication links as a key research gap.
    \item Finally, the review outlines a forward-looking research agenda for robust, transferable, and receiver-aware wavelet denoising. It identifies the need for multi-objective and uncertainty-aware optimisation, leakage-validated decomposition-level selection, dependency-aware shrinkage, task-aligned benchmarking, and reproducible reporting. It further highlights deep unfolding and trainable wavelet priors as promising directions for learning the transform, decomposition depth, and shrinkage behaviour directly from communication-relevant data while preserving interpretability.
\end{enumerate}

\subsection{ORGANISATION OF THE PAPER}

The outline of the paper is presented below. Section I (Introduction) highlights the rationale of wavelet denoising across different signal processing domains, discusses the existing reviews and presents the contributions of the current review. Section II (Methodology) describes the process employed for conducting the systematic literature search. The databases, keywords, time frame 2020-2025, and inclusion/exclusion criteria used in the study are stated in detail. Section III (Wavelet Denoising Theory) provides an overview of the theoretical basis of wavelets and wavelet-based denoising. Section IV (Wavelet Denoising Workflow and Evaluation Measures) describes the denoising workflow, wavelet variants, and analyzes the identified papers by three parameter selection problems: mother wavelet choice, fixed decomposition level, and threshold selection strategies. Based on this discussion, a decision framework is proposed for the wavelet parameter selection in the telecommunication sector. Limitations and directions for future work are also identified. Conclusion remarks are provided in Section V.  

\section{METHODOLOGY}
\label{sec:method}

This survey follows a structured and auditable literature review workflow aligned with PRISMA style reporting practice in order to ensure transparency and reproducibility of study identification, screening, and coding. PRISMA denotes the Preferred Reporting Items for Systematic Reviews and Meta Analyses. The methodology is designed to capture wavelet based denoising parameter selection literature comprehensively, while also enabling a clear accounting of how much of the evidence base relates to telecommunications and networking applications.

\subsection{SCOPE, TIME WINDOW, AND NOTATION}
\label{subsec:scope_notation}
The review covers peer reviewed primary research published between 2020 and 2025. Let us define the set of publication years as $\mathcal{Y}=\{2020,2021,\ldots,2025\}$.
The temporal window defines the primary corpus used for PRISMA screening, coding, heat-map construction, and temporal trend analysis; it is not intended to define the historical boundary of the field. Seminal pre-2020 contributions on wavelet shrinkage, multiresolution/filter-bank analysis, wavelet-packet representations, and wavelet-based multicarrier communication are retained as foundational anchors, whereas the quantitative trend synthesis is restricted to recent peer-reviewed primary studies published during 2020-2025.

Let us define the set of bibliographic sources queried as $\mathcal{D}$, with $|\mathcal{D}|=10$ in this study.
Let us define $\mathcal{R}$ as the set of retrieved records prior to screening, including duplicates, with $|\mathcal{R}|=3569$. Let us define $\mathcal{R}_{\mathrm{scr}}$ as the set of records carried into title and abstract screening after de-duplication and removal of ineligible record types, with $|\mathcal{R}_{\mathrm{scr}}|=2585$, and let us define $\mathcal{F}$ as the set of reports retained at the end of full-text screening and carried into the coding and reliability stage, with $|\mathcal{F}|=164$. Let us define $\mathcal{S}\subseteq\mathcal{F}$ as the final set of eligible studies that were coded along both axes after adjudication against the complete inclusion criteria of Table~\ref{tab:incl_excl}. The final coded corpus size is $|\mathcal{S}|=124$. The distinction matters because these four quantities are frequently conflated: $2585$ is the number of records screened, not the number of eligible studies, and $164$ is the number of reports reaching the reliability stage, not the number finally coded. The complete stage-by-stage reconciliation is given in Table~\ref{tab:reconcillation}.

Each eligible study is coded along two orthogonal axes. The first axis is a parameter focus axis that captures the dominant methodological emphasis within wavelet denoising. The second axis is an application domain axis that captures the dominant usage context, with a dedicated category for telecommunications and networking. Let us define $c_{\mathrm{P}}(s)$ as the parameter focus label assigned to study $s\in\mathcal{S}$, and let us define $c_{\mathrm{D}}(s)$ as the application domain label assigned to the same study. For clarity, acronyms used later in the methodology are as follows. GNSS denotes Global Navigation Satellite System. GPS denotes Global Positioning System. OFDM denotes Orthogonal Frequency Division Multiplexing. MIMO denotes Multiple Input Multiple Output. CSI denotes Channel State Information. TDOA denotes Time Difference of Arrival. AOA denotes Angle of Arrival. IoT denotes Internet of Things. WSN denotes Wireless Sensor Network. BER denotes bit error rate. EVM denotes error vector magnitude.

\subsection{LITERATURE SEARCH}
\label{subsec:search}

A structured search was conducted to characterize recent developments in wavelet based denoising with particular emphasis on parameter selection. Searches were performed across Google Scholar, IEEE Xplore, the ACM Digital Library, Springer Nature, Wiley Online Library, Taylor and Francis Online, the SPIE Digital Library, the Directory of Open Access Journals, PubMed, and the Cochrane Library. Where supported, queries were restricted to title, abstract, and author keywords; otherwise, platform default search was used and later refined during screening.

A two stage search strategy was employed. In Stage one, the search targeted wavelet denoising of signals with parameter selection as the central contribution. The Stage one concept set covered wavelet denoising in signals, mother wavelet selection or optimization, decomposition level or scale selection, and thresholding strategies. Boolean operators, phrase matching, and truncation were used where supported, with a publication year filter restricting results to $\mathcal{Y}$. In Stage two, the Stage one concept set was expanded using telecommunications and networking keywords to enrich coverage of receiver and networked sensing pipelines. The Stage two enrichment terms included channel estimation, Channel State Information, Orthogonal Frequency Division Multiplexing, Multiple Input Multiple Output, spectrum sensing, cognitive radio, interference mitigation, localization, Global Navigation Satellite System, and underwater acoustic communications. This enrichment step provides a repeatable mechanism to increase retrieval of studies that evaluate denoising choices through communications and networking tasks and metrics, without weakening the core methodological focus on wavelet parameter selection.

\begin{figure}[htbp]
\centerline{\includegraphics[width=3.5in]{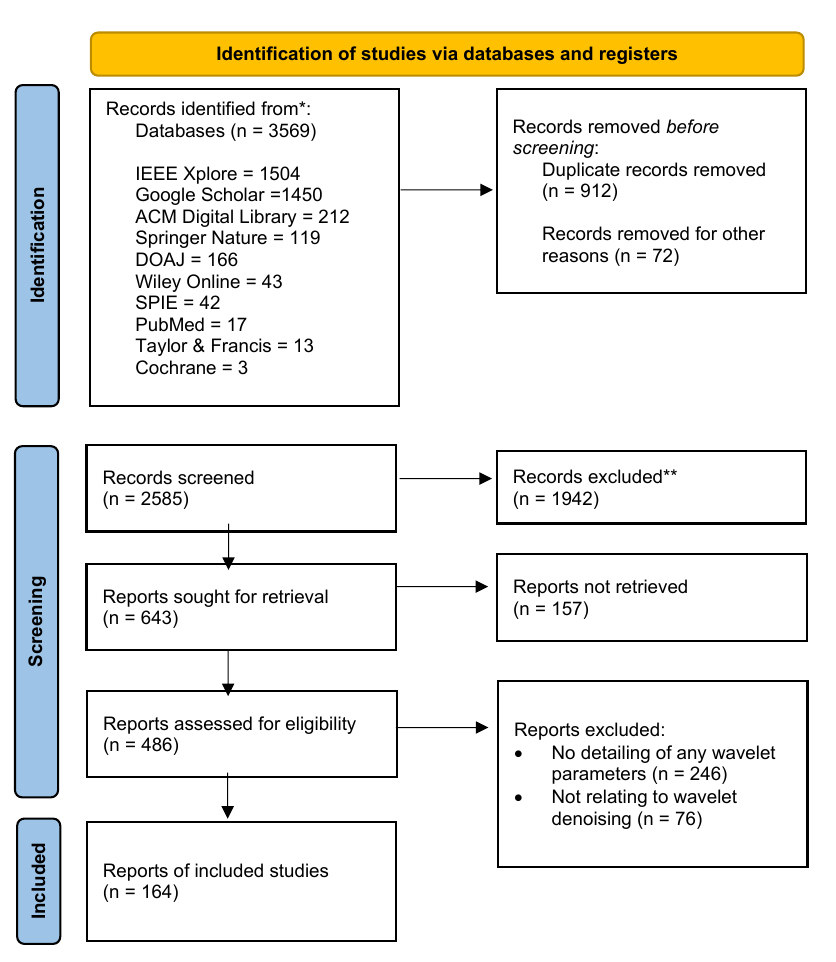}}
\caption{PRISMA flow diagram}
\label{fig:PRISMA}
\vspace{-4mm}
\end{figure}

Only English language full text publications were considered. Inclusion was restricted to peer reviewed primary research articles and peer reviewed full conference papers that reported sufficient methodological and results detail to support evidence synthesis. Review articles, tutorials, editorials, letters, theses, book chapters, and incomplete records such as abstracts without full papers were excluded. Table~\ref{tab:incl_excl} outlines the inclusion and exclusion criteria adopted in this study. Fig.~\ref{fig:PRISMA}  depicts the PRISMA flow diagram followed in this review.

\begin{table*}[!t]
\caption{Inclusion and Exclusion Criteria}
\label{tab:incl_excl}
\centering
\footnotesize
\setlength{\tabcolsep}{3pt}
\renewcommand{\arraystretch}{1.08}
\begin{tabular}{p{68pt} p{185pt} p{190pt}}
\toprule
\textbf{Domain} &
\textbf{Inclusion criteria} &
\textbf{Exclusion criteria} \\
\midrule

Time window &
Publications dated 2020--2025 &
Publications before 2020 or after 2025 \\

Language &
Full text available in English &
Non-English publications or records without accessible English full text \\

Peer review status &
Articles published in peer-reviewed journals or peer-reviewed full conference proceedings &
Non-peer-reviewed material including unrefereed preprints, technical reports, white papers, and non-scholarly web content \\

Study type &
Primary research reporting original methods, experiments, simulations, datasets, or algorithmic developments &
Review articles, meta-analyses, tutorials, editorials, letters, commentaries, theses, dissertations, and book chapters \\

Topical focus,\par core theme &
Studies whose central contribution is wavelet-based denoising of one-dimensional signals &
Studies in which wavelet denoising is tangential, briefly mentioned, not evaluated, or not applied to signal processing \\

Topical focus,\par parameter selection &
Studies addressing at least one of the following with explicit methodology or evaluation: mother wavelet selection, decomposition level selection, or thresholding rule design or tuning &
Studies that do not examine wavelet parameter selection or thresholding in a substantive manner \\

Telecom\-munications and networking\par relevance tag &
Studies that evaluate wavelet denoising choices in relation to communications or networking tasks or metrics such as synchronisation, channel estimation, equalisation, detection performance, BER, EVM, spectrum sensing probability of detection, spectrum sensing probability of false alarm, localisation error for GNSS, TDOA or AOA, robustness to interference or jamming, or underwater or optical link processing &
Studies that do not connect denoising performance to communications or networking tasks or metrics when the telecommunications and networking tag is applied \\

Data and application\par domain &
Signal-based applications including telecommunications and networking, biomedical signals, industrial condition monitoring, geophysical and environmental signals, and other measurable sensor signals &
Non-signal domains such as text processing, purely image-only work without a signal framing, and unrelated fields \\

Publication venue and\par indexing &
Articles indexed in at least one of the targeted sources including Google Scholar, PubMed, Taylor \& Francis Online, the Cochrane Library, Wiley Online Library, IEEE Xplore, Springer Nature, the Directory of Open Access Journals, the SPIE Digital Library, and the ACM Digital Library &
Publications not retrievable from the above databases and platforms including grey literature, local reports, and unindexed outlets \\

Document\par completeness &
Full-text papers with complete methodology and results sections enabling extraction of denoising setup and outcomes &
Conference abstracts without full papers, posters only, extended abstracts, and incomplete or irreproducible records \\

\bottomrule
\end{tabular}
\end{table*}

\vspace{-4mm}

\subsection{SCREENING, ELIGIBILITY AND CODING}
\label{subsec:screening_coding}

All retrieved records were exported where possible, merged into a master list, and de duplicated by matching Digital Object Identifier when available and otherwise by title, author, and venue similarity. Screening proceeded in three stages comprising title screening, abstract screening, and full text screening against predefined eligibility criteria.

Eligibility required that wavelet denoising is a central component of the study and that the study addresses at least one parameter selection subtopic, namely mother wavelet choice, decomposition depth or level selection, or thresholding rule selection or tuning. In addition, studies were tagged as telecommunications and networking relevant when the denoising objective is explicitly tied to communications or networking tasks or metrics such as synchronization, channel estimation, equalization, detection performance, bit error rate, error vector magnitude, spectrum sensing probability of detection and probability of false alarm, localization error for Global Navigation Satellite System or Time Difference of Arrival or Angle of Arrival, robustness to interference or jamming, or underwater or optical link processing.

Each eligible study $s\in\mathcal{S}$ was coded along two axes. Let us define the parameter focus label set as $\mathcal{C}_{\mathrm{P}}=\{\mathrm{G},\mathrm{W},\mathrm{T},\mathrm{L}\}$, where $\mathrm{G}$ denotes general wavelet denoising studies without a dominant parameter selection contribution, $\mathrm{W}$ denotes mother wavelet selection or optimisation, $\mathrm{T}$ denotes thresholding strategy design or tuning, and $\mathrm{L}$ denotes decomposition level or scale selection. Let us define the application domain label set as $\mathcal{C}_{\mathrm{D}}=\{\mathrm{TEL},\mathrm{BIO},\mathrm{IND},\mathrm{GEO},\mathrm{OTH}\}$, where $\mathrm{TEL}$ denotes telecommunications and networking, $\mathrm{BIO}$ denotes biomedical signals, $\mathrm{IND}$ denotes industrial condition monitoring signals, $\mathrm{GEO}$ denotes geophysical or environmental signals not explicitly tied to communications links, and $\mathrm{OTH}$ denotes other signal domains. When a study spans multiple domains, the dominant domain was taken to be the domain associated with the primary experimental validation and performance metrics reported.
Inter-rater reliability was assessed over the 164 records that reached the full-text reliability stage. The two reviewers agreed on 150 records, giving \(91.5\%\) observed agreement and Cohen's \(\kappa = 0.65\), indicating substantial agreement. Disagreements were resolved by discussion before the full inclusion criteria were applied to obtain the final coded corpus.

\begin{figure}
\centerline{\includegraphics[width=3.5in]{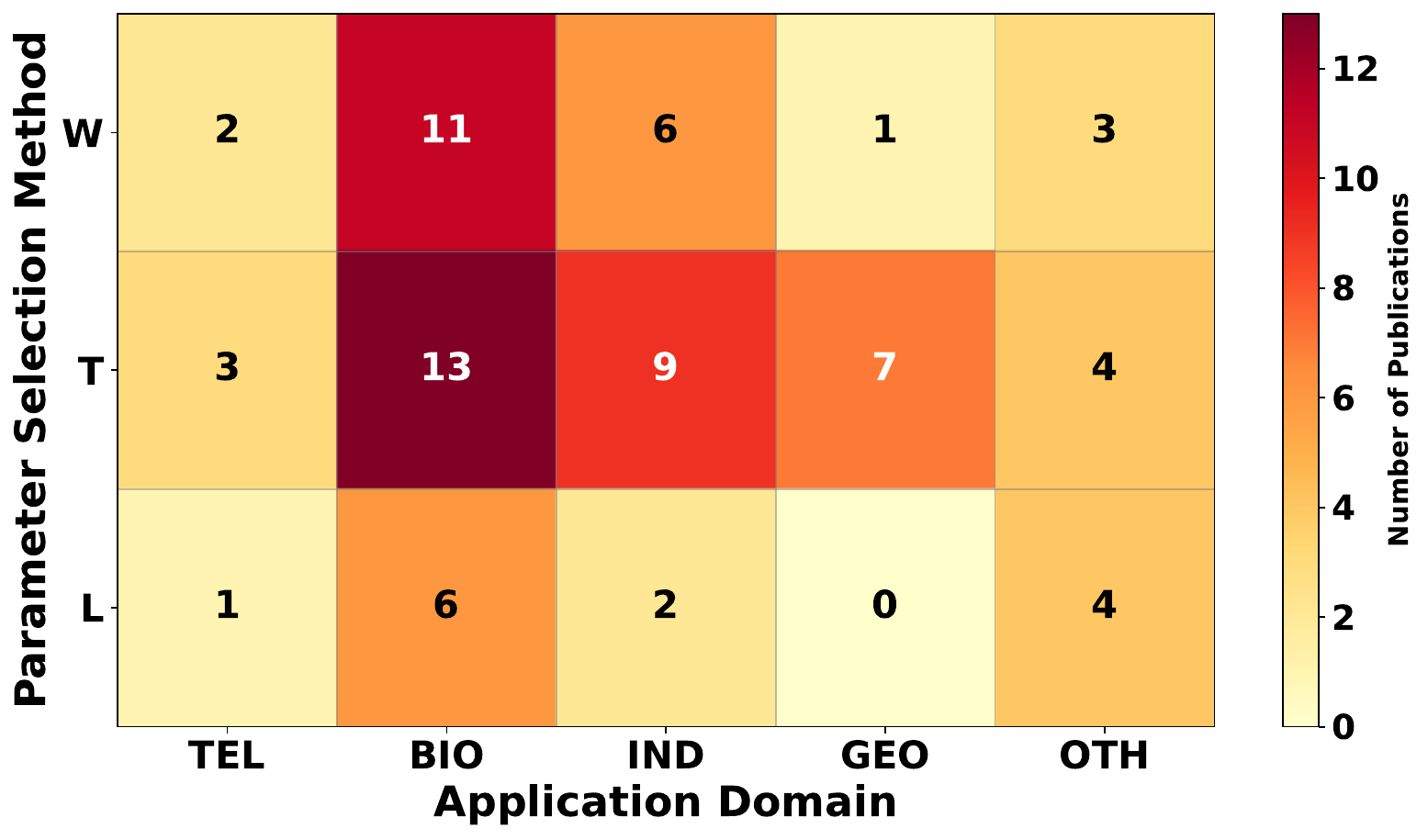}}
\caption{Heatmap of Parameter methods across various Application domains}
\label{fig:heat_map}
\end{figure}

Fig.~\ref{fig:heat_map} displays the heat map showing the relationship between parameter selection methods and application domains in the literature reviewed. It is apparent from the heat map that there is a high concentration of studies in the BIO domain, where the highest number of researches have been conducted in all three categories of parameter selection methods, particularly in wavelet denoising for biomedical applications. On the other hand, TEL and GEO are less represented. 
The cell counts sum to 72, corresponding to the studies in categories W, T, and L; the 52 studies in category G are excluded from this figure by construction, since they carry no dominant parameter-selection label.
In summary, the heat map illustrates how the focus of the methods studied relates to application areas. It is evident that more attention is paid to biomedical and industrial applications, whereas telecommunications, especially, is poorly covered.

Table~\ref{tab:reconcillation} reconciles the record counts across every PRISMA stage and connects them to the coded categories, so that each figure reported elsewhere in the paper can be traced to a defined stage. Three points are worth stating explicitly. First, the $2585$ records of Fig.~\ref{fig:PRISMA} are the records entering title and abstract screening; they are not eligible studies. Second, the $164$ reports labelled as included in Fig.~\ref{fig:PRISMA} are the reports that survived full-text screening and were carried into the coding and reliability stage; adjudication of the $14$ inter-rater disagreements followed by application of the complete inclusion criteria of Table~\ref{tab:incl_excl} at full-text level removed a further $40$ reports, giving the final coded corpus of $|\mathcal{S}|=124$ studies. Third, the heat map of Fig.~\ref{fig:heat_map} displays only the $72$ studies whose dominant contribution is a parameter-selection decision, that is, the categories $\mathrm{W}$, $\mathrm{T}$, and $\mathrm{L}$. The remaining $52$ studies fall in category $\mathrm{G}$, which denotes general wavelet-denoising work without a dominant parameter-selection contribution; category $\mathrm{G}$ is now reported in Table~\ref{tab:reconcillation} rather than left implicit, since $23+36+13=72$ and $72+52=124$.

\begin{figure}
\centerline{\includegraphics[width=3.5in]{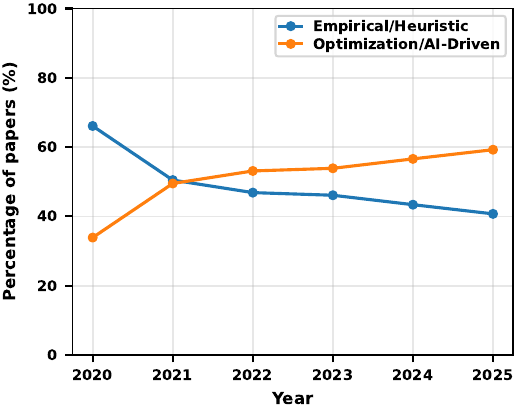}}
\caption{Shift in Design Paradigms (2020-2025). Shares are computed within each publication year over the screened corpus, with per-year denominators $n=(2020),(2021),(2022),(2023),(2024),(2025)$}
\label{fig:Shift in Design}
\end{figure}

Fig.~\ref{fig:Shift in Design} quantifies the temporal trend across the screened corpus over the 2020-2025 window. The plotted quantities are proportions of the studies screened in each year, not of the final coded corpus $\mathcal{S}$, and they are reported as indicative shares rather than as exact ratios; they are intended to show the direction of the shift rather than to support year-by-year inference, the per-year denominators are given in the caption of Fig.~\ref{fig:Shift in Design}.
In 2020, empirical or heuristic parameter‑selection methods accounted for approximately 66\% of the studies, with optimization‑ and data‑driven methods constituting the remaining 34\%. By 2025 this balance had reversed, with optimization‑ and data‑driven methods rising to approximately 59\% and empirical methods declining to approximately 41\%. The methodological drivers behind this trend, and its limited penetration into communication‑specific tasks, are examined in Sections IV‑E and IV‑G.

\begin{table}[h!]
\centering
\caption{Reconciliation of record counts across PRISMA stages and coded categories.}
\label{tab:reconcillation}
\small
\begin{tabular}{lcc}
\toprule
\textbf{Stage or category} & \textbf{Symbol} & \textbf{Count} \\
\midrule
\multicolumn{3}{l}{\textit{Identification}} \\
Records retrieved from the ten sources & $|\mathcal{R}|$ & 3569 \\
Duplicate records removed & & 912 \\
Records removed for other reasons & & 72 \\
\midrule
\multicolumn{3}{l}{\textit{Screening}} \\
Records screened on title and abstract & $|\mathcal{R}_{\mathrm{scr}}|$ & 2585 \\
Records excluded at this stage & & 1942 \\
Reports sought for retrieval & & 643 \\
Reports not retrieved & & 157 \\
\midrule
\multicolumn{3}{l}{\textit{Eligibility}} \\
Reports assessed at full text & & 486 \\
Excluded, no wavelet parameter detail & & 246 \\
Excluded, not wavelet denoising & & 76 \\
Reports entering coding and reliability & $|\mathcal{F}|$ & 164 \\
Excluded after adjudication against Table~3 & & 40 \\
\midrule
\multicolumn{3}{l}{\textit{Included and coded}} \\
Final coded corpus & $|\mathcal{S}|$ & \textbf{124} \\
\midrule
\multicolumn{3}{l}{\textit{Parameter-focus breakdown} $c_{\mathrm{P}}$} \\
General, no dominant parameter focus & G & 52 \\
Mother-wavelet selection & W & 23 \\
Thresholding strategy & T & 36 \\
Decomposition level or scale & L & 13 \\
Subtotal shown in Fig.~2 (W, T, L) & & 72 \\
\midrule
\multicolumn{3}{l}{\textit{Application-domain breakdown} $c_{\mathrm{D}}$\textit{, categories W, T, L}} \\
Telecommunications and networking & TEL & 6 \\
Biomedical & BIO & 30 \\
Industrial condition monitoring & IND & 17 \\
Geophysical and environmental & GEO & 8 \\
Other signal domains & OTH & 11 \\
\bottomrule
\end{tabular}
\end{table}

\section{WAVELET DENOISING THEORY}
\label{sec:theory}

Wavelet denoising is a widely used technique for suppressing noise while preserving signal structures that are localised in time and frequency. In many communication receivers, the observed waveform is conveniently modelled as the sum of a desired baseband signal and an additive disturbance, that is $p(t)=q(t)+g(t)$, where $p(t)$ is the received signal, $q(t)$ is the underlying signal of interest, and $g(t)$ denotes noise and interference. Classical denoising theory often assumes that $g(t)$ is additive white Gaussian noise (AWGN) \cite{Donoho1995SoftThresh,misiti2013wavelets}. However, practical channels in telecommunications and networking frequently exhibit coloured noise, narrowband interferers, and impulsive components, particularly in industrial Internet of Things (IoT) deployments, power line communications, and underwater acoustic links. This section summarises only the transform relations required for the remainder of the paper and then provides a tutorial perspective on why wavelets are attractive in communication systems beyond textbook Fourier based processing.

\subsection{FROM CONTINUOUS TO DISCRETE WAVELET ANALYSIS}
\label{subsec:cwt_dwt}

The continuous wavelet transform (CWT) of a signal $p(t)$ is written as
\begin{equation}
\mathrm{CWT}(a,b)=\int_{-\infty}^{+\infty} p(t)\,\frac{1}{\sqrt{|a|}}\,\psi^{*}\!\left(\frac{t-b}{a}\right)\,dt
\label{eq:cwt}
\end{equation}
where $a$ denotes the scale parameter, $b$ denotes the translation parameter, and $\psi(\cdot)$ denotes the mother wavelet. The CWT provides a highly redundant representation because coefficients are computed across a continuum of scales and shifts. For implementation, the discrete wavelet transform (DWT) adopts dyadic sampling, typically using $a=2^{j}$ and $b=k2^{j}$, which yields a nonredundant multiresolution decomposition that is computationally efficient \cite{misiti2013wavelets}. Under this dyadic grid, the DWT coefficients can be expressed as
\begin{equation}
\mathrm{DWT}(j,k)=\frac{1}{\sqrt{|2^{j}|}}
\int_{-\infty}^{+\infty} p(t)\,
\psi^{*}\!\left(\frac{t-2^{j}k}{2^{j}}\right)\,dt
\label{eq:dwt}
\end{equation}
where $j$ denotes the decomposition level and $k$ denotes the location index.

In practice, the DWT is implemented through a two channel analysis filter bank with downsampling, as introduced in the multiresolution framework~\cite{S1}. At each level, a low pass analysis filter and a high pass analysis filter split the spectrum into a low frequency approximation component and a high frequency detail component. Let us define $f_{\mathrm{N}}$ as the Nyquist frequency of the sampled input. After $J$ levels, the approximation subband spans approximately $[0, f_{\mathrm{N}}/2^{J}]$ and the collection of approximation and detail subbands forms a complete representation of the signal over $[0,f_{\mathrm{N}}]$. Let us define $wA_{j,k}$ and $wD_{j,k}$ as the approximation and detail coefficients at level $j$ and index $k$, respectively. Let us also define $l(\cdot)$ and $h(\cdot)$ as the low pass and high pass analysis filters associated with the chosen mother wavelet. The recursion at the next level can be written compactly as
\begin{align}
wA_{j+1,k} &= \sum_{m} l(m-2k)\,wA_{j,m} \label{eq:wa_rec}\\
wD_{j+1,k} &= \sum_{m} h(m-2k)\,wA_{j,m}. \label{eq:wd_rec}
\end{align}
This representation is sufficient for the subsequent sections, since the survey focuses on selecting the mother wavelet, decomposition level, and thresholding rule that govern $l(\cdot)$, $h(\cdot)$, and $J$.

\subsection{THRESHOLDING AS A DENOISING MECHANISM}
\label{subsec:thresholding_basic}

Wavelet denoising is commonly performed by shrinking detail coefficients that are dominated by noise and then reconstructing the signal. Let us define $w$ as a generic wavelet coefficient and let us define $\lambda$ as a threshold. In soft thresholding, the coefficient is attenuated toward zero according to
\begin{equation}
\hat{w}=\mathrm{sign}(w)\max(|w|-\lambda,0)
\label{eq:soft}
\end{equation}
while in hard thresholding, coefficients below the threshold are set to zero and the remainder are kept unchanged \cite{Donoho1995SoftThresh}. The practical challenge is that $\lambda$ should depend on the noise statistics and the communication task, and the most suitable form of shrinkage may differ under Gaussian noise versus heavy tailed impulsive noise. These issues motivate the parameter selection strategies surveyed later in this paper.

\subsection{TUTORIAL PERSPECTIVE: WAVELETS IN COMMUNICATION SYSTEMS}
\label{subsec:tutorial_comms}

\begin{figure}[htbp]
\centering
\subfloat[]{%
    \includegraphics[width=0.48\linewidth]{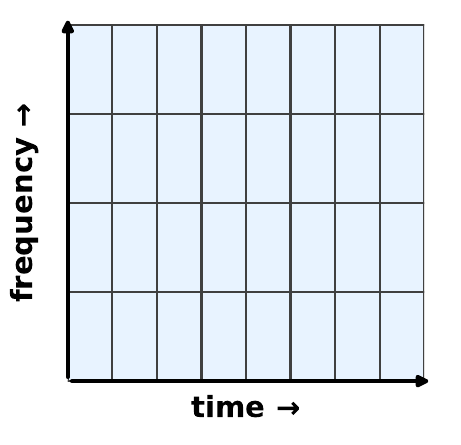}%
    \label{fig:stft_tiling}%
}\hfil
\subfloat[]{%
    \includegraphics[width=0.48\linewidth]{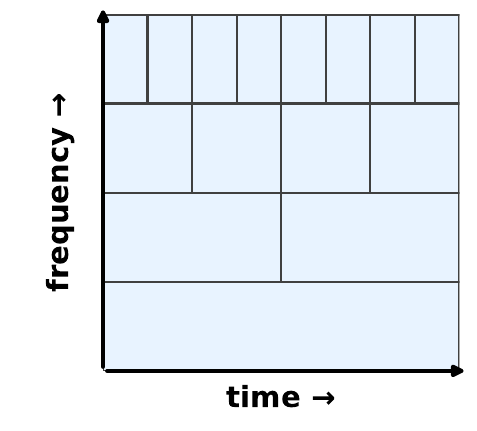}%
    \label{fig:wavelet_tiling}%
}
\caption{Time-Frequency tiling comparison of (a) Short-Time Fourier Transform (b) Wavelet transform}
\label{fig:stft_vs_wavelet}
\end{figure}

Fourier analysis and the fast Fourier transform (FFT) provide an efficient basis for stationary or quasi stationary sinusoids, and they underpin many standard receiver operations. Nevertheless, communication waveforms and channel impairments are frequently nonstationary. Burst transmissions, sporadic access in IoT and random access networks, time varying multipath, and impulsive interference produce transients that are sparse in time. A wavelet basis provides joint time frequency localisation through multiresolution analysis, which lets short time structures be captured by high frequency detail subbands while slower variations are captured by low frequency approximation subbands. 
This distinction is clearly evident in Fig.~\ref{fig:stft_vs_wavelet}, where the fixed-resolution tiling in the case of short-time Fourier transform (STFT) is compared to the multiresolution tiling in the case of wavelets. The STFT employs a fixed window and provides fixed resolution in all frequency bands, an approach that can be used for stationary signals but not suited for nonstationary signals with both slow dynamics and fast transients. On the other hand, the wavelets provide multiresolution tiling, whereby high-frequency components have better time resolution and low-frequency components have better frequency resolution\cite{11422959}. This facilitates precise localisation of transients, symbols, and disturbances in time while maintaining spectral characteristics of smooth components.

This property is particularly useful when the disturbance is not well represented as stationary AWGN, for example when interference occurs as intermittent spikes or short bursts. In such cases, thresholding in the wavelet domain can suppress large isolated coefficients associated with impulsive events while preserving the underlying modulation envelope or symbol transitions, provided that the wavelet family, decomposition depth, and thresholding rule are selected to match the receiver bandwidth and sampling rate.

Telecommunications receivers often operate on sampled complex baseband signals. Let us define the sampling frequency as $f_{\mathrm{s}}$ and the corresponding Nyquist frequency as $f_{\mathrm{N}}=f_{\mathrm{s}}/2$. In multicarrier systems, let us define $\Delta f$ as the subcarrier spacing. The DWT decomposition depth $J$ controls the effective subband bandwidths, since the approximation subband bandwidth decreases roughly as $f_{\mathrm{N}}/2^{J}$. This makes $J$ a communications relevant design variable because it determines whether impulsive components occupy a small number of detail subbands or spread across many bands, which directly affects threshold selection and the risk of distorting the desired signal. Similarly, the mother wavelet determines the analysis filters $l(\cdot)$ and $h(\cdot)$, hence it influences the time domain shape of the basis functions used to represent bursts, symbol transitions, and channel induced transients.

A further communications specific connection arises in wavelet based multicarrier modulation, often termed wavelet based orthogonal frequency division multiplexing. In wavelet based multicarrier systems, the synthesis and analysis filter banks replace sinusoidal subcarriers, and the wavelet family and filter length govern spectral containment, time frequency localisation, and sensitivity to synchronisation and channel dispersion. Consequently, parameter selection is not only a denoising question but also a waveform design question in multicarrier communication, and the choice of wavelet family and decomposition depth affects both out of band emissions and robustness to time and frequency selectivity. This motivates the paper's later focus on mother wavelet selection, decomposition level choice, and thresholding strategies in settings where performance is measured through communications metrics such as bit error rate (BER) and error vector magnitude (EVM), and where interference can be impulsive or non Gaussian.

Wavelet-based OFDM should also be interpreted within the broader filter-bank multicarrier (FBMC) evolution of multicarrier waveform design. In conventional cyclic-prefix OFDM (CP-OFDM), the subcarrier waveform is effectively shaped by a rectangular time-domain pulse, which enables simple FFT-based implementation and orthogonality under synchronous conditions, but produces limited spectral localization. FBMC generalizes this architecture by replacing the rectangular pulse with a spectrally localized prototype filter and by implementing modulation and demodulation through synthesis and analysis filter banks. In the widely studied FBMC/OQAM formulation, the transmitted discrete-time baseband signal can be represented as 
\begin{equation} x_{\mathrm{FBMC}}[n] = \sum_{\mu}\sum_{r=0}^{M-1} a_{r,\mu}\eta[n-\mu M/2]\, e^{j2\pi r n/M}e^{j\phi_{r,\mu}},
\label{eq:fbmc_oqam_signal} 
\end{equation} 
where \(M\) denotes the number of subcarriers, \(a_{r,\mu}\in\mathbb{R}\) denotes the real-valued OQAM symbol on subcarrier \(r\) and time index \(\mu\), \(\eta[n]\) is the prototype filter, and \(\phi_{r,\mu}\) is the phase-staggering term that preserves orthogonality in the real field~\cite{R1,R2} From this perspective, wavelet-OFDM and wavelet-packet multicarrier modulation are not isolated variants of OFDM, but belong to the same filter-bank waveform-design lineage in which the basis functions, subband filters, and time-frequency localization properties are explicitly engineered. This connection is important for the present review because the parameters that govern wavelet denoising, namely the mother wavelet \(\psi(\cdot)\), the analysis filters \(l(\cdot)\) and \(h(\cdot)\), and the decomposition depth \(J\), play roles analogous to prototype-filter and subband-resolution choices in FBMC systems. Both wavelet-based multicarrier modulation and FBMC seek to improve time-frequency localization and spectral containment relative to rectangularly windowed OFDM. However, these gains are accompanied by design trade-offs. FBMC can provide improved out-of-band emission behavior and is attractive in fragmented-spectrum and asynchronous-access scenarios, but FBMC/OQAM also introduces intrinsic imaginary interference and complicates channel estimation, equalization, and MIMO integration~\cite{R2,R3,R4}. Therefore, the FBMC literature clarifies that the wavelet-OFDM discussion is part of a broader contemporary waveform-design trajectory. It also reinforces the central argument of this review: wavelet-family selection, decomposition-level selection, and thresholding should be treated as coupled receiver and waveform-design decisions when performance is evaluated using communication-level metrics such as bit error rate (BER), error vector magnitude (EVM), synchronization accuracy, and channel-estimation fidelity. A related OFDM synchronization study formulates the receiver signal space in Hilbert space and uses wavelet analysis to extract symbol-rate and frame components, with BER and FER evaluated under AWGN and frequency-offset conditions~\cite{R38}.

The same logic extends to related wavelet-domain multicarrier schemes. Filtered-orthogonal wavelet division multiplexing swaps the Fourier transform of filtered-OFDM for an inverse DWT, dropping the cyclic prefix and improving peak-to-average power ratio (PAPR), BER, and out-of-band leakage, with biorthogonal bases favouring PAPR and BER and discrete-Meyer favouring spectral containment~\cite{R4}. FBMC channel-estimation surveys~\cite{R3}] and comparative multicarrier studies~\cite{R5} situate these schemes in the 5G-and-beyond waveform landscape. In each case the wavelet or prototype filter is the modulation basis, so comparing families and levels is a waveform-design question, not a denoising one.

Recent MIMO DWT-OFDM work further extends this waveform-design perspective by performing equalization and co-CFO compensation directly in the wavelet domain, showing that wavelet-domain receiver design can influence BER, processing complexity, and data-rate performance~\cite{R34}.A related underwater RF implementation used wavelet-OFDM with loop-shaped antennas for AUV communication, shifting operation to lower-frequency compression modes to counter seawater attenuation and achieving Mbps-level short-range links ~\cite{R33}. The same pattern appears in optical and optical-RF-convergence links. In visible-light communication, wavelet-OFDM has been offered in place of FFT-OFDM, for instance a non-Hermitian-symmetry wavelet-transform OFDM scheme for indoor MIMO visible-light systems with an imaging receiver~\cite{R5}, and a Meyer-prototype wavelet-OFDM visible-light MIMO system ~\cite{R6}. For optical-RF convergence, a 2×2 Alamouti-coded wavelet-OFDM-over-fibre link with adaptation reported lower side-lobes and PAPR and better BER than FFT-OFDM under optical nonlinearity, with low-order orthogonal wavelets aiding power efficiency and biorthogonal wavelets aiding spectral containment ~\cite{R7}. Underwater RF experiments similarly show wavelet-OFDM being used as the modulation basis for short-range AUV communication in seawater, with operation in the 125 kHz–1.75 MHz band and antenna-dependent Mbps-level transmission performance ~\cite{R37}. Underwater RF AUV communication provides a further example of this waveform-design category, where compressed low-frequency wavelet-OFDM modes and loop-shaped antennas enable meter-scale TCP/UDP communication and Mbps-level video/control links in seawater ~\cite{R33}. As before, the wavelet here is the modulation basis, which is separate from the denoising parameter selection surveyed in Section IV.

\begin{figure*}[!t]
\centering

\subfloat[\label{fig:impulsive_time}]{
    \includegraphics[width=0.98\textwidth]{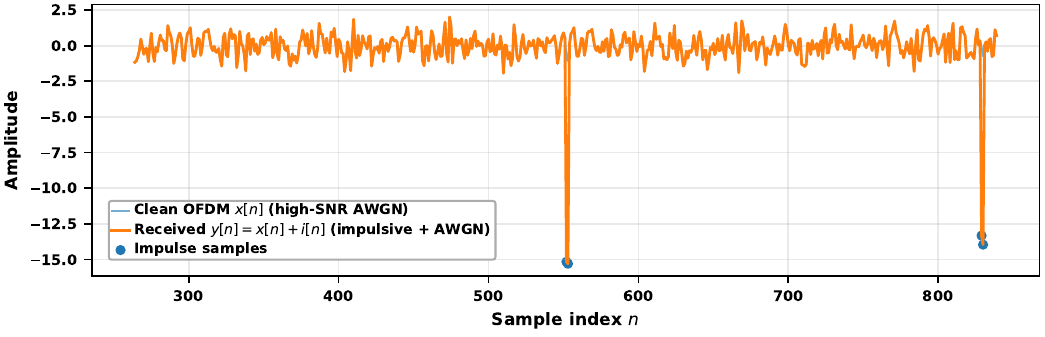}
}
\vspace{1mm}

\subfloat[\label{fig:impulsive_frequency}]{
    \includegraphics[width=0.98\textwidth]{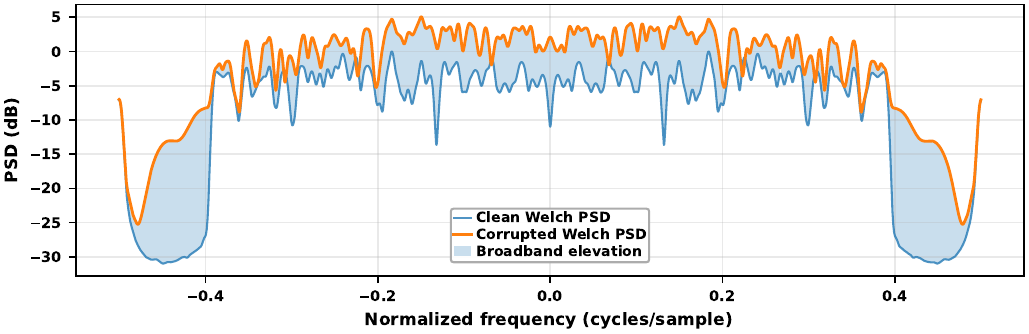}
}
\vspace{1mm}

\subfloat[\label{fig:impulsive_wavelet}]{
    \includegraphics[width=0.98\textwidth]{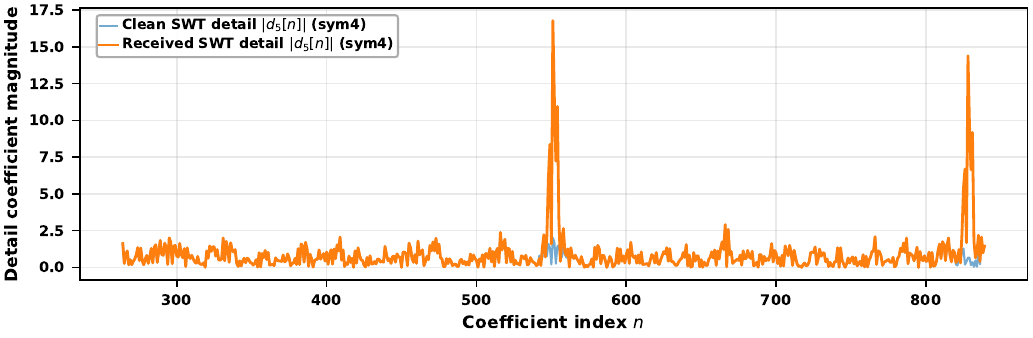}
}

\caption{Illustration of impulsive noise in an OFDM receiver across (a) the time domain, (b) the frequency domain, and (c) the wavelet domain.}
\label{fig:impulsive_noise_ofdm}
\end{figure*}

 Fig.~\ref{fig:impulsive_noise_ofdm} provides a communication-oriented illustration of impulsive noise in an OFDM receiver across the time, frequency, and wavelet domains. In the time domain, impulsive interference appears as sparse, high-amplitude excursions superimposed on the received waveform. Because these events are highly localized in time, their spectral effect is broadband, leading to an elevated power spectral density over a wide frequency range rather than a narrowband distortion. In the wavelet domain, however, the same impulses are concentrated into a relatively small number of large detail coefficients, while the useful OFDM structure remains distributed more smoothly across the remaining coefficients. This separation explains why wavelet denoising is effective for impulsive environments; given an optimal mother wavelet and decomposition level, interference-dominated detail coefficients can be effectively eliminated by thresholding the wavelet coefficients without disturbing the original signal's modulation and transition structures. The use of stationary wavelet transform (SWT) coefficients is especially relevant in burst-based receivers, since their shift-invariant construction preserves temporal alignment and yields more stable localisation of impulsive events across arbitrary sample offsets. 
\par Wavelet based denoising is typically implemented as a three stage procedure that combines multiresolution analysis with coefficient shrinkage and perfect reconstruction. The canonical three-stage denoising workflow is detailed in Table ~\ref{tab:denoising_steps} under Section IV. In many communications and networking scenarios, translation robustness is important because packets and bursts may arrive with uncertain timing and because impulsive interference may occur at arbitrary sample offsets. The SWT addresses this requirement by providing a shift-invariant extension of the DWT; its construction and the associated computational trade-offs are detailed in Section IV alongside the quantitative comparison in Tables ~\ref{tab:receiver_level_performance} and ~\ref{tab:complexity_realtime}. Complex wavelet extensions, most notably the dual-tree complex wavelet transform (DTCWT)~\cite{1550194}, provide approximate shift invariance and improved directional selectivity; their construction and relevance to communication signals are discussed in Section IV. The wavelet packet transform (WPT) further generalizes the DWT by decomposing both
approximation and detail subbands at each level, thereby enabling finer spectral partitions while retaining perfect reconstruction \cite{cody1994wavelet},~\cite{S2}. Its construction and denoising implications are
detailed in Section IV alongside the DWT and SWT comparison.

\begin{table}
\caption{Canonical Steps in Wavelet Based Denoising and Key Design Choices}
\label{tab:denoising_steps}
\centering
\footnotesize
\renewcommand{\arraystretch}{1.05}
\begin{tabular}{p{0.16\linewidth}p{0.77\linewidth}}
\toprule
\textbf{Step} & \textbf{Description and required selections} \\
\hline
Decomposition &
Compute the discrete wavelet transform of the noisy signal. Select the mother wavelet $\psi(\cdot)$ and the maximum decomposition level $J$. The output comprises approximation coefficients at level $J$ and detail coefficients across levels $1$ through $J$. \\
Thresholding &
Apply shrinkage to detail coefficients to attenuate noise and interference. Select a threshold value $\lambda$ and a shrinkage rule such as hard thresholding or soft thresholding. Popular threshold selection heuristics include \emph{rigrsure}, \emph{heursure}, \emph{sqtwolog}, and \emph{minimaxi}.\\
Reconstruction &
Reconstruct the denoised signal using the inverse discrete wavelet transform. Let us denote this synthesis stage as IDWT, which combines the level $J$ approximation coefficients with the thresholded detail coefficients under perfect reconstruction. \\
\hline
\end{tabular}
\end{table}

\section{WAVELET DENOISING WORKFLOW AND EVALUATION MEASURES}

Wavelet based denoising is typically implemented as a structured three stage workflow comprising multiresolution decomposition, coefficient shrinkage, and perfect reconstruction. This workflow is common across application domains, but the choice of wavelet parameters and the interpretation of performance measures depend strongly on the downstream task. In the present survey, this workflow is used to motivate the subsequent parameter selection sections, since mother wavelet choice, decomposition depth, and thresholding strategy jointly determine the denoising outcome.
Because this paper is a review that also reports original communication-domain experimentation, we state which results are our own. The following report experiments conducted for this paper: Figs.~\ref{fig:impulsive_noise_ofdm},~\ref{fig:BER vs $E_b/N_0$},~\ref{fig:shape_matching_concept},~\ref{fig:sparsity_levels},~\ref{fig:Thresholding performance under AWGN and Impulsive Noise}, and ~\ref{fig:Thresholding_functions}, together with Tables ~\ref{tab:receiver_level_performance},~\ref{tab:complexity_realtime},~\ref{tab:friedman_results},~\ref{tab:radioml_fom_efi_wavelet},~\ref{tab:synthetic_ofdm_holdout}, and ~\ref{tab:measured_channel_holdout}, and
Supplementary Figs.~S1 to~S3. All remaining figures and tables either document the review protocol and corpus, consolidate definitions and trade-offs drawn from the reviewed studies, or are
conceptual and framework diagrams. As a general convention, quantities reported with confidence
intervals are our own measurements, and quantities attributed to a citation are as reported by the
cited study.

\begin{figure*}[!t]
\centering
\includegraphics[width=\linewidth]{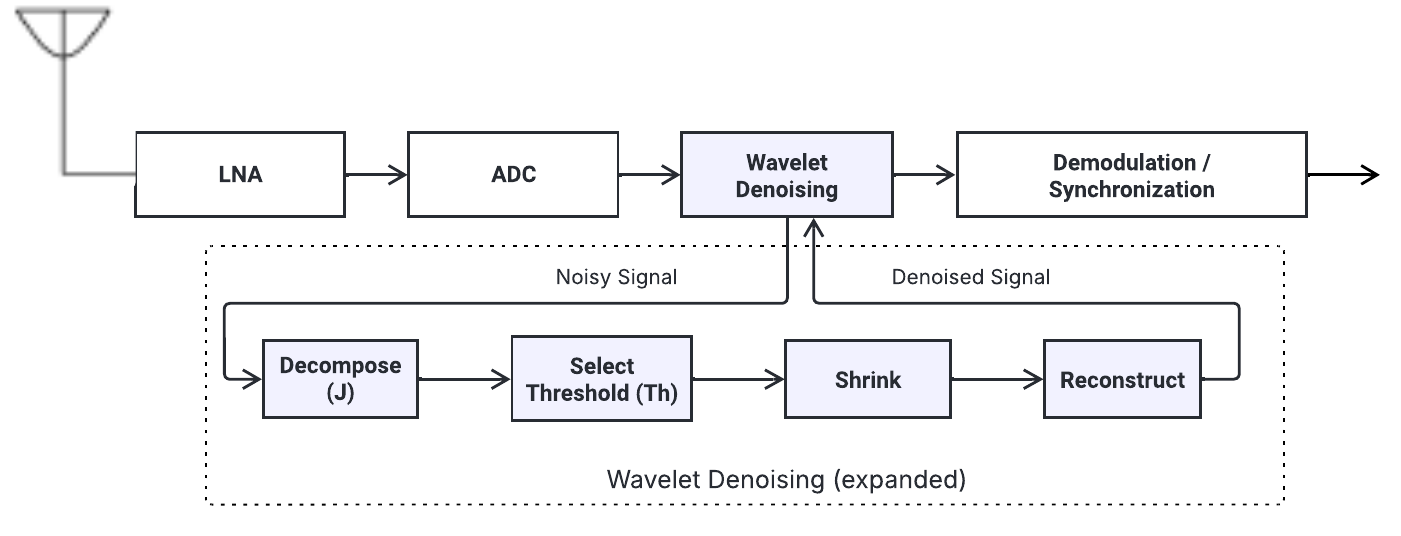}
\caption{Wavelet-Enhanced Receiver Pipeline}
\label{fig:Wavelet-Enhanced Receiver}
\end{figure*}

Fig.~\ref{fig:Wavelet-Enhanced Receiver} illustrates a wavelet-enhanced receiver pipeline in which denoising is inserted after analog front-end amplification and digitization, but before downstream synchronization and demodulation. This placement is critical: noise and interference corrupt the sampled waveform before symbol timing, carrier recovery, and detection are performed. Reducing disturbance at this stage therefore improves the quality of all subsequent receiver operations. The process follows a canonical sequence. The noisy digital signal is first decomposed into multiresolution subbands up to level \(J\); a thresholding rule is then selected to identify interference-dominated coefficients; shrinkage is applied to attenuate those coefficients; and the denoised waveform is finally reconstructed. This process exploits the fact that short-lived disturbances and abrupt transitions tend to concentrate in specific detail coefficients, whereas the useful signal energy is distributed more systematically across approximation and detail subbands. Consequently, appropriate selection of the decomposition depth and threshold governs the trade-off between interference suppression and preservation of modulation structure, elevating wavelet denoising from generic preprocessing to a deliberate receiver-level design choice.

Let us model the sampled noisy signal as $p[n]=q[n]+g[n]$ for $n=0,1,\ldots,N-1$, where $p[n]$ denotes the observed noisy signal, $q[n]$ denotes the unknown clean signal, and $g[n]$ denotes additive disturbance. Let us define $q_{1}[n]$ as the reconstructed denoised signal obtained after wavelet shrinkage and inverse transformation. Let us define $\mu_{q}$ and $\mu_{q_{1}}$ as the sample mean values of $q[n]$ and $q_{1}[n]$, respectively.

In the first stage, the noisy signal $p[n]$ is decomposed using DWT. This stage requires selecting a mother wavelet $\psi(\cdot)$ and fixing the maximum decomposition level $J$. The selected wavelet determines the associated analysis filters, while $J$ controls the multiresolution partitioning of the spectrum. The second stage applies thresholding to the detail coefficients in order to suppress noise and interference while retaining salient signal structures. This stage requires selecting a threshold value and a thresholding rule, most commonly hard or soft thresholding. Widely used threshold selection heuristics in the wavelet denoising literature include \emph{rigrsure}, \emph{heursure}, \emph{sqtwolog}, and \emph{minimaxi}. In the third stage, the denoised signal $q_{1}[n]$ is reconstructed from the level $J$ approximation coefficients and the thresholded detail coefficients using the inverse discrete wavelet transform (IDWT). Since IDWT corresponds to the perfect reconstruction synthesis filter bank associated with the chosen wavelet, this reconstruction stage preserves components that are not suppressed by shrinkage.

Several wavelet variants are also employed in the reviewed literature to improve robustness or to provide more flexible frequency partitions. The stationary wavelet transform (SWT) is a shift-invariant extension of the DWT in which the down-sampling step intrinsic to dyadic decomposition is suppressed. Rather than decimating the approximation and detail sequences at each level, SWT uses scale-dependent up-sampling of the low-pass and high-pass analysis filters. This construction preserves the full temporal resolution of the original signal at every decomposition level, so the resulting approximation and detail coefficients remain the same length as the input signal. The key practical consequence is that small time shifts in the input produce more consistent coefficient alignment across scales, yielding more stable thresholding decisions and improved interpretability when burst timing or impulsive events vary across observations. These properties make SWT particularly relevant in burst-based receivers and in scenarios where impulsive interference occurs at arbitrary sample offsets. 
The complex wavelet transform extends real-valued decompositions to the complex domain to improve analytic properties that are desirable in denoising and feature extraction. A widely used implementation is the dual-tree complex wavelet transform (DT-CWT), which employs two parallel real-valued DWTs to form the real and imaginary parts of complex coefficients. The DT-CWT offers approximate shift invariance and improved directional selectivity in two and higher dimensions, while retaining fast filter-bank implementations with limited redundancy. Although often discussed in image processing, these properties are also relevant to communication signals and channel representations, for example when processing time-frequency representations, multi-antenna measurements, or two-dimensional delay–Doppler grids, where stable phase-aware coefficient behaviour can improve suppression of interference while preserving structured signal components.
A concrete communication instance is the use of a DT-CWT with twin support-vector regression for channel estimation in dense 5G millimetre-wave indoor-hotspot scenarios, with BER evaluated across mobility and receiver-sensitivity configurations~\cite{R8}.

The wavelet packet transform (WPT) generalizes the DWT by applying the decomposition recursively to both the approximation and the detail subbands at each level, rather than to the approximation subband alone. This produces a complete binary tree of subspaces and a richer library of candidate orthonormal bases, which can be exploited to obtain representations better matched to the spectral occupancy of the signal and the disturbance. Because the WPT retains the perfect reconstruction property, it supports denoising pipelines
that select a best basis from the packet tree and then apply shrinkage in the selected subspace \cite{cody1994wavelet}.

\begin{table*}[!t]
\caption{Receiver-level denoising performance across transform families.}
\label{tab:receiver_level_performance}
\centering
\footnotesize
\setlength{\tabcolsep}{3pt}
\renewcommand{\arraystretch}{1.08}
\begin{tabular}{p{45pt} p{95pt} p{120pt} p{120pt} p{90pt}}
\hline
\textbf{Method} &
\begin{tabular}[c]{@{}l@{}}
\textbf{$\Delta\mathrm{SNR}$ /}\\
\textbf{denoising gain}
\end{tabular} &
\textbf{NMSE} &
\textbf{BER outcome} &
\textbf{EVM} \\
\hline

DWT &
$\displaystyle 1.088 \pm 0.291~\mathrm{dB}$ &
$\displaystyle (7.437 \pm 1.060)\times 10^{-2}$ &
$\displaystyle (2.982 \pm 0.848)\times 10^{-3}$ &
$\displaystyle 22.359 \pm 1.582\%$ \\

SWT &
$\displaystyle 1.644 \pm 0.307~\mathrm{dB}$ &
$\displaystyle (6.597 \pm 0.962)\times 10^{-2}$ &
$\displaystyle (1.962 \pm 0.578)\times 10^{-3}$ &
$\displaystyle 21.017 \pm 1.538\%$ \\

WPT &
$\displaystyle 0.991 \pm 0.332~\mathrm{dB}$ &
$\displaystyle (7.492 \pm 1.050)\times 10^{-2}$ &
$\displaystyle (2.781 \pm 0.793)\times 10^{-3}$ &
$\displaystyle 22.669 \pm 1.530\%$ \\

\hline
\end{tabular}
\end{table*}

\begin{table*}[!t]
\caption{Computational complexity across transform families. The final column reports the smallest frame duration $T_{\mathrm{frame}}$ for which the real-time indicator of Table~9 satisfies $\mathcal{F}_{\mathrm{RT}}=1$, which by definition equals the measured $t_{\mathrm{frame}}$. A deployment is therefore real-time feasible for a given
transform precisely when its frame budget exceeds the tabulated value, so the entry can be checked directly against any target numerology without reference to the particular one used here.}
\label{tab:complexity_realtime}
\centering
\footnotesize
\setlength{\tabcolsep}{3pt}
\renewcommand{\arraystretch}{1.08}
\begin{tabular}{p{70pt} p{160pt} p{130pt} p{120pt}}
\hline
\textbf{Method} &
\textbf{Mean time} &
\textbf{FLOP proxy} &
\begin{tabular}[c]{@{}l@{}}
\textbf{Minimum frame budget}\\
\textbf{$T_{\mathrm{frame}}$}
\end{tabular} \\
\hline

DWT &
$\displaystyle 0.171 \pm 0.002~\mathrm{ms/frame}$ &
$\displaystyle 62{,}720$ &
0.171 ms \\

SWT &
$\displaystyle 0.382 \pm 0.005~\mathrm{ms/frame}$ &
$\displaystyle 215{,}040$ &
0.382 ms \\

WPT &
$\displaystyle 0.266 \pm 0.001~\mathrm{ms/frame}$ &
$\displaystyle 143{,}360$ &
0.266 ms \\

\hline
\end{tabular}
\end{table*}

\begin{table*}[!t]
\caption{Friedman test results across receiver-level and complexity metrics.}
\label{tab:friedman_results}
\centering
\footnotesize
\setlength{\tabcolsep}{3pt}
\renewcommand{\arraystretch}{1.08}
\begin{tabular}{p{75pt} p{170pt} p{245pt}}
\hline
\textbf{Metric} &
\textbf{Friedman result} &
\textbf{Interpretation} \\
\hline

$\Delta\mathrm{SNR}$ &
$\displaystyle \chi^{2}(2)=99.633,\quad p=2.32\times 10^{-22}$ &
Transform choice significantly affects denoising gain. \\

MSE/NMSE &
$\displaystyle \chi^{2}(2)=100.800,\quad p=1.29\times 10^{-22}$ &
Transform choice significantly affects reconstruction error. \\

BER &
$\displaystyle \chi^{2}(2)=64.212,\quad p=1.14\times 10^{-14}$ &
Transform choice significantly affects receiver error rate. \\

EVM &
$\displaystyle \chi^{2}(2)=103.333,\quad p=3.64\times 10^{-23}$ &
Transform choice significantly affects constellation distortion. \\

Mean time &
$\displaystyle \chi^{2}(2)=120.000,\quad p=8.76\times 10^{-27}$ &
Transform choice significantly affects runtime. \\

\hline
\end{tabular}
\end{table*}

Because the OFDM baseband is complex valued and the transforms compared here are real-valued filter banks, the in-phase and quadrature components are decomposed, thresholded, and reconstructed independently by the same real-valued transform, using a common threshold derived
from the median absolute deviation of the pooled first-level detail coefficients of both
components. No dual-tree or complex-wavelet construction is used in Tables~\ref{tab:receiver_level_performance},~\ref{tab:synthetic_ofdm_holdout}, and ~\ref{tab:measured_channel_holdout}. This is
a deliberate choice, since it isolates the effect of the transform family, but it is also the
specific limitation on naive cross-domain transfer identified in Section~IV-F: per-component
real-valued processing does not enforce phase coherence between the I and Q branches, and a
phase-coherent extension such as the DTCWT is the natural remedy.

Tables~\ref{tab:receiver_level_performance} and ~\ref{tab:complexity_realtime} report the communication-oriented comparison of the best DWT-, SWT-, and WPT-based denoising configurations obtained from the declared parameter search space for the OFDM--Middleton Class A impulsive-noise benchmark. The results show that SWT provides the strongest waveform- and receiver-level performance. This behaviour is physically consistent with the undecimated and approximately shift-invariant structure of SWT. Middleton Class A interference is temporally impulsive and spectrally broadband; hence, preserving full-length detail coefficients across scales improves the localization of impulsive events and stabilizes coefficient-domain thresholding. As a result, SWT suppresses interference-dominated components more effectively while preserving the constellation structure required for reliable demodulation. By contrast, DWT is critically sampled and therefore computationally efficient, but the downsampling operation makes the coefficient representation more sensitive to the sample-level position of impulsive disturbances. WPT provides a richer packet-tree decomposition, but in this benchmark the additional spectral granularity does not translate into a reliable receiver-level gain, suggesting that shift robustness is more important than finer subband partitioning for broadband impulsive-noise suppression.

The statistical analysis shown in Table~\ref{tab:friedman_results} supports these interpretations. Paired Friedman tests show that transform choice significantly affects denoising gain, MSE/NMSE, BER, EVM, and processing time. The post-hoc Wilcoxon signed-rank tests with Holm correction further show that SWT significantly outperforms both DWT and WPT in denoising gain, reconstruction error, BER, and EVM. The DWT-WPT BER comparison, however, is not statistically significant, indicating that the small numerical BER difference between these two transforms should not be interpreted as a reliable receiver-level advantage. This reinforces the need to report both waveform-domain and receiver-domain metrics: BER captures hard-decision reliability, whereas EVM and NMSE reveal residual constellation and waveform distortion that may affect higher-order modulation, synchronization, channel estimation, or coding performance.

The computational results follow the expected structural ordering. DWT has the lowest measured runtime and the smallest FLOP proxy because its critically sampled filter-bank structure reduces the number of processed coefficients across successive scales. WPT incurs a higher arithmetic burden than DWT because both approximation and detail branches are decomposed at every level, and SWT incurs the same per-level burden because it avoids downsampling and retains full-length coefficient sequences at every scale. Here, SWT attains the largest tabulated proxy because the configuration selected for it uses one decomposition level more than that selected for WPT. Feasibility against a given receiver numerology can be read directly from the minimum frame budgets reported in Table~\ref{tab:complexity_realtime}. Overall, the results demonstrate a clear accuracy-complexity trade-off: SWT is preferable when receiver robustness under impulsive noise is prioritized, whereas DWT is more suitable when latency, memory, and implementation cost dominate the receiver design. Summing up, Tables~\ref{tab:receiver_level_performance},\ref{tab:complexity_realtime} and ~\ref{tab:friedman_results} establish that selecting an appropriate transform is not a single default design choice and should be treated as an application-dependent optimization problem.

\begin{figure*}[!t]
\centering
\includegraphics[width=\textwidth]{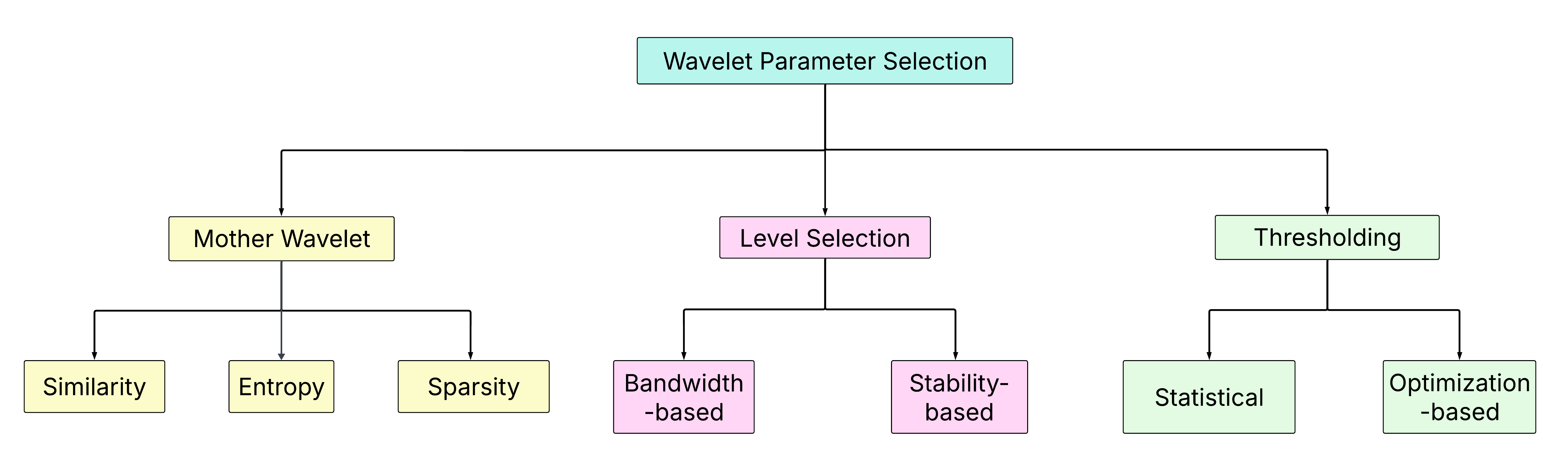}
\captionsetup{justification=centering}
\caption{Taxonomy of wavelet-parameter selection strategies discussed in this review.}
\label{fig:Taxonomy Tree}
\end{figure*}

Two aspects of this benchmark are stated explicitly so that the reported figures can be checked. First, the complexity results are a software benchmark on a general-purpose processor
and establish the relative ordering of the transform families rather than absolute embedded performance. Second, the real-time indicator $\mathcal{F}_{\mathrm{RT}}$ of Table~\ref{tab:perf_metrics} compares the
measured per-frame latency $t_{\mathrm{frame}}$ against the frame budget
$T_{\mathrm{frame}}=N/f_{\mathrm{s}}$ set by the receiver numerology. Rather than report a
feasibility flag tied to one numerology, Table~\ref{tab:complexity_realtime} reports the minimum frame budget at which each
transform becomes feasible, which by definition equals its measured latency. A reader can
therefore evaluate feasibility for any target numerology directly, and the ordering DWT, WPT, SWT
indicates which family violates a tightening budget first.

The FLOP proxy of Table~\ref{tab:complexity_realtime} is a deterministic multiply-accumulate count rather than a
measured instruction count, and is therefore reproducible from the configuration alone. For a
frame of $N$ samples, an analysis filter of length $L_{f}$, and a decomposition depth $J$, the
low-pass and high-pass branches at each stage give
\begin{equation*}
{\color{black}
C_{\mathrm{frame}}^{\mathrm{DWT}} = 2 L_{f} N \sum_{j=1}^{J} 2^{-j},
\qquad
C_{\mathrm{frame}}^{\mathrm{SWT}} = C_{\mathrm{frame}}^{\mathrm{WPT}} = 2 L_{f} N J ,}
\end{equation*}
the decimated form reflecting the halving of the coefficient count at each successive DWT scale and the undecimated and full-tree forms both retaining $N$ coefficients per level. The
values tabulated correspond to $N=4480$ samples, $L_{f}=8$, and decomposition depths $J=3$ for
DWT, $J=3$ for SWT, and $J=2$ for WPT, these being the unique parameter set consistent with all three tabulated proxy values.

To compare wavelet choices and to support parameter selection, many studies report quantitative measures of denoising performance. Fig.~\ref{fig:Taxonomy Tree} summarizes the parameter-selection taxonomy adopted in this review. When a reference signal $q[n]$ is available or when a trusted surrogate reference is constructed, fidelity and similarity measures such as signal to noise ratio, peak signal to noise ratio, correlation coefficients, and percentage root mean square difference are commonly used. Table~\ref{tab:perf_metrics} summarises these measures and their mathematical definitions. In scenarios where a clean reference is not available, performance is typically assessed through task level criteria, and the parameter selection sections that follow therefore distinguish between reference based waveform metrics and task performance based evaluation.

\subsection{PROPOSED UNIFIED DECISION FRAMEWORK FOR WAVELET DENOISING}

\begin{figure*}[!t]
\centering
\includegraphics[width=\textwidth,height=0.9\textheight,keepaspectratio]{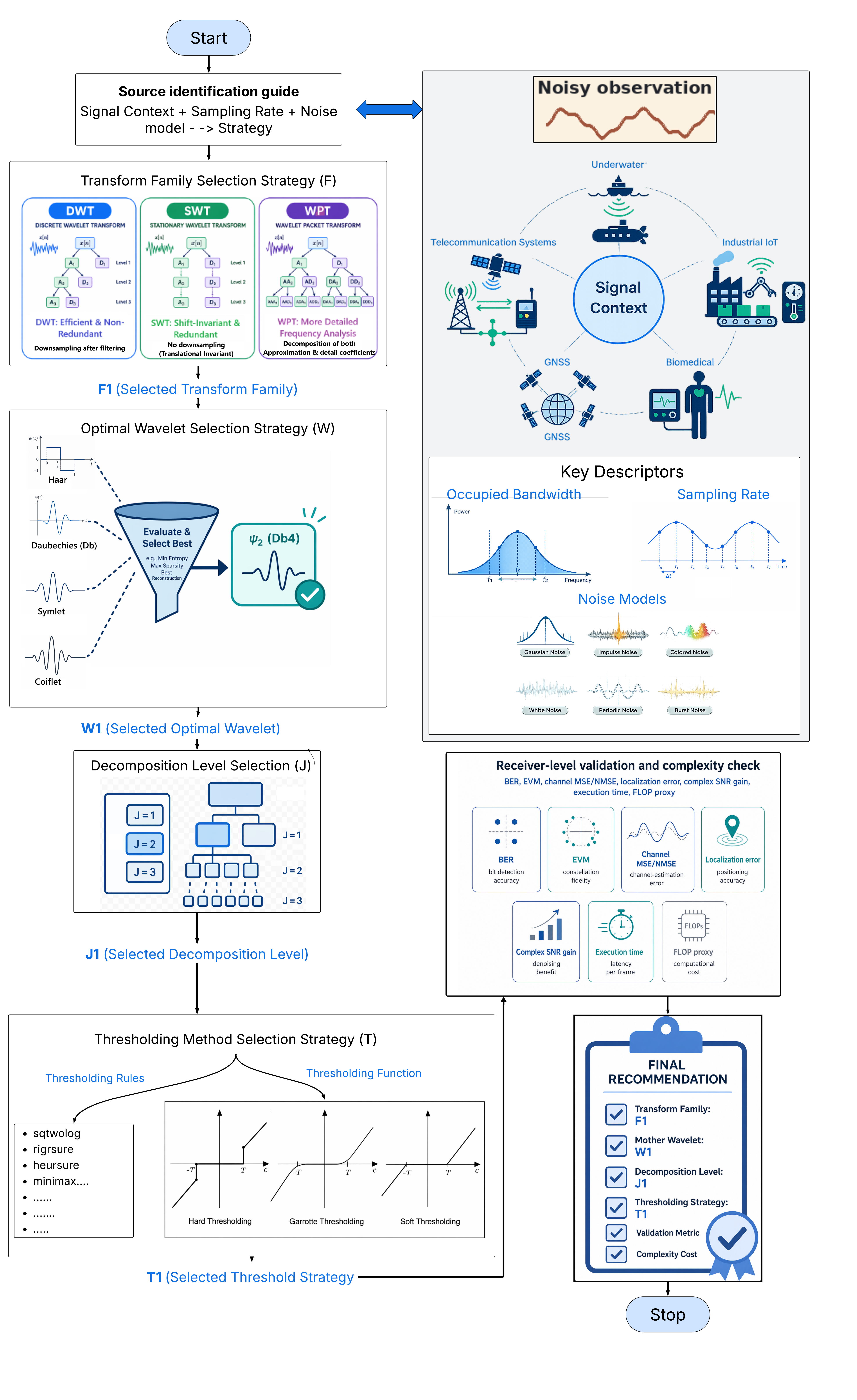}
\caption{Unified Decision Framework for Selection of Wavelet Parameters in Communication Systems}
\label{fig:unified_decision_framework}
\end{figure*}

Wavelet denoising is often presented as a generic preprocessing operation, yet in practical communication and sensing systems its effectiveness is governed by a set of tightly coupled design decisions rather than by the mere use of a wavelet transform. Fig.\ref{fig:unified_decision_framework} proposes a decision framework for implementing wavelet denoising in communication systems. The proposed decision framework treats denoising as a receiver- and signal-dependent optimization problem in which the final outcome depends jointly on four parameters: the transform family, the mother wavelet, the decomposition level, and the thresholding strategy. Rather than selecting these quantities independently or by precedent, the framework organizes their selection around measurable descriptors of the observed signal, namely signal context, sampling rate, occupied bandwidth, and the governing disturbance model. In this sense, the framework is not an abstract taxonomy alone, but a structured decision pathway that maps signal and system knowledge into a reproducible denoising configuration. 

\par{The motivation for such a framework is especially strong in communication-oriented settings. In burst-mode receivers, synchronization pipelines, OFDM systems, GNSS processing chains, underwater acoustic links, industrial IoT platforms, and biomedical instrumentation, the disturbance rarely appears in a form that is equally well handled by a single default wavelet choice. Impulsive interference, for example, is highly localized in time yet broadband in frequency, whereas in the wavelet domain it tends to concentrate into a small number of large-magnitude coefficients. This is precisely why the denoising benefit depends not only on thresholding itself, but on whether the selected transform preserves temporal alignment, whether the chosen basis matches the signal morphology, whether the decomposition depth isolates the correct dyadic scales, and whether the shrinkage rule is appropriate for the assumed noise statistics. Because these conditions are interdependent, the parameters are best resolved jointly rather than by precedent.}

An important feature of the proposed framework is that the design variables should not be interpreted as independent. In particular, mother-wavelet selection and decomposition-level selection are mutually coupled. The mother wavelet determines the analysis filters, support length, regularity, symmetry, and boundary behaviour, all of which influence the approximation and detail coefficients obtained at each level. Conversely, the decomposition level determines the dyadic scale range over which a candidate wavelet is evaluated. Therefore, when no reliable prior wavelet or decomposition level is known, these two quantities should be resolved as a coupled pair rather than as two independent sequential choices. In such cases, the admissible search space should be defined over (($\psi$,J)) pairs, where $\psi$ denotes a candidate mother wavelet and J denotes a valid decomposition level constrained by signal length, filter length, sampling frequency, occupied bandwidth, and computational budget.

At a high level, Fig.\ref{fig:unified_decision_framework} is read as a flow bounded by a Start and a Stop terminator. The pathway opens with a source-identification step that fuses the signal context, sampling rate, and noise model into an initial strategy. These inputs are supplied by the signal and descriptor column on the right of the figure, which links a representative noisy observation to its application domain, whether telecommunication, GNSS and localization, underwater acoustic, industrial IoT, or biomedical, and to the key descriptors of occupied bandwidth, sampling rate, and assumed noise model. From this characterization, the framework proceeds through four sequential selection stages, each emitting an explicit intermediate decision: the transform family (F1), the optimal mother wavelet (W1), the decomposition level (J1), and the thresholding strategy (T1). The resulting configuration \(\{F1,W1,J1,T1\}\) is then submitted to the receiver-level validation and complexity check, where it is assessed against task-oriented endpoints such as BER, EVM, channel-estimation error (MSE/NMSE), localization error, and complex SNR gain, alongside implementation-cost measures such as execution time and a FLOP proxy. The validated choices are consolidated into a final recommendation before the
pathway terminates. The per-stage selection logic, that is, how each of F1, W1, J1, and T1 is derived from the signal and receiver descriptors, is developed in Section~IV-G in a single consolidated  (Fig.~\ref{fig:Decision_Rule}).

\subsection{TECHNIQUES FOR SELECTION OF THE OPTIMAL WAVELET}

Across the reviewed wavelet based denoising literature, predominantly published between 2020 and 2025, techniques for selecting an optimal mother wavelet can be grouped into three broad categories. The first category is empirical or task performance driven selection, in which a small candidate set of wavelets is evaluated end to end and the wavelet that yields the best task outcome is retained. The second category is metric driven selection, in which the wavelet is chosen by optimizing an explicit criterion computed either from the wavelet coefficients or from a reference and reconstructed signal pair. The third category is optimization oriented selection, in which wavelet choice is embedded in a broader tuning problem that may also include decomposition depth and threshold parameters. This subsection focuses first on empirical selection, since it remains common in practice and it directly motivates the need for more systematic criteria.
\vspace{-4mm}
\subsubsection{Empirical or task performance driven}
\label{subsubsec:mw_empirical}

Empirical selection is widely used in application oriented studies because it is simple to implement and it directly targets the objective of interest. The typical procedure is to specify a candidate set of wavelets, often drawn from widely used compactly supported families, and then to select the wavelet that maximizes performance after denoising. In \cite{ElBouny2020QRS}, stationary wavelet transform denoising was combined with the Teager energy operator and Coiflet 1 was retained from a small pool because it provided maximal resemblance with the dominant transient structure of interest. In \cite{Lowast2020WristPulse}, variants of Daubechies, Symlet, and biorthogonal wavelets were compared and the selected wavelets were those yielding improved signal to noise ratio, lower root mean square error, and smaller absolute error. In \cite{Potdar2021PCG}, candidate wavelets spanning Haar, Daubechies, Symlet, and Coiflet families were evaluated and sym20 was selected because it maximized the signal to noise ratio while minimizing the root mean square error. In another study involving the detection of hoist rope faults, sym8 emerged as the optimal wavelet as it maximises SNR~\cite{li2022hoist}. 

A similar empirical approach appears in studies involving neural signals, where wavelets from families such as Daubechies, Symlet, Coiflet, biorthogonal, discrete Meyer, and Haar are evaluated using denoising performance measures including signal to noise ratio, root mean square error, percentage root mean square difference, and correlation based criteria \cite{Suhail2020EEGCognitive,Atangana2020EEGMotherWavelet,Baldazzi2020NeuralDenoising,daud2022denoising,cheng2020optimal}. In these works, the wavelet is typically selected by ranking candidates according to a combination of metrics computed after reconstruction. Empirical selection has also been adopted in application settings where downstream processing is correlation based. For example, in a study on time delay estimation in underwater reverberant environments, candidate wavelets were restricted to those yielding high cross correlation with the signal of interest \cite{ranjeet2011retained}. In another study aimed at real time wave filtering for dynamic positioning of marine vessels, sym6 was selected empirically because it provided superior real time filtering of wave frequency components \cite{Liu2022MarineVessel}. For structural health monitoring strain measurements, the mother wavelet was chosen according to the most favourable combination of high signal to noise ratio, low root mean square error, and strong correlation between the denoised and reference signals \cite{Sharie2020GPSFOM}.

\begin{figure}[!t]
\centerline{\includegraphics[width=3.5in]{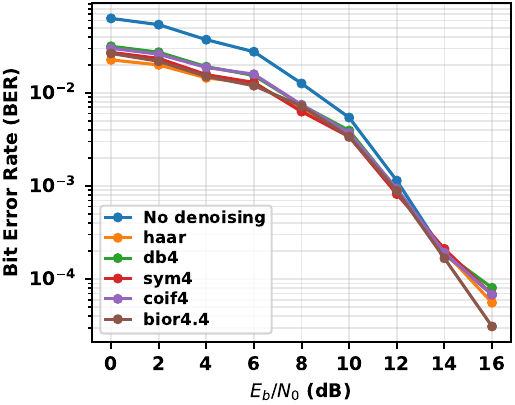}}
\caption{BER versus $E_b/N_0$ for representative mother-wavelet
families under Middleton Class A impulsive noise. The decomposition
level and thresholding configuration are fixed in this experiment so
that the wavelet-family effect can be isolated.}
\label{fig:BER vs $E_b/N_0$}
\end{figure}

Fig.~\ref{fig:BER vs $E_b/N_0$} shows the bit error rate (BER) of an OFDM receiver as a function of $E_b/N_0$ when the received waveform is corrupted by Middleton Class~A impulsive noise and then processed by a wavelet-based denoising block prior to demodulation. Relative to the no-denoising baseline, all tested mother wavelets improve BER because the impulsive disturbance produces sparse, high-magnitude wavelet coefficients that can be isolated and suppressed more effectively than in the time domain or the Fourier domain. Although the differences among wavelet families are modest, they are not negligible, indicating that the choice of mother wavelet influences how accurately impulsive events are localized in the wavelet domain and, consequently, how effectively the interference can be removed without distorting the useful OFDM signal structure.

It should be noted that Fig.~\ref{fig:BER vs $E_b/N_0$} is a controlled mother-wavelet sensitivity experiment rather than a complete wavelet-parameter optimization study. In this simulation, the decomposition level and thresholding configuration are fixed so that the effect of changing the mother wavelet can be examined without simultaneously varying other wavelet parameters. The improvement observed for Haar, Daubechies, Symlet, Coiflet, and biorthogonal wavelets indicates that the qualitative BER reduction under Middleton Class A impulsive noise is not specific to a single mother-wavelet family. Nevertheless, the relative ordering among the candidate wavelets should not be interpreted as universal, since it depends on the candidate set, decomposition level, thresholding rule, threshold scale, OFDM numerology, sampling rate, impulsive-noise parameters, and receiver performance metric. Thus, Fig.~~\ref{fig:BER vs $E_b/N_0$} supports the robustness of the qualitative denoising trend across representative
compact wavelets, while the final mother-wavelet choice remains a
task-dependent design variable.

A canonical communication setting where wavelet denoising acts directly on a receiver estimate is pilot-aided channel estimation in OFDM systems. Here the least-squares (LS) or DFT-based channel estimate is shrunk in the wavelet domain before interpolation and equalisation. A recent OFDM channel-estimation study compared seven wavelet families for denoising the LS channel-impulse-response estimate and found the Meyer (dmey) wavelet to be the strongest across four standard channel models and four modulation orders~\cite{R9}. Earlier IoT-oriented work denoised the LS estimate with an improved threshold function and reported BER and MSE gains over plain LS, DFT-domain denoising, and conventional soft-threshold wavelet denoising~\cite{R10}. That work was later extended with a two-level DWT denoising stage and a distance-decision analysis evaluated over a TDL-A channel~\cite{R11}. A complementary line couples wavelet thresholding with compressed-sensing recovery for sparse OFDM channels, with the threshold tied to the estimated noise level~\cite{R12}. In each case the wavelet basis is chosen by linking the candidate transform to estimation- and detection-level endpoints such as channel-estimation MSE and BER rather than to waveform fidelity alone, which is exactly the receiver-oriented selection criterion advocated here.

Although empirical comparisons provide valuable application specific evidence, they do not by themselves constitute a general methodology for wavelet selection because the outcome depends on the candidate set, the decomposition depth, the thresholding configuration, and the evaluation metric. This limitation is particularly relevant when transferring designs across domains. In telecommunications and networking receivers, the denoising objective is often coupled to downstream tasks such as synchronization, channel estimation, detection, and interference mitigation. In such cases, empirical selection is most informative when it links the chosen wavelet to receiver oriented objectives, and when it reports sensitivity to sampling rate, signal bandwidth, and non Gaussian disturbances. These considerations motivate the metric driven and optimization oriented approaches reviewed in the subsequent subsections. 

A further receiver-oriented example is cognitive-radio spectrum sensing, a well-established communication use of wavelet denoising. Because energy detection alone cannot reliably separate a weak primary-user signal from noise at low SNR, wavelet denoising is applied to the received signal before detection. Representative 2020–2025 studies include a two-stage model that performs wavelet denoising prior to energy detection to improve primary-user detection at low SNR~\cite{R13}, and a lifting-wavelet-assisted expectation-maximisation method for joint parameter estimation and detection in cooperative sensing~\cite{R14}.Together these put the wavelet-basis and threshold-selection decisions organised in this review to work in the spectrum-sensing receiver, evaluated through detection-theoretic endpoints such as probability of detection and probability of false alarm rather than waveform fidelity alone. 
Table~\ref{tab:perf_metrics} summarizes commonly used metrics employed in the reviewed literature to compare candidate mother wavelets and decomposition levels when a clean reference signal is available.

\begin{table*}[!t]
\caption{Performance metrics and their mathematical definitions.}
\label{tab:perf_metrics}
\centering
\footnotesize
\setlength{\tabcolsep}{3pt}
\renewcommand{\arraystretch}{1.08}
\begin{tabular}{p{36pt} p{130pt} p{322pt}}
\hline
\textbf{Sl.\ no.} &
\textbf{Performance metric} &
\textbf{Equation} \\
\hline

1 &
Input signal-to-noise ratio \par $\mathrm{SNR}_{\mathrm{in}}$ (dB) &
$\displaystyle
\mathrm{SNR}_{\mathrm{in}}(\mathrm{dB})=
10\log_{10}\!\left(
\frac{\sum_{n=0}^{N-1} q^{2}[n]}
{\sum_{n=0}^{N-1} (p[n]-q[n])^{2}}
\right)
$ \\

2 &
Output signal-to-noise ratio \par $\mathrm{SNR}_{\mathrm{out}}$ (dB) &
$\displaystyle
\mathrm{SNR}_{\mathrm{out}}(\mathrm{dB})=
10\log_{10}\!\left(
\frac{\sum_{n=0}^{N-1} q^{2}[n]}
{\sum_{n=0}^{N-1} (q[n]-q_{1}[n])^{2}}
\right)
$ \\

3 &
Peak signal-to-noise ratio \par $\mathrm{PSNR}$ (dB) &
$\displaystyle
\begin{aligned}
q_{\max} &= \max_{0 \le n \le N-1}|q[n]|, \\[1pt]
\mathrm{MSE} &= \frac{1}{N}\sum_{n=0}^{N-1}(q[n]-q_{1}[n])^{2}, \quad
\mathrm{PSNR}(\mathrm{dB}) =
10\log_{10}\!\left(\frac{q_{\max}^{2}}{\mathrm{MSE}}\right)
\end{aligned}
$ \\

4 &
Cross-correlation coefficient \par $r_{q,q_{1}}$ (Pearson) &
$\displaystyle
\begin{aligned}
\mu_{q} &= \frac{1}{N}\sum_{n=0}^{N-1} q[n], \\[1pt]
\mu_{q_{1}} &= \frac{1}{N}\sum_{n=0}^{N-1} q_{1}[n], \\[1pt]
r_{q,q_{1}} &=
\frac{\sum_{n=0}^{N-1}(q[n]-\mu_{q})(q_{1}[n]-\mu_{q_{1}})}
{\sqrt{\left(\sum_{n=0}^{N-1}(q[n]-\mu_{q})^{2}\right)
\left(\sum_{n=0}^{N-1}(q_{1}[n]-\mu_{q_{1}})^{2}\right)}}
\end{aligned}
$ \\

5 &
Normalised correlation coefficient \par $\gamma_{q,q_{1}}$ (energy normalised) &
$\displaystyle
\gamma_{q,q_{1}}=
\frac{\sum_{n=0}^{N-1} q[n]\,q_{1}[n]}
{\sqrt{\left(\sum_{n=0}^{N-1} q^{2}[n]\right)
\left(\sum_{n=0}^{N-1} q_{1}^{2}[n]\right)}}
$ \\

6 &
Percentage root mean square difference \par $\mathrm{PRD}$ &
$\displaystyle
\mathrm{PRD}(\%)=
100\times
\sqrt{
\frac{\sum_{n=0}^{N-1}(q[n]-q_{1}[n])^{2}}
{\sum_{n=0}^{N-1} q^{2}[n]}
}
$ \\

7 &
Bit error rate \par $\mathrm{BER}$ &
$\displaystyle
\mathrm{BER} =
\frac{1}{N_{b}}\sum_{i=1}^{N_{b}}
\mathbf{1}\!\left\{\hat{b}_{i}\ne b_{i}\right\}
$ \\

8 &
Error vector magnitude \par $\mathrm{EVM}_{\mathrm{rms}}$ (\%) &
$\displaystyle
\mathrm{EVM}_{\mathrm{rms}}(\%) =
100
\sqrt{
\frac{\sum_{m=1}^{M}\left|s_{m}-\hat{s}_{m}\right|^{2}}
{\sum_{m=1}^{M}\left|s_{m}\right|^{2}}
}
$ \\

9 &
Computational cost per frame \par $C_{\mathrm{frame}}$ &
$\displaystyle
C_{\mathrm{frame}} =
N_{\mathrm{mult}} + N_{\mathrm{add}} + N_{\mathrm{comp}}
$ \\

10 &
Latency per frame \par $t_{\mathrm{frame}}$ &
$\displaystyle
t_{\mathrm{frame}} =
t_{\mathrm{end}} - t_{\mathrm{start}}
$ \\

11 &
Real-time feasibility \par $\mathcal{F}_{\mathrm{RT}}$ &
$\displaystyle
\mathcal{F}_{\mathrm{RT}} =
\begin{cases}
1, & t_{\mathrm{frame}} \le T_{\mathrm{frame}}, \\[2pt]
0, & t_{\mathrm{frame}} > T_{\mathrm{frame}} .
\end{cases}
$ \\

12 &
Normalised mean square error \par $\mathrm{NMSE}$ &
$\displaystyle
\mathrm{NMSE} =
\frac{\sum_{m=1}^{M}\bigl|H_{m}-\hat{H}_{m}\bigr|^{2}}
     {\sum_{m=1}^{M}\bigl|H_{m}\bigr|^{2}}
$ \\

13 &
Complex SNR gain \par $\Delta\mathrm{SNR}_{\mathbb{C}}$ &
$\displaystyle
\Delta\mathrm{SNR}_{\mathbb{C}} =
\mathrm{SNR}^{\mathbb{C}}_{\mathrm{out}} - \mathrm{SNR}^{\mathbb{C}}_{\mathrm{in}}
$ \\

\hline
\end{tabular}
\vspace{0.5ex}
\begin{minipage}{0.96\textwidth}
\footnotesize
\textit{Note:} In entry 12, $H_m$ denotes the true complex channel coefficient on subcarrier $m$, $\hat{H}_m$ its estimate, and $M$ the number of subcarriers; NMSE is scale invariant and is reported per frame, then averaged. Entry 13 is the complex-valued counterpart of entries 1 and 2, the modulus replacing the real square so that in-phase and quadrature error contribute jointly. It is evaluated on the complex baseband waveform before and after denoising, so it measures the disturbance removed from the received signal and is distinct from the channel-estimation error of entry 12.Where NMSE is reported for the transform-family benchmark of Tables 6 and 8, it denotes the corresponding waveform-domain reconstruction error, computed with $q[n]$ and $q_{1}[n]$ in place of $H_m$ and $\hat{H}_m$.
\end{minipage}
\end{table*}

\subsubsection{Metrics in Time, Frequency, or Wavelet Domain}
\label{subsubsec:mw_metrics}

A more systematic group of techniques selects the mother wavelet by optimizing explicit signal level or coefficient level criteria that quantify how effectively a candidate basis separates structured components from noise. These criteria provide an objective alternative to purely empirical ranking, and they are also convenient when a clean reference signal is unavailable. In telecommunications and networking pipelines, this class of methods is particularly attractive because the receiver often has access to some form of structural prior, such as a known preamble, a known pulse shaping filter, or a parametric model of the channel impulse response. In such cases, wavelet selection can be framed as selecting the basis that yields a compact representation of the desired structure, while spreading noise and interference across coefficients.

A practical tutorial interpretation is as follows. Let us consider $\psi_k[n]$ as the discrete time samples of a candidate mother wavelet $k$, and $s[n]$ as a known or partially known reference structure in the receiver, such as a pulse shaping waveform, a preamble, or an estimated channel impulse response. A similarity based wavelet selection rule can then be implemented by maximizing a normalized correlation measure
\begin{equation}
\rho_k = \frac{\sum_{n} s[n]\psi_k[n]}{\sqrt{\left(\sum_{n} s^2[n]\right)\left(\sum_{n} \psi_k^2[n]\right)}},
\label{eq:wavelet_similarity_comms}
\end{equation}
or by maximizing a task specific criterion computed from wavelet coefficients that are expected to be sparse for the desired structure.

\begin{figure*}[!t]
\centering
\subfloat[]{%
    \includegraphics[width=0.49\textwidth]{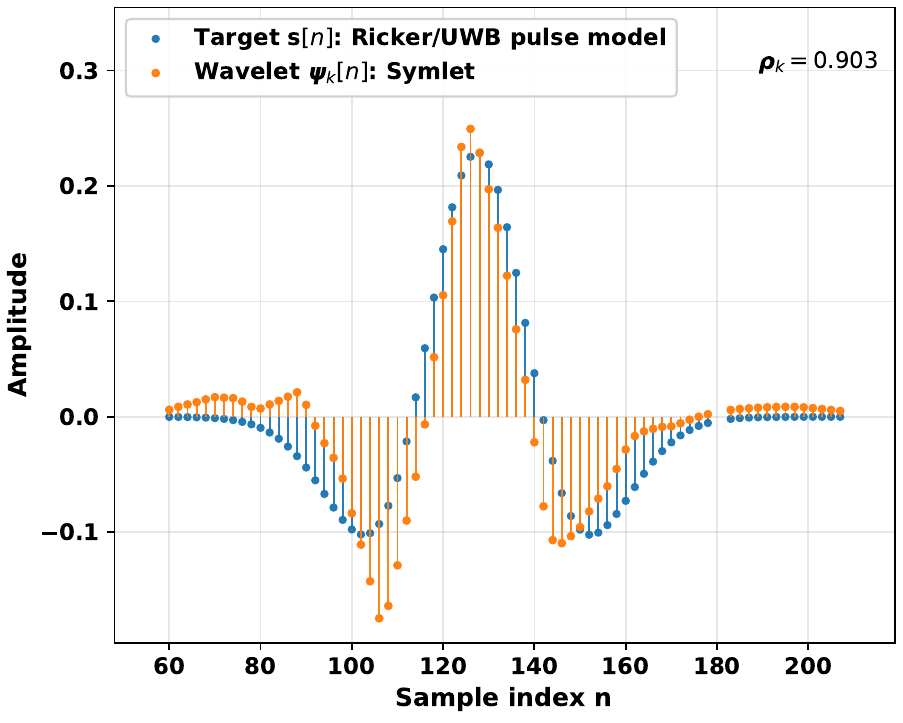}%
    \label{fig:shape_matching_high}%
}\hfill
\subfloat[]{%
    \includegraphics[width=0.49\textwidth]{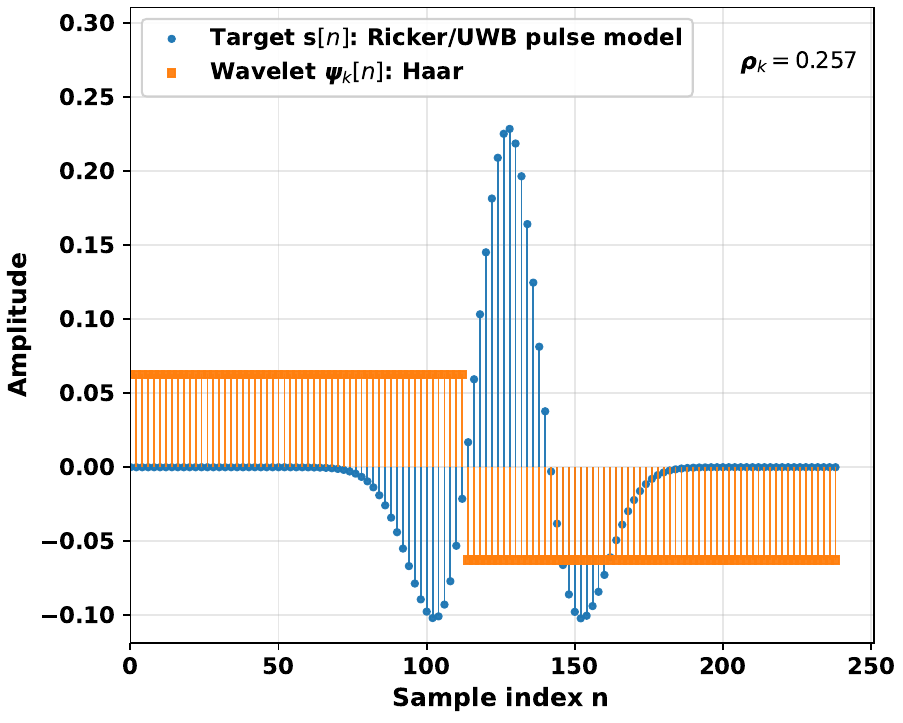}%
    \label{fig:shape_matching_low}%
}
\caption{Conceptual illustration of shape-matching-based selection of mother wavelet: (a) high similarity and (b) low similarity.}
\label{fig:shape_matching_concept}
\end{figure*}

Fig.\ref{fig:shape_matching_concept} illustrates the principle of shape-matching-based mother-wavelet selection. Using the normalized similarity measure in Eqn.(6), a Symlet wavelet is seen to align much more closely with the target transient $s[n]$ than a Haar wavelet, resulting in a substantially larger similarity score. The figure therefore demonstrates that wavelets whose shapes better match the underlying signal structure are more suitable candidates for sparse and faithful representation. This viewpoint motivates why compactly supported wavelet families such as Daubechies and Symlet frequently appear as candidates in practice. Many communication waveforms and channel induced transients are dominated by localized changes, smooth transitions, and finite duration structures, and wavelets with appropriate regularity and vanishing moments can improve concentration of energy into a small subset of coefficients. The studies reviewed below primarily apply these ideas in non communication domains as well, but the metrics themselves translate directly to communication receiver settings because they quantify representation compactness and structural match.

\paragraph{Similarity and energy based metrics}

An important class of wavelet selection criteria employs similarity derived measures. A representative example is the figure of merit (FOM) framework used for weak GPS signal denoising in \cite{Sharie2020GPSFOM}, where DWT based denoising was evaluated across a broad set of candidate wavelet families and inter signal similarity was quantified by three criteria denoted $P$, $Q$, and $MS$. In that formulation, $P$ evaluates resemblance between global and local time domain deviations, $Q$ evaluates similarity via a norm defined in Sobolev space, and $MS$ models a fictitious system linking the signals of interest and estimates similarity via the group delay and magnitude of the resulting frequency response. The composite FOM is defined as \cite{Sharie2020GPSFOM}
\begin{equation}
\mathrm{FOM} = \frac{P \times MS}{Q}.
\end{equation}
The mother wavelet that maximizes this composite criterion is deemed optimal. A related GPS study uses an FOM formulation based on $P$ and $Q$ criteria to select the optimal wavelet \cite{Moradi2023WeakGPS}. These examples are particularly relevant to communication receiver settings because weak signal acquisition and tracking are often correlation driven, and similarity criteria directly reward wavelets that preserve the signal structure required by the downstream correlator. These weak-signal acquisition pipelines are themselves a representative communication use of wavelet denoising. Because the received GNSS signal sits below the noise floor after long propagation, wavelet denoising is applied to the post-correlation and integration output to raise the effective SNR and improve acquisition sensitivity. A recent study in this setting applies wavelet denoising to improve receiver acquisition sensitivity for a regional navigation receiver~\cite{R15}. Together with the FOM-based studies above [74], [75], these pipelines put the wavelet-basis, decomposition-level, and threshold-selection decisions of this review to work in the receiver correlation domain, evaluated through SNR, detection probability, and acquisition sensitivity.

A related correlation-based criterion has been applied to the received underwater acoustic waveform itself. Using a sparse tunable-Q wavelet transform, the subbands kept for reconstruction are identified by a correlation-coefficient criterion that favours the components most similar to the signal of interest, with the transform Q-factor and this similarity criterion together governing the noise reduction~\cite{R16}. As in the weak-signal acquisition case above, this is a correlation-driven selection rule, applied here to a communication-adjacent underwater link.

In a study on seismic signals, He et al.\ computed a decomposition stability index from wavelet packet decomposed signals \cite{he2022novel}. The stability index is expressed as
\begin{equation}
D = \frac{\mathrm{mean}_{\mathrm{relevance}}}{\left(\mathrm{variance}_{\mathrm{relevance}}\right)^{2}},
\end{equation}
where the relevance term is evaluated from correlation and variance contribution rates between the original signal and reconstructed versions obtained through inverse wavelet packet transform. Although demonstrated on seismic data, the same logic applies to receiver pipelines in which stability across reconstructions is desirable, for example when denoising precedes estimation tasks that are sensitive to distortions.

Energy based criteria are also widely used. In partial discharge denoising studies \cite{Li2020PDNovelShrinkage,Zhou2020PDBlockThresh}, signal energy and subband signal to noise ratio are used to identify wavelets that concentrate discharge energy into a small subset of levels while reducing noise elsewhere. In rotating machinery denoising, Brusa et al.\ evaluated candidate wavelets using kurtosis and the visual clarity of localized defect signatures in the time frequency representation \cite{Brusa2023BearingDenoise}. A related energy surrogate measure is the noise envelope area (NEA) proposed in \cite{Chen2020HeartSoundNEA}. In that work, a clean reference is not available, so classical signal to noise ratio is not used as a primary metric. Instead, the residual noise envelope area after denoising is integrated and the wavelet configuration yielding the smallest NEA is selected. The candidate family is restricted to Daubechies wavelets (db2, db3, \ldots) and db2 is reported as optimal \cite{Chen2020HeartSoundNEA}.

To incorporate computational constraints, Jang et al.\ introduced an efficiency index $\mathrm{EFI}(j,k)$ in the context of denoising DCG signals \cite{Jang2021DCG}. The wavelet is selected by maximizing $\mathrm{EFI}(j,k)$ over a wavelet space of 115 wavelets derived from Daubechies, Fej\'er-Korovkin, biorthogonal spline, reverse biorthogonal spline, and Symlet families, under the constraint that the approximation band covers 0 to 5. The index is defined as \cite{Jang2021DCG}
\begin{equation}
\mathrm{EFI}(j,k) = \frac{\mathrm{SNR}_{j,k}}{\eta_c(k)},
\end{equation}
where $\mathrm{SNR}_{j,k}$ is calculated as per Table~\ref{tab:perf_metrics} and $\eta_c(k)$ represents execution complexity for wavelet index $k$,
\begin{equation}
\eta_c(k) = \log_{10}\Bigl(\bigl(q_l-\mathrm{len}(m[k])\bigr)\times \mathrm{len}(m[k])\Bigr),
\end{equation}
with $q_l$ denoting the signal length and $\mathrm{len}(m[k])$ denoting the mother wavelet length. 

Table~\ref{tab:radioml_fom_efi_wavelet} illustrates the variation of FOM and EFI across different mother wavelets evaluated on RadioML 2018.01A dataset~\cite{R39}. FOM evaluates the fidelity of the denoised signal relative to the reference waveform through time- and frequency-domain similarity components, whereas EFI evaluates the denoising SNR obtained per wavelet-complexity cost. The FOM-based ranking is dominated by higher-order Daubechies and Coiflet wavelets, with db37, coif16, coif14, db24, and db26 occupying the leading positions. These wavelets also provide high complex SNR gains, generally above 6.5 dB for the leading FOM-ranked candidates. This trend indicates that, for RadioML complex baseband waveforms, smoother and longer-support wavelets preserve the modulation-dependent I/Q trajectory and spectral structure more effectively than very compact low-order bases. Physically, this is expected because communication waveforms are not merely isolated time-domain impulses; they contain pulse-shaped transitions, phase-bearing constellation trajectories, and spectrally structured components. A wavelet with greater regularity and a larger number of vanishing moments can represent these smooth baseband structures with lower reconstruction distortion after coefficient shrinkage.

The EFI-based ranking exhibits a different pattern. The leading EFI candidates are shorter or moderate-length wavelets such as bior3.3, bior3.5, bior3.7, bior3.9, db8, sym11, db10, sym9, sym10, and sym8. These wavelets achieve higher EFI because the EFI criterion penalizes the computational burden associated with longer wavelet filters. Consequently, EFI favours wavelets that provide a favourable SNR-efficiency balance rather than the highest absolute waveform fidelity. The complex SNR gains of the EFI-leading wavelets remain positive, but they are generally lower than those of the FOM-leading wavelets. This confirms that EFI is best interpreted as an efficiency-oriented design criterion, particularly relevant for low-latency or resource-constrained receiver implementations.

\begin{table*}[!t]
\centering
\caption{FOM- and EFI-based mother-wavelet evaluation on RadioML 2018.01A complex I/Q communication waveforms.}
\label{tab:radioml_fom_efi_wavelet}
\renewcommand{\arraystretch}{1.08}
\scriptsize
\begin{tabular}{c c l c c c c}
\hline
\textbf{Selection criterion} &
\textbf{Rank} &
\textbf{Mother wavelet} &
\textbf{Filter length} &
\textbf{FOM} &
\textbf{EFI} &
\textbf{Complex SNR gain (dB)} \\
\hline
\multicolumn{7}{l}{\textit{FOM-based mother-wavelet ranking}} \\
FOM & 1  & db37   & 74 & $4.123 \pm 0.094$ & $3.271 \pm 0.031$ & $6.830 \pm 0.158$ \\
FOM & 2  & coif16 & 96 & $4.041 \pm 0.097$ & $3.208 \pm 0.031$ & $6.840 \pm 0.164$ \\
FOM & 3  & coif14 & 84 & $4.017 \pm 0.098$ & $3.229 \pm 0.032$ & $6.777 \pm 0.168$ \\
FOM & 4  & db24   & 48 & $4.007 \pm 0.096$ & $3.333 \pm 0.035$ & $6.566 \pm 0.173$ \\
FOM & 5  & db26   & 52 & $4.006 \pm 0.094$ & $3.325 \pm 0.034$ & $6.632 \pm 0.170$ \\
FOM & 6  & dmey   & 62 & $3.985 \pm 0.096$ & $3.309 \pm 0.032$ & $6.792 \pm 0.162$ \\
FOM & 7  & db35   & 70 & $3.980 \pm 0.096$ & $3.274 \pm 0.031$ & $6.775 \pm 0.161$ \\
FOM & 8  & coif12 & 72 & $3.979 \pm 0.097$ & $3.248 \pm 0.033$ & $6.676 \pm 0.172$ \\
FOM & 9  & db28   & 56 & $3.963 \pm 0.095$ & $3.315 \pm 0.033$ & $6.686 \pm 0.166$ \\
FOM & 10 & db15   & 30 & $3.956 \pm 0.095$ & $3.392 \pm 0.039$ & $6.192 \pm 0.187$ \\
\hline
\multicolumn{7}{l}{\textit{EFI-based mother-wavelet ranking}} \\
EFI & 1  & bior3.3 & 8  & $3.205 \pm 0.056$ & $3.555 \pm 0.034$ & $4.964 \pm 0.141$ \\
EFI & 2  & bior3.5 & 12 & $3.367 \pm 0.063$ & $3.531 \pm 0.036$ & $5.480 \pm 0.160$ \\
EFI & 3  & bior3.7 & 16 & $3.206 \pm 0.066$ & $3.467 \pm 0.035$ & $5.626 \pm 0.160$ \\
EFI & 4  & bior3.9 & 20 & $3.361 \pm 0.067$ & $3.426 \pm 0.035$ & $5.764 \pm 0.160$ \\
EFI & 5  & db8     & 16 & $3.424 \pm 0.089$ & $3.415 \pm 0.044$ & $5.390 \pm 0.200$ \\
EFI & 6  & sym11   & 22 & $3.625 \pm 0.092$ & $3.414 \pm 0.041$ & $5.849 \pm 0.192$ \\
EFI & 7  & db10    & 20 & $3.415 \pm 0.089$ & $3.412 \pm 0.042$ & $5.702 \pm 0.194$ \\
EFI & 8  & sym9    & 18 & $3.198 \pm 0.088$ & $3.410 \pm 0.043$ & $5.538 \pm 0.196$ \\
EFI & 9  & sym10   & 20 & $3.648 \pm 0.089$ & $3.408 \pm 0.043$ & $5.684 \pm 0.198$ \\
EFI & 10 & sym8    & 16 & $3.413 \pm 0.084$ & $3.408 \pm 0.044$ & $5.357 \pm 0.200$ \\
\hline
\end{tabular}
\vspace{0.5ex}
\begin{minipage}{0.96\textwidth}
\footnotesize
\textit{Note:} Values are reported as mean $\pm$ 95\% confidence interval. 
The FOM block is sorted in descending FOM and the EFI block is sorted in descending EFI. 
Complex SNR gain is reported as a communication-domain validation metric and is not used as a ranking score.
\end{minipage}
\end{table*}

The contrast between the FOM and EFI rankings is technically important. FOM identifies wavelets that best preserve the complex I/Q waveform morphology after denoising, whereas EFI identifies wavelets that provide the best denoising efficiency after accounting for filter-length complexity. Thus, the preferred mother wavelet depends on the receiver objective. If waveform fidelity and complex-I/Q error reduction are prioritized, the FOM-leading high-order Daubechies and Coiflet wavelets are more suitable. If computational efficiency is the dominant constraint, the EFI-leading biorthogonal, Daubechies, and Symlet candidates are more appropriate.

\paragraph{Sparsity based criteria}

Sparsity in the wavelet domain provides a domain agnostic selection principle with direct relevance to communication receiver signals that exhibit sparse transients, burst edges, or impulsive interference. Bekerman and Srivastava introduced the use of sparsity measures for wavelet denoising parameter optimization \cite{Bekerman2021SparsityPlot}. A systematic use of sparsity for mother wavelet selection is presented by Sahoo et al.\ \cite{Sahoo2024OptimalWavelet}, where a wavelet sample space spanning biorthogonal, Coiflet, Daubechies, reverse biorthogonal, and Symlet families is evaluated using the mean sparsity change measure $\mathrm{mean}_{\mathrm{sc}}$,
\begin{equation}
\mathrm{mean}_{\mathrm{sc}} = \frac{sP_{J+1}-sP_{1}}{J-1},
\end{equation}
where $J$ is the decomposition level used in the evaluation and $sP_j$ denotes a sparsity proxy for the detail coefficients at level $j$,
\begin{equation}
sP_j = \frac{\max\lvert wD_j\rvert}{\sum_{k=1}^{q_j} \lvert wD_j[k]\rvert},
\qquad 1 \le j \le J,
\end{equation}
with $wD_j$ denoting the detail coefficient vector at level $j$ and $q_j$ denoting its length. The key observation in \cite{Sahoo2024OptimalWavelet} is that noise alone yields nearly constant $\mathrm{mean}_{\mathrm{sc}}$ across wavelets, whereas structured signals yield wavelet dependent values. The wavelet maximizing $\mathrm{mean}_{\mathrm{sc}}$ is selected as optimal. The same criterion was later adopted for wavelet selection in dispersion compensation of chirped fiber Bragg gratings \cite{yan2025enhancing}. 

It is important to note that the mean sparsity-change criterion is conditional on the decomposition level used during evaluation. Therefore, $\mathrm{mean}_{sc}$ should be interpreted more precisely as a pairwise score $M_{sc}(\psi,J)$, rather than as a criterion depending on the mother wavelet alone. This dependence arises because the coefficient vector $w_{D_j}$ is produced by the analysis filters associated with the selected wavelet $\psi$, whereas the number of evaluated scales and the dyadic subbands included in the sparsity profile are determined by $J$. Consequently, selecting a mother wavelet by maximizing $\mathrm{mean}_{sc}$ at an arbitrary fixed level may favour a wavelet whose coefficients are sparse only under that particular scale partition.

A practical resolution is to evaluate sparsity-based criteria over admissible $(\psi,J)$ pairs. Let $\mathcal{W}$ denote the candidate wavelet set and let $\mathcal{J}(\psi)$ denote the valid levels for wavelet $\psi$, after accounting for signal length, filter length, boundary extension, sampling frequency, and any occupied-bandwidth constraints. The sparsity-based selection can then be expressed as a joint or profiled search over $\psi \in \mathcal{W}$ and $J \in \mathcal{J}(\psi)$. If a task-level metric is available, the selected pair should be the one that optimizes the downstream objective. If no such metric is available, the selected wavelet may be compared at its profiled level, namely the level at which its sparsity-change curve exhibits a stable and physically interpretable transition between noise-dominated and signal-dominated scales.

\paragraph{Entropy and energy to entropy metrics}

Entropy based measures are widely used for selecting wavelets that yield compact representations. In \cite{Wronka2023SoundTextures}, an automated entropy based optimization over a discrete wavelet packet transform is proposed for stochastic resynthesis of sound textures. A candidate bank of 62 compactly supported orthogonal or biorthogonal wavelets is constructed, spanning Haar, Daubechies, Symlet, Coiflet, biorthogonal, reverse biorthogonal, discrete Meyer, and Fej\'er--Korovkin families. For each candidate basis, the terminal node with highest relative entropy is identified and the mother wavelet associated with the decomposition yielding the largest entropy is selected \cite{Wronka2023SoundTextures}. In an SWT based framework for denoising ECG signals contaminated by EMG \cite{Raggi2025SWTTemplateECG}, a subject specific template is decomposed across wavelet level combinations and Shannon entropy of detail coefficients is computed. The wavelet and level combination that minimizes entropy is selected, consistent with the idea that a better matched wavelet yields a more compact coefficient distribution. The Shannon entropy for a given wavelet and level pair $(\psi,j)$ is written as,
\begin{equation}
\mathrm{Entropy}(\psi,j) = -\sum_{i=1}^{N} pr_i \log_{2}(pr_i),
\end{equation}
where $pr_i$ denotes the probability associated with the detail coefficient distribution.

Building on these ideas, the energy to entropy index (ESE) has been proposed as a combined criterion. For detail coefficients, ESE is computed as \cite{Ahmed2024ECGThresholding}
\begin{equation}
\mathrm{ESE} = \frac{\mathrm{Energy}(wD_j)}{\mathrm{Entropy}(wD_j)}.
\end{equation}
In \cite{Ahmed2024ECGThresholding}, this ratio along with cross correlation has been employed to identify the mother wavelet suited for robust thresholding and reconstruction. The same metric is utilised to identify the optimal wavelet for the denoising of inertial sensor signals in wearable human activity recognition \cite{Nematallah2024HARWavelet}.  

The ESE ratio is illustrated in Supplementary Fig.~S3, which compares how efficiently different mother wavelets represent inertial-sensor activity signals from the WISDM dataset. That figure is retained as a cross-domain illustration of the criterion rather than as communication-domain evidence, and it has been moved to the supplementary material for that reason: the underlying signals are human-activity accelerometer traces, not communication waveforms, so the specific wavelet ordering it reports carries no receiver-level implication. Its purpose here is only to show that the ESE ratio discriminates between signal classes with different transient content, higher values indicating that more of the signal energy is concentrated in a smaller number of large coefficients and hence a more compact representation with lower coefficient entropy. Dynamic activities yield markedly higher ESE ratios than stationary ones, and bior3.5 and rbio1.3 rank highest among the tested wavelets, consistent with
a closer match to the local transient structure of those signals. The communication-domain counterpart of this analysis is Table~\ref{tab:radioml_fom_efi_wavelet}, which evaluates mother wavelets on RadioML 2018.01A complex I/Q waveforms by modulation class using FOM, EFI, and complex SNR gain, and it is that table, not Supplementary Fig.~S3, that supports the wavelet-selection guidance given for communication receivers.

Whereas the ESE ratio represents compactness before denoising, FOM and EFI provide complementary measures after denoising by assessing reconstruction fidelity and overall denoising efficiency.
More recently, a Complex Proportional Assessment (COPRAS) based decision framework was introduced that combines the maximum energy to Shannon entropy ratio and the maximum cross correlation coefficient to rank candidate mother wavelets for speech signals recorded through face masks and shields \cite{Bharathi2025COPRAS}. The same multi criteria structure is directly applicable to receiver side selection problems when multiple objectives are present, such as coefficient compactness, distortion control, and computational cost, although the downstream criteria in communication systems are typically expressed in terms of estimation or detection performance rather than waveform similarity alone. Table~\ref{tab:mw_metrics_comms} summarizes the dominant metric families used for mother-wavelet selection and recasts them in a communication-oriented interpretation. The comparison emphasizes the trade-off between computational burden, robustness under bursty and heavy-tailed disturbances, and suitability for waveform types encountered in practical receivers.

\begin{table*}[!t]
\centering
\caption{Comparison of mother-wavelet selection metrics for communication-oriented signals.}
\label{tab:mw_metrics_comms}
\footnotesize
\setlength{\tabcolsep}{3.5pt}
\renewcommand{\arraystretch}{1.05}
\begin{tabular}{p{0.11\textwidth}p{0.18\textwidth}p{0.15\textwidth}p{0.20\textwidth}p{0.25\textwidth}}
\toprule
\textbf{Metric family} & \textbf{Typical computation} & \textbf{Computation complexity} & \textbf{Robustness to noise and outliers} & \textbf{Best suited communication signal types and use cases} \\
\midrule

Correlation and similarity based &
Correlation with a reference structure such as a preamble, pilot, pulse shaping waveform, or estimated channel impulse response; includes normalised correlation and figure-of-merit style composites &
Low to medium, depending on whether correlation is evaluated in time domain, wavelet domain, or repeatedly over candidates &
Medium; sensitive to reference mismatch; robustness improves when correlation is computed on denoised or band-limited components &
Burst preambles and training sequences; synchronization and time-delay estimation; short packets in Internet of Things; channel sounding snapshots; underwater time-delay estimation pipelines \\

Energy concentration and subband SNR based &
Energy packing into a small number of subbands; subband SNR and related concentration indices; may incorporate level constraints &
Low; dominated by wavelet transform plus per-band energy computation &
Medium; sensitive to heavy-tailed interference unless robust statistics are used; improves when applied band-wise or with local noise estimation &
Wideband transients and impulsive interference scenarios; interference localisation; industrial IoT measurements; power line and underwater channels where energy bursts dominate \\

Sparsity based &
Coefficient sparsity measures and sparsity-change curves across levels; selects wavelet that yields sparse structured coefficients relative to noise &
Low to medium; requires wavelet transform and simple norms; increases if evaluated across many candidate wavelets and levels &
High for impulsive noise removal when the desired signal is compressible; lower if the desired waveform is not sparse in the chosen basis &
Ultra-wideband impulses; chirp-like transients; sparse multipath channel impulse responses; impulsive noise suppression prior to detection \\

Entropy and energy-to-entropy based &
Shannon or relative entropy of coefficients; energy-to-entropy ratios; selects wavelet yielding compact coefficient distributions &
Medium; requires coefficient histograms or probability estimates plus entropy computation; repeated evaluation across candidates increases cost &
Medium to high; entropy can be destabilised by rare extreme outliers unless robust binning or trimming is used &
Complex and mixed spectral occupancy signals; adaptive receivers under varying noise floors; scenarios needing compact representations without a clean reference \\

Composite multi-criteria decision metrics &
Weighted combinations of correlation, entropy, energy, and smoothness measures; decision frameworks that rank candidate wavelets &
Medium to high; cost is dominated by repeated computation of all component metrics across candidates &
High if the composite explicitly includes robust metrics and uses validation across conditions; otherwise depends on metric design &
Heterogeneous deployments; mixed interference conditions; settings where multiple objectives must be traded such as low distortion and strong impulsive-noise rejection \\

Receiver-task and end-to-end performance based &
Candidate mother wavelets are ranked using downstream communication metrics such as BER, EVM, channel-estimation MSE, synchronization error, acquisition sensitivity, localization error, probability of detection, probability of false alarm, or throughput &
Medium to high; requires receiver simulation or measurement across candidate wavelets, SNRs, channels, and noise models &
High when validated across multiple channel/noise conditions; lower if tuned to a single modulation, channel, or operating point &
OFDM channel estimation, spectrum sensing, GNSS acquisition, synchronization pipelines, localization receivers, underwater and optical links, and end-to-end communication receiver evaluation \\

\bottomrule
\end{tabular}
\end{table*}

\subsection{TECHNIQUES FOR SELECTION OF THE OPTIMAL DECOMPOSITION LEVEL}
\label{subsec:dl_selection}

Decomposition-level selection is likewise conditional on the chosen mother wavelet. Although $J$ determines the nominal dyadic partition of the frequency axis, the practical coefficient distribution at each level depends on the wavelet filter length, support, regularity, symmetry, phase behaviour, and boundary response. Therefore, criteria such as energy concentration, entropy, sparsity change, approximation--detail energy ratios, reconstruction error, or task-level performance should not be interpreted as functions of $J$ alone. More precisely, a level-selection score should be regarded as $Q(\psi,J)$, where the same decomposition depth may produce different denoising behaviour for different wavelet families.

In the absence of strong prior knowledge, practitioners should avoid selecting $J$ using a single default wavelet and then transferring that level to all candidate wavelets. A more robust procedure is to define an admissible $(\psi,J)$ search space and evaluate the selected criterion over that space. If receiver-level validation is available, such as BER, EVM, synchronization error, channel-estimation error, localization error, or probability of detection, the preferred configuration should be selected by optimizing that metric. If no clean reference or receiver-level metric is available, a profiled strategy can be used: for each candidate wavelet, identify the most stable level using sparsity-breakpoint, entropy-minimisation, energy-concentration, or band-alignment criteria, and then compare wavelets at their corresponding profiled levels. Sensitivity over neighbouring levels should be reported because a reliable configuration should not depend on an isolated and unstable optimum.

In DWT and SWT based denoising, the decomposition level $J$ specifies the number of dyadic scales used to represent the signal. At small values of $J$, the decomposition is shallow and provides limited scale separation, so noise and interference remain mixed with structured components. At larger values of $J$, the representation becomes increasingly coarse, with more energy transferred into the approximation coefficients and into coarser detail subbands. If $J$ is too small, thresholding remains ineffective because noise persists across scales. If $J$ is too large, over smoothing can occur and informative signal structure may be attenuated. Selecting $J$ is therefore equivalent to selecting a multiresolution partition in which signal dominated and noise dominated components are sufficiently disentangled to enable robust thresholding and faithful reconstruction.

A communications and networking perspective is to link $J$ to sampling and bandwidth. Let us define the sampling frequency as $f_{\mathrm{s}}$ and the Nyquist frequency as $f_{\mathrm{N}}=f_{\mathrm{s}}/2$. After $J$ levels, the approximation subband spans approximately $0$ to $f_{\mathrm{N}}/2^{J}$, while the detail subbands occupy successive dyadic bands up to $f_{\mathrm{N}}$. This relationship implies that the selection of $J$ is fundamentally a sampling dependent choice that determines which frequency regions are treated as coarse structure and which are treated as details for thresholding. In software defined radio (SDR) settings, where $f_{\mathrm{s}}$ is explicitly configured and may differ across deployments, the same denoising task can require different $J$ values because the dyadic subband edges scale with $f_{\mathrm{s}}$. In multicarrier systems, where the signal energy is distributed across subcarriers and where subcarrier spacing determines the spectral granularity, selecting $J$ can be interpreted as selecting the dyadic resolution at which noise and interference are isolated relative to the occupied bandwidth. This interpretation does not replace empirical tuning, but it provides an explanatory basis for why fixed integer choices of $J$ are not generally transferable across sampling rates and bandwidths.

A substantial number of early and application oriented studies address decomposition depth by fixing $J$ to a small integer based on prior practice or qualitative experimentation, without invoking a formal selection rule. Such choices provide empirical reassurance that moderate depths may be adequate for typical sampling rates, but they do not yield transferable guidance beyond the specific setting. For example, an ECG denoising approach based on the lifting wavelet transform and total variation minimisation used a fixed two level decomposition \cite{Talbi2020LWTTV}, whereas another study on threshold selection for wavelet denoising constrained the decomposition depth to five levels so that optimisation could focus exclusively on tuning threshold values \cite{Xiao2022ThresholdOpt}. A similar fixed-depth choice shows up in a communication receiver, where an OFDM channel-estimation denoiser used a two-level decomposition on purpose, trading finer scale separation for lower per-frame cost~\cite{R9}.
\vspace{-4mm}
\subsubsection{Frequency band alignment and sampling based rules}
\label{subsubsec:dl_band_alignment}

An alternative principle-based heuristic seeks to determine $J$ such that the subbands are aligned with the known frequency support of the underlying process. In communications and networking, this support can be specified by the occupied baseband bandwidth, by the bandwidth of a preamble or pilot structure used for synchronization and channel estimation, or by the spectral granularity induced by a multicarrier configuration. Under this viewpoint, the objective is to choose $J$ such that the approximation band covers the slow varying components that should be preserved, while detail bands at higher frequencies capture predominantly noise and interference that can be attenuated by thresholding. The explanatory value of this heuristic is that it makes $J$ an explicit function of $f_{\mathrm{s}}$ and of the bandwidth region that should remain undistorted.

The same frequency alignment idea is widely used in biomedical and structural sensing applications, where the prescribed support is defined by domain specific frequency bands rather than by occupied communication bandwidth. In studies on EEG analysis \cite{Atangana2020EEGMotherWavelet,Suhail2020EEGCognitive}, the decomposition depth is chosen so that dyadic detail bands approximate the canonical EEG rhythms, given the sampling frequency. The general aim is to select the smallest $J$ for which the highest detail subband still covers the upper end of the physiologically relevant frequency range, while the approximation band captures low frequency drift. A similar strategy has been applied to vibration based monitoring of bridge structures, where wavelet bands are aligned with characteristic modal frequency ranges \cite{Silik2024WaveletSHM}. Related reasoning appears in work on Doppler cardiogram and phonocardiogram denoising, where the decomposition level is constrained so that the approximation band spans a prescribed low frequency corridor, for example 0 to 5 Hz for Doppler cardiogram, with higher frequency components and noise allocated to detail bands. Although these works do not optimise $J$ explicitly, they provide a clear instance of sampling and band support guided level selection, and the same logic applies when the prescribed band support is determined by receiver bandwidth and sampling configuration.
\vspace{-8mm}
\subsubsection{Task performance driven metrics}
\label{subsubsec:dl_task_metrics}

A further increase in methodological sophistication is to treat the decomposition level as a discrete hyperparameter to be explored systematically, with each candidate value assessed using task specific performance or denoising metrics. In a study on time delay estimation in underwater reverberant environments \cite{Rajan2020UnderwaterTDE}, multiple decomposition depths were evaluated and the level that minimised both the bias and variance of the maximum likelihood delay estimate was selected. This approach links $J$ directly to the accuracy of the final estimation task rather than to intermediate measures alone. A comparable task-driven choice appears in satellite navigation. For Global Navigation Satellite System multipath-error extraction and mitigation, the wavelet decomposition level and threshold are refined jointly so that the dyadic scales isolating the slowly varying multipath error are separated from the noise-dominated detail, which links level selection directly to positioning accuracy~\cite{R17}.

In another work on DWT based phonocardiogram denoising \cite{Lowast2020WristPulse}, the decomposition level was tuned jointly with the mother wavelet and the thresholding rule by scanning a range of levels, for example 4 to 10, and selecting the configuration that provided a favourable compromise between signal to noise ratio and root mean square error. The recommendation of a ten level decomposition emerged from this systematic hyperparameter search in that application.Similarly, while carrying out the task of denoising fetal and maternal heart sounds, the level of decomposition was also tuned with mother wavelet and thresholding rule with an objective to maximize the SNR values~\cite{minani2023fetal}.

\subsubsection{Metrics in time, frequency, or wavelet domain}
\label{subsubsec:dl_metrics}
Efficiency and sparsity measures have also been used as criteria for determining the decomposition level. An efficiency index, $\mathrm{EFI}(j,k)$, was used to guide decomposition level selection in \cite{Jang2021DCG}. In that study, decomposition levels from three to eight were evaluated over 115 candidate wavelets and $\mathrm{EFI}(j,k)$ was computed for each level wavelet pair. The decomposition level that maximised this index under the imposed bandwidth constraint was selected \cite{Jang2021DCG}.

\begin{figure}[!t]
\centering
\includegraphics[width=\linewidth]{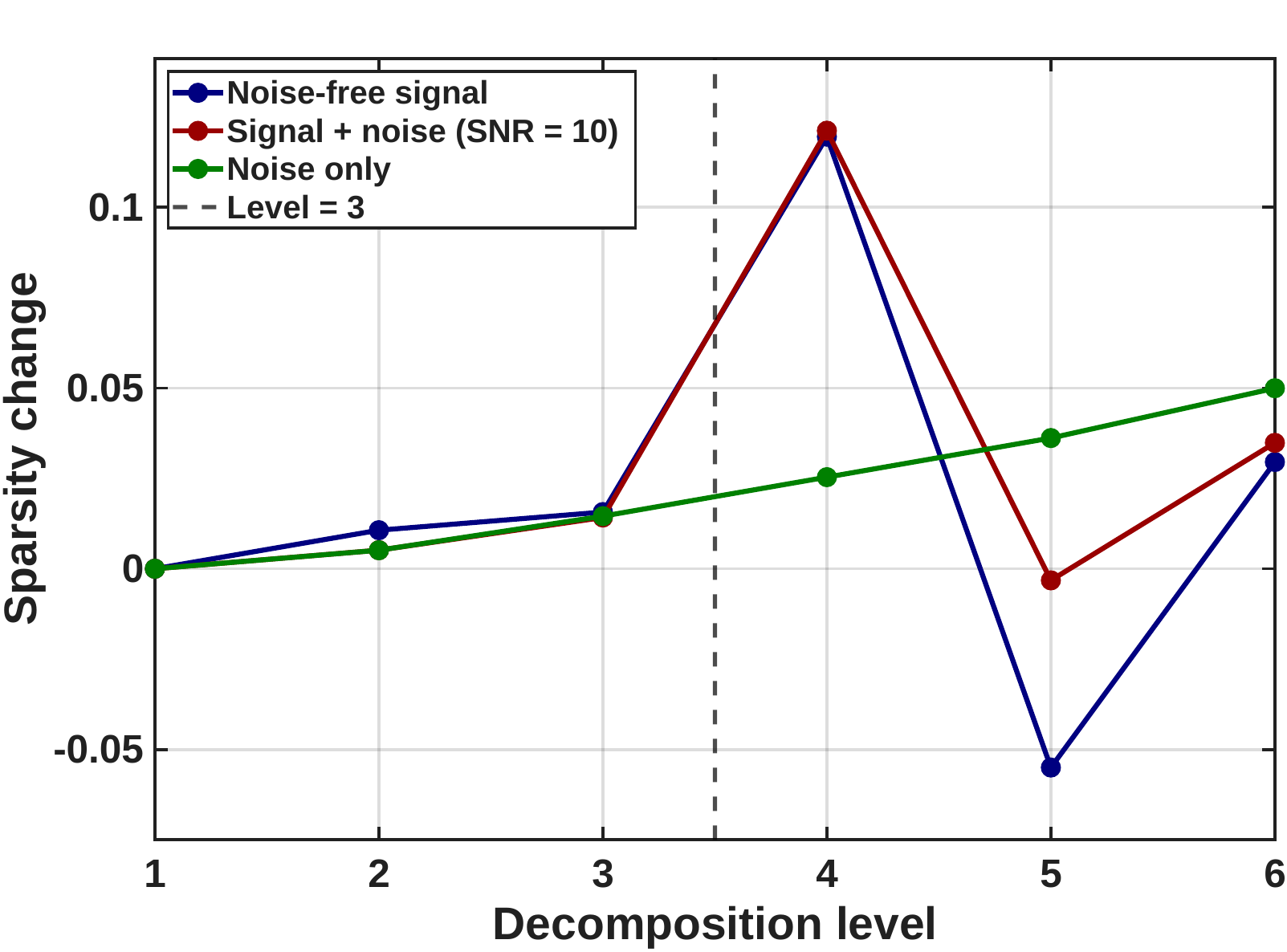}
\caption{Sparsity change for different decomposition levels.}
\label{fig:sparsity_levels}
\end{figure}

The sparsity-based framework introduced in \cite{Bekerman2021SparsityPlot} and formalized in \cite{Sahoo2024OptimalWavelet} also provides a principled
criterion for decomposition-level selection, as detailed in Section IV-B2. By monitoring the sparsity change $\Delta sP_{j}$ across levels, the method identifies the structural break-point at which noise-dominated detail coefficients give way to signal-dominated structure; Fig.~\ref{fig:sparsity_levels} illustrates this transition on a simulated ESR signal, with level 3 selected as the boundary between coefficients appropriate for thresholding and those that should be
preserved~\cite{Bekerman2021SparsityPlot}.

Energy based criteria are also used to determine the decomposition level. In a study on partial discharge denoising in large generators \cite{Li2020PDNovelShrinkage}, the appropriate level was selected by examining both the energy percentage and the signal to noise ratio of the approximation coefficients, and the level at which the partial discharge energy was most concentrated and the signal to noise ratio was maximal was selected. Machine learning techniques such as decision trees have also been employed in combination with energy spectral density measures to determine an optimum level \cite{soltani2020decision}.
\vspace{-4mm}
\subsubsection{Distribution based criteria: Jarque Bera and composite indices}
\label{subsubsec:dl_distribution}

A framework for choosing the decomposition level using the Jarque Bera test was presented in \cite{Zhang2025JBDecomp}. The scheme focuses on signals affected by Gaussian white noise, where the presence of a normal distribution is tested initially using the Jarque-Bera criterion. After confirming the noise distribution, wavelet denoising is performed at multiple decomposition levels to generate candidate reconstructions. For each level, a composite index is calculated considering different performance metrics such as root mean square error and smoothness using predefined weights. The optimal decomposition level is considered the one where the composite index is minimal\cite{Zhang2025JBDecomp}.
\vspace{-4mm}
\subsubsection{Application specific metrics: Noise envelope area and entropy}
\label{subsubsec:dl_app_specific}

Two notable works deploy application specific scalar metrics to optimise decomposition level alongside wavelet selection. In \cite{Chen2020HeartSoundNEA}, noise envelope area (NEA) was used as a joint criterion for selecting both the mother wavelet and the decomposition level. For each combination of Daubechies wavelet and decomposition depth from one to ten levels, wavelet threshold denoising was applied and NEA was computed. Plots of NEA as a function of decomposition level showed that db2 at ten levels yielded the smallest noise envelope area across heart sound categories, with only marginal improvement beyond nine levels, and the decomposition level was fixed at ten based on minimisation of this noise metric \cite{Chen2020HeartSoundNEA}. In \cite{Raggi2025SWTTemplateECG}, a subject specific ECG template was constructed and decomposed using multiple candidate wavelets across several decomposition levels. For each wavelet level configuration, the Shannon entropy of the detail coefficients was computed, and the combination yielding the lowest entropy was selected as optimal. This strategy is based on the assumption that an appropriate wavelet representation of a structured template yields a compact coefficient distribution and therefore lower entropy \cite{Raggi2025SWTTemplateECG}. Table~\ref{tab:dl_pros_cons} consolidates decomposition-level determination approaches into a set of method families and contrasts their latency profile, transferability across sampling configurations, and the overhead introduced by additional tuning parameters. This comparison highlights when fixed heuristics are sufficient and when systematic selection is necessary.

\begin{table*}[!t]
\centering
\caption{Pros and cons of decomposition-level determination methods.}
\label{tab:dl_pros_cons}
\footnotesize
\setlength{\tabcolsep}{3.5pt}
\renewcommand{\arraystretch}{1.05}
\begin{tabular}{p{0.13\textwidth}p{0.2\textwidth}p{0.17\textwidth}p{0.22\textwidth}p{0.15\textwidth}}
\toprule
\textbf{Method family} & \textbf{Core idea} & \textbf{Suitability for real-time and low latency} & \textbf{Accuracy and transferability} & \textbf{Parameter overhead} \\
\midrule

Empirical fixed-level selection &
Select a small integer depth by precedent or quick qualitative testing &
High; minimal computation and no search &
Low to medium; may not transfer across sampling rates or bandwidths; sensitive to implementation details &
Low; requires only a fixed $J$ \\

Frequency-band alignment and sampling-based rules &
Choose $J$ so dyadic sub-bands align with occupied bandwidth or known spectral support given $f_{\mathrm{s}}$ &
High; computation is negligible once $f_{\mathrm{s}}$ and the target band are specified &
Medium; transfers across sampling rates if band support is correctly specified; can be confounded by boundary effects and leakage &
Low; requires knowledge of $f_{\mathrm{s}}$ and a band-support target \\

Metric-driven selection using waveform fidelity or task measures &
Scan candidate $J$ values and select the one optimizing SNR, RMSE, estimation bias, detection performance, or related metrics &
Medium; requires repeated transforms over candidate levels; can be reduced with coarse-to-fine search &
High when the metric matches the operational task; lower if fidelity metrics are used for task endpoints &
Medium; requires a metric definition and a search range for $J$ \\

Sparsity-based level selection &
Select $J$ at a sparsity-change breakpoint where noise-dominated details transition to structured coefficients &
Medium to high; computation is modest; requires computing sparsity curves across levels &
Medium to high; robust when the signal is compressible; sensitive if the signal is not sparse in the basis &
Low to medium; requires sparsity threshold and breakpoint rule \\

Entropy and distribution-based level selection &
Select $J$ using entropy or distribution tests and composite indices across candidate reconstructions &
Low to medium; entropy and distribution tests add cost and may need multiple reconstructions &
Medium; can work without a clean reference but may require stabilization under outliers &
Medium; requires entropy settings or distribution-test thresholds \\

Optimization and metaheuristic selection &
Treat $J$ and other parameters as decision variables and optimize jointly under an objective function &
Low for strict real-time; high offline; online feasible only with constrained candidate sets &
High if objective aligns with task and validation is robust; risk of overfitting without cross-condition testing &
High; introduces optimizer hyper-parameters and a larger search space \\

\bottomrule
\end{tabular}
\end{table*}

\subsection{TECHNIQUES FOR SELECTION OF THE OPTIMAL THRESHOLDING TECHNIQUE}
\label{subsec:threshold_selection}

In studies published between 2020 and 2025, thresholding strategies are typically characterized along two coupled dimensions. The first dimension is the thresholding function, which specifies how each wavelet coefficient is attenuated and includes hard, soft, Garrote type, block, and hybrid shrinkage schemes. The second dimension is the thresholding rule, which specifies how the threshold value is selected and includes universal, Stein's unbiased risk estimate (SURE) based, minimax, and BayesShrink rules, as well as false discovery rate (FDR) based and optimization driven approaches. In telecommunications and networking settings, thresholding is frequently motivated by the need to suppress intermittent and non Gaussian disturbances that manifest as impulsive events. Such heavy tailed disturbances are observed in several practical channels and networked sensing deployments, including underwater acoustic links, power line communication environments, and industrial Internet of Things measurements. 

Power-line communication (PLC) is often cited as a canonical impulsive-noise channel, yet a targeted 2020–2025 search shows that recent PLC work on impulsive noise is rarely framed as a wavelet-thresholding denoising problem with explicit parameter selection. Three other lines dominate instead. First, wavelets are used to define the multicarrier waveform itself, a waveform-design use of the kind discussed in Section III-C, here in a bandpass wavelet-OFDM PLC system~\cite{R18}. Second, impulsive-noise mitigation is most often handled by nonlinear time-domain preprocessing such as clipping and blanking, sparse Bayesian learning, or cyclostationary filtering, none of which are wavelet-thresholding methods. Third, learning-based denoisers have begun to appear for PLC channel estimation under impulsive and Gaussian noise~\cite{R19}. Mature wavelet-thresholding parameter-selection work in the wider power-systems domain targets power-quality disturbance monitoring rather than communication links, so it is not classified here as a telecommunications study. This relative absence of dedicated wavelet parameter-selection studies for PLC denoising is consistent with the under-representation visible in Fig.~\ref{fig:heat_map}, and it motivates the receiver-aligned framework of Section IV-F as a route for carrying wavelet-denoising design principles into impulsive-noise links.
The nonlinear time-domain schemes noted above, rather than an absence of processing constitute the practical standard of comparison for impulsive channels, and the communication experiments reported in this paper do not
include them; this is stated as a limitation in Section~IV-G-1 and as a research priority in Section~IV-I.

In these conditions, threshold design must attenuate sporadic large magnitude wavelet coefficients that correspond to impulsive interference while preserving waveform components required for downstream estimation and detection. The reviewed literature provides two complementary responses to this challenge. The first response retains classical thresholding rules as repeatable baselines and embeds them into more adaptive pipelines. The second response modifies the shrinkage function itself or optimizes its parameters so that attenuation is smoother and more robust to outliers.

\subsubsection{Classical hard and soft thresholding with standard rules}
\label{subsubsec:thr_classical}

The foundational group uses Donoho-Johnstone style shrinkage together with analytical threshold rules. Several studies compare hard and soft thresholding functions under canonical rules and report their relative effectiveness in the corresponding application settings \cite{Baldazzi2020NeuralDenoising,Suhail2020EEGCognitive,Potdar2021PCG,malleswari2021improved,Ghosh2022PCGParametric,daud2022denoising}. Let us denote the detail coefficient at level $j$ and index $k$ as $wD_{j,k}$ and let us denote the threshold value at level $j$ as $Th_{j}$. The hard thresholding operator is
\begin{equation}
\overline{wD}_{j,k} =
\begin{cases}
wD_{j,k}, & \text{if } |wD_{j,k}| \ge Th_{j},\\
0,        & \text{otherwise},
\end{cases}
\end{equation}
and the soft thresholding operator is
\begin{equation}
\overline{wD}_{j,k} =
\begin{cases}
\operatorname{sgn}(wD_{j,k})\left(|wD_{j,k}|-Th_{j}\right),
& \text{if } |wD_{j,k}| \ge Th_{j},\\
0, & \text{otherwise}.
\end{cases}
\end{equation}

A widely used analytical rule is the universal rule, often termed VisuShrink~\cite{S3}, expressed as
\begin{equation}
Th_{\mathrm{univ}} = \sigma \sqrt{2\log N},
\end{equation}
where $N$ is the number of samples and $\sigma$ is an estimate of the noise standard deviation, typically obtained via the median absolute deviation of first level detail coefficients. 
The same MAD noise estimate underpins adaptive thresholding in communication receivers. In OFDM channel estimation, an adaptive MAD-based threshold applied to the LS channel estimate requires no prior channel statistics or noise-variance knowledge, yet yields \(70.8\)--\(84.3\%\) MSE reduction and \(7.9\)--\(32.2\%\) BER improvement over plain LS while remaining \(6.3\)--\(6.7\times\) faster than the MMSE estimator at \(O(N\log N)\) complexity~\cite{R9}. SURE based rules, including rigrsure and heursure, select the threshold by minimizing Stein's unbiased risk estimate for soft thresholding~\cite{S4}. Minimax thresholds are derived to minimize the maximum risk over a specified function class~\cite{S3,S6}. BayesShrink implements a level dependent strategy in which a separate threshold is computed for each detail scale $j$ based on Bayesian estimation of the signal and noise variance at that level~\cite{S5}. In \cite{Potdar2021PCG}, soft thresholding with BayesShrink is identified as the most effective configuration for the reported study, whereas \cite{Ghosh2022PCGParametric} reports that soft thresholding combined with the rigrsure rule is preferred for the reported framework.This waveform-design use of wavelets is illustrated in power-line communications, where bandpass wavelet-OFDM has been proposed as an alternative to FFT-OFDM with reduced adjacent-subchannel interference~\cite{R18}. As with FBMC, such systems concern the modulation transform rather than the denoising parameter-selection problem surveyed here.

\begin{figure*}[!t]
\centering
\subfloat[]{%
    \includegraphics[width=0.49\textwidth]{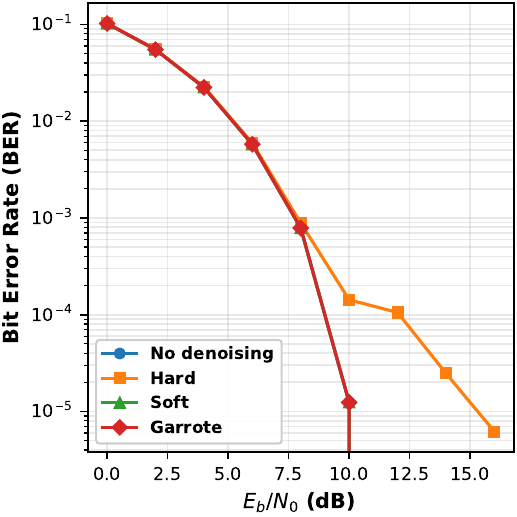}%
    \label{fig:threshold_awgn}%
}\hfill
\subfloat[]{%
    \includegraphics[width=0.49\textwidth]{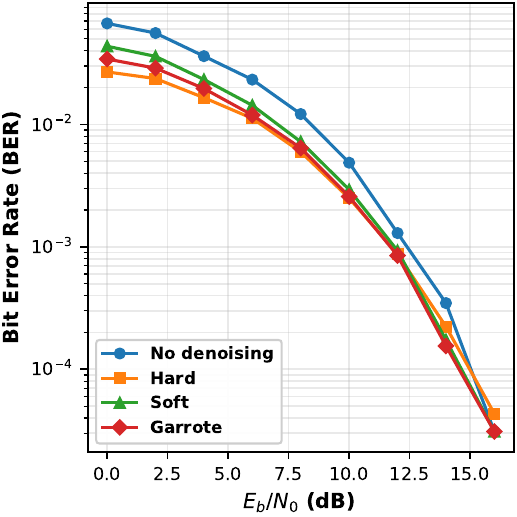}%
    \label{fig:threshold_middleton}%
}
\caption{Thresholding performance using a fixed OFDM/QPSK receiver setting, fixed mother wavelet $\psi=\mathrm{db4}$, fixed decomposition level $J=3$, and fixed threshold scale, in the presence of (a) AWGN and (b) Middleton Class A impulsive noise. The fixed wavelet configuration is used only to isolate the effect of the thresholding rule; it is not intended to imply that $\mathrm{db4}$ and $J=3$ are universally optimal.}
\label{fig:Thresholding performance under AWGN and Impulsive Noise}
\end{figure*}

Fig.\ref{fig:Thresholding performance under AWGN and Impulsive Noise} compares the BER performance of an OFDM receiver under AWGN and Middleton Class A impulsive noise for a fixed wavelet setting (db4, level 3, threshold scale 4.0), when wavelet denoising is performed using hard, soft, and Garrote thresholding. In the AWGN scenario, the BER curves corresponding to no denoising, soft thresholding, and hard thresholding are nearly overlapping, implying that wavelet thresholding provides little advantage when the noise is broadly distributed rather than sparse. However, all three thresholding methods outperform the no-denoising baseline in the presence of impulsive noise. This suggests that thresholding effectiveness is highly sensitive to the noise model. 

To determine whether the thresholding trends in Fig.~\ref{fig:Thresholding performance under AWGN and Impulsive Noise} are specific to the fixed \(\psi=\mathrm{db4}\), \(J=3\) configuration, an additional sensitivity analysis was conducted over the wavelet families \(\{\mathrm{haar}, \mathrm{db4}, \mathrm{sym4}, \mathrm{coif4}, \mathrm{bior4.4}\}\) and decomposition levels \(J \in \{2,3,4,5\}\). The resulting AWGN and Middleton Class A impulsive-noise BER grids are provided in Supplementary Figs.~S1 and S2, respectively. The AWGN results show that the no-denoising, soft-thresholding, and Garrote-thresholding curves remain nearly coincident across the tested wavelet families and decomposition levels, whereas hard thresholding can introduce a high-SNR penalty at larger decomposition depths. This indicates that, under AWGN, wavelet thresholding provides little systematic BER improvement across the tested configurations. In contrast, under Middleton Class A impulsive noise, thresholding remains beneficial across representative compact wavelet families and moderate decomposition levels, although the exact BER values and the ordering among hard, soft, and Garrote thresholding vary with $\psi$ and $J$. Soft and Garrote thresholding exhibit the most stable behaviour across the tested grid, while hard thresholding may become less favourable at higher $E_b/N_0$ and larger $J$. Thus, the trends reported in Fig.\ref{fig:Thresholding performance under AWGN and Impulsive Noise} are not unique to the selected $\mathrm{db4}$, $J=3$ setting; rather, the qualitative noise-model-dependent behaviour persists across the tested wavelet families and decomposition depths. Although the additional sensitivity analysis confirms that the qualitative AWGN-versus-impulsive-noise thresholding behaviour persists across the tested wavelet families and decomposition levels, broader generalization would require evaluation over additional OFDM numerologies, modulation orders, channel models, impulsive-noise parameters, and receiver tasks. Together with the mother-wavelet comparison in Fig.~\ref{fig:BER vs $E_b/N_0$}, the additional
sensitivity grids show that the principal conclusion is not tied to a
single wavelet family; however, the numerical BER values and the ranking of wavelet configurations remain parameter-dependent.

\vspace{-5mm}
\subsubsection{Classical rules as baselines in later optimization}
\label{subsubsec:thr_baselines}

Classical thresholding rules are frequently retained as baseline comparators in optimization oriented studies, which is important for assessing whether performance gains arise from the optimizer or from the shrinkage design itself. In \cite{Xiao2022ThresholdOpt}, the universal rule family comprising sqtwolog, heursure, rigrsure, and minimaxi is used as the reference set, while metaheuristic algorithms including Aquila optimizer, gradient based optimizer, and modified grey wolf optimizer are used to optimize threshold values. In \cite{Nguyen2025ICEEMDAN}, combinations of heursure, minimaxi, sqtwolog, and rigrsure are evaluated within an ICEEMDAN-wavelet denoising pipeline and the preferred configuration is selected based on combustion pressure reconstruction accuracy. In \cite{Yang2025PowerDenoising}, the thresholding rule itself, including universal, SURE based, minimax, FDR based, and Bayes type options, is treated as a discrete optimization variable alongside the wavelet family, decomposition level, and transform type.
\vspace{-5mm}
\subsubsection{Classical rules in structurally adaptive settings}
\label{subsubsec:thr_structural}

Some structurally adaptive methods preserve a universal type rule at their core but apply it locally in time or scale so that the effective threshold becomes dependent on local noise conditions. This localization is relevant when disturbances are intermittent and impulsive, since global noise scale estimates may be dominated by bursts. In \cite{Hermawan2024IntervalSWT}, interval and scale dependent thresholds of the form
\begin{equation}
Th_{j,k} = \sigma_{j,k} \sqrt{2\log N_{j,k}}
\end{equation}
are used, where $\sigma_{j,k}$ denotes a local noise standard deviation estimate for level $j$ and time interval $k$, and $N_{j,k}$ denotes the number of samples within that interval. This construction can be interpreted as applying the universal rule after localizing noise estimation, while retaining a closed form threshold structure. In these works, the threshold function remains classical and the methodological novelty lies in where and how the standard rules are applied. A related SWT-based bioacoustic method uses signal autocorrelation and impulsive-noise structure to select level-dependent thresholds for dolphin whistles under low SNR, combining a correlation metric with sliding-window suppression of impulsive coefficients to preserve whistle continuity while reducing underwater noise~\cite{R36}.
\vspace{-5mm}
\subsubsection{New or modified threshold functions}
\label{subsubsec:thr_new}

A large family of works proposes new functional forms for the shrinkage operator, often with tunable parameters that interpolate between hard and soft behavior or that enforce continuity and differentiability at the threshold. Such modifications are relevant in impulsive disturbance settings because they can attenuate large magnitude outliers while avoiding discontinuities that may introduce artifacts or distortions in the reconstructed waveform.

\paragraph{Block and variance based schemes}

Variance based and block thresholding schemes appear in \cite{Lowast2020WristPulse,Zhou2020PDBlockThresh}. Block
scheme implies grouping wavelet coefficients into blocks, estimating a risk for each block by means of SURE, and applying
commonly the same shrinkage factor to each coefficient within
the block. This approach yields smoother attenuation and can reduce artifacts compared to pointwise hard or soft thresholding because it exploits local dependence across coefficients. In \cite{Lowast2020WristPulse}, a variance thresholding strategy is introduced in which local variance computed over sliding windows is used to determine whether coefficients are predominantly noise, thereby defining a structural threshold based on variance comparisons rather than relying exclusively on a global scalar threshold.

\paragraph{Parametric threshold functions}

Parametric thresholding schemes introduce tunable shape parameters that allow the shrinkage function to interpolate between hard and soft behavior. In \cite{Gong2021PowerThreshold}, a parametric threshold function is proposed and the threshold value is made level dependent via scale factors,
\begin{equation}
Th_{j} = \sigma_{j} {\frac{\sqrt{2 \ln N}}{e^{\left(\frac{j}{3}-1\right)}}},
\end{equation}
which yields a flexible level dependent shrinkage operator. Jiang et al.\ proposed a differentiable threshold function for substation condition signals including partial discharge pulses \cite{jiang2022wavelet}. In surface electromyography, an improved Garrote type threshold function with parameters controlling continuity at the threshold and asymptotic decay is proposed in \cite{ouyang2023improved}. In radar applications, \cite{Qiu2024RadarComplex} proposed a continuous, radar specific threshold function tailored to complex clutter environments and designed to preserve weak targets while suppressing noise more smoothly than hard thresholding. In chromatography, \cite{Zhu2023Chromatography} constructed a high order differentiable threshold function whose parameters are tuned using a modified genetic particle swarm optimization procedure. The resulting shrinkage operator is written as \cite{Zhu2023Chromatography}

\[
\overline{wD}_{j,k} =
\begin{cases}
&\operatorname{sgn}(wD_{j,k}) \left(
\left| wD_{j,k} \right|
 - \dfrac{Th}{\alpha^{\beta\left(\sqrt{\lvert wD_{j,k} \rvert / Th}-1\right)}}
\right), \\
& ~~~~~~~~~~~~~~~~~~~~~~~\text{for}~~\lvert wD_{j,k} \rvert \ge Th, \\[2ex]
&0,~~\text{for}~~\lvert wD_{j,k} \rvert < Th ,                         
\end{cases}
\tag{21}
\]
where $\alpha$ and $\beta$ are tuning parameters. When $\alpha=1$, this function reduces to soft thresholding, and when $\alpha \rightarrow +\infty$, it approaches hard thresholding.

\begin{figure}[!t]
\centerline{\includegraphics[width=3.5in]{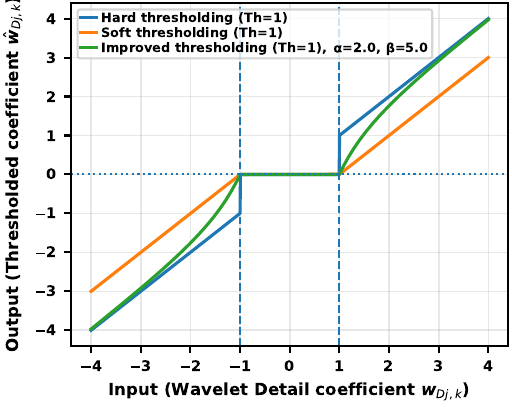}}
\caption{Thresholding function family}
\label{fig:Thresholding_functions}
\end{figure}

Fig.~\ref{fig:Thresholding_functions} compares the coefficient mappings produced by hard (Eq.16), soft (Eq.17), and improved thresholding (Eq.21) for a common threshold. Hard thresholding preserves retained coefficients exactly but is discontinuous at the threshold, while soft thresholding is continuous at the cost of a constant magnitude bias. The improved thresholding function provides a smoother intermediate behaviour, reducing threshold discontinuity while preserving large coefficients more faithfully. This improves the balance between suppressing noise-dominated coefficients and retaining salient signal structures in the reconstructed waveform.
Wang et al.\ modified the regulatory term of a semisoft thresholding function with tunable threshold parameters that can be optimized to suppress Gaussian noise \cite{wang2020novel}. Tian et al.\ proposed an improved wavelet threshold function for bridge temperature monitoring, with a functional form designed to avoid excessive attenuation of slow thermal trends while still reducing high frequency measurement noise \cite{Tian2025BridgeTemp}. An analogous domain-specific construction has been used in satellite navigation, where an improved wavelet-threshold scheme denoises the satellite-signal correlation output and reports SNR gains and stronger correlation-peak detection for weak-signal positioning, navigation, and timing~\cite{R20}.

\paragraph{Hybrid and domain specific shrinkage rules}

Several recent studies embed thresholding within hybrid denoising operators. In \cite{Ahmed2024ECGThresholding}, a hybrid shrinkage function is defined that combines wavelet domain variance estimates with robust time domain statistics, including local median absolute deviation. At each scale, the threshold is shaped by both sets of statistics, yielding an operator that adapts to signal morphology while remaining resilient to transient noise bursts. A related hybrid construction has been applied to underwater acoustic signals. A balanced wavelet-threshold function that interpolates between hard and soft shrinkage through a tunable balance parameter is embedded in a complete ensemble empirical-mode-decomposition (CEEMDAN) pipeline and applied to the noise-dominated modes, using a sym4 basis at five decomposition levels~\cite{R31}. Similarly, Yang et al. used Spearman VMD and SdrSampEn to separate noise-dominated and signal-dominated IMFs, applying improved wavelet-threshold denoising to the former and Savitzky–Golay filtering to the latter for ship-radiated-noise suppression~\cite{R30}. As with the parametric functions discussed above, the balance parameter controls the trade-off between faithful coefficient preservation and smooth attenuation.
In \cite{Lin2024PipelineDents}, alongside a universal type threshold used for standard DWT, a dictionary based mode is introduced in which denoising is performed by iteratively selecting the largest magnitude dictionary coefficients under an energy based stopping criterion rather than by applying a single scalar threshold.
\vspace{-5mm}
\subsubsection{Optimisation and metaheuristic tuning of thresholds}
\label{subsubsec:thr_metaheuristic}

A second major class of approaches treats the threshold values as explicit optimization variables, while typically keeping the thresholding function fixed, most commonly soft thresholding. This class is of practical interest in impulsive and non stationary environments because analytical thresholds derived under Gaussian assumptions may be suboptimal when disturbances are heavy tailed or when noise statistics vary over time and scale.

\paragraph{Optimisation of level dependent thresholds}

In \cite{Xiao2022ThresholdOpt}, the mother wavelet (Daubechies 3), the decomposition depth (five levels), and the use of soft thresholding are fixed a priori, and three metaheuristic optimizers are used to search over level dependent thresholds $Th_{j}$ with the objective of maximizing signal to noise ratio and or minimizing root mean square error. Classical rules serve only as baseline comparators. A similar design is adopted in \cite{Hassan2022AcousticAE}, in which soft thresholding is combined with particle swarm optimization to tune either the thresholds themselves or the parameters of a chosen threshold function, with the goal of improving the signal to noise ratio of acoustic emission signals. A study on acoustic signal denoising proposed a good point set and dynamic elite group guidance simulated annealing selection artificial bee colony algorithm using mean square error as the objective to tune the parameters of a newly proposed threshold function for suppressing Gaussian noise \cite{wang2020novel}. Variants of this optimization centered threshold tuning strategy appear in 2023, including \cite{Ming2023PDHybridPSO,Hua2023PDUltrasonic,Li2023Magnetotelluric}, where hybrid particle swarm optimization, genetic algorithms, and grey wolf optimization are used to adjust SWT based or DWT based thresholds for partial discharge and magnetotelluric signals. In \cite{Zhu2023Chromatography}, the parameters of a smooth, high order differentiable threshold function are tuned using a modified genetic particle swarm optimization procedure. Yang et al.\ proposed a generalized cross validation based objective function to select the wavelet threshold in a seismic random noise suppression framework. The objective function is optimized using an improved chaotic fruit fly optimization algorithm with Kent mapping and adaptive search coefficients, avoiding a requirement for prior noise statistics \cite{yang2024suppressing}.

\paragraph{Threshold rules as discrete optimisation variables}

Several studies treat the type of thresholding rule as a discrete design variable within a broader optimization framework. In \cite{Chen2024SwitchgearML}, metaheuristic algorithms including particle swarm optimization, Energy Valley Optimization, and Subtraction Average Based Optimization are allowed to select among different threshold rules, for example sqtwolog and SURE type criteria, in addition to choosing the wavelet family and decomposition level. In \cite{Yang2025PowerDenoising}, the threshold rule itself, universal, SURE based, minimax, FDR based, or Bayesian, is treated as a categorical optimization variable alongside the wavelet family, decomposition depth, and transform type. In \cite{Brusa2023BearingDenoise}, a more constrained discrete search over Heursure, Minimaxi, Sqtwolog, and Rigrsure in combination with different wavelets and decomposition levels is performed and the preferred configuration is selected based on error metrics.
\vspace{-5mm}
\subsubsection{Structurally adaptive and time varying thresholding}
\label{subsubsec:thr_timevarying}

Several recent approaches adapt the threshold in time, scale, or spectral structure rather than relying on a single global rule. In \cite{Hermawan2024IntervalSWT}, the ECG is partitioned into segments using a change point criterion and a noise variance is estimated for each wavelet level and interval $(j,k)$. Interval dependent thresholds are then applied, making the denoising process adaptive across both scales and time. In \cite{Wang2024MAWDUSCA}, thresholds are distributed across multiple spectral bands, with band specific adaptation guided by an unsupervised source counting procedure rather than a single global threshold. In the context of non stationary signals in electrical power systems, \cite{Yang2025PowerDenoising} incorporates FDR based thresholding as one of its candidate rules, interpreting wavelet coefficients as test statistics with associated $p$ values and choosing thresholds such that the expected proportion of false discoveries is controlled.

\subsection{OPTIMISATION-BASED TECHNIQUES}
\label{subsec:opt_based}

With the proliferation of advanced denoising frameworks, a number of optimisation-based approaches have been proposed in which the mother wavelet, the decomposition level, and the thresholding strategy are treated as joint design variables. In these methods, the design problem is posed explicitly as a search over a coupled parameter space, so that the selected configuration reflects interaction effects between wavelet choice, dyadic depth, and shrinkage behaviour, rather than treating each component as independently optimisable.

A representative example is partial-discharge denoising using the stationary wavelet transform (SWT) in \cite{Ming2023PDHybridPSO}, where a hybrid particle swarm optimisation (PSO) scheme is used to adaptively select both the mother wavelet and the thresholds. In that framework, the threshold component is not restricted to a fixed analytical rule, but is optimised in conjunction with a smooth, multi-order differentiable threshold function derived from Stein's unbiased risk estimate (SURE), thereby coupling the threshold function form and the threshold values to the optimisation objective. In \cite{Ahmed2024ECGThresholding}, an Automatic Wavelet Frequency Band Selection Algorithm explores combinations of wavelet families and decomposition levels and identifies those for which the approximation coefficients exhibit minimal fluctuation while preserving ECG morphology. This procedure yields an automated selection of both the wavelet and the level within the proposed denoising pipeline. Another ECG study employs PSO for jointly searching over mother wavelet family and order, decomposition level, threshold rule, and thresholding type \cite{azzouz2024efficient}. The objective combines SNR improvement and percentage root mean square difference (PRD), so that the PSO selects a configuration that increases SNR while limiting distortion. The method retains standard hard and soft shrinkage functions and classical threshold rules such as sqtwolog, heursure, and rigrsure, illustrating that optimisation-based gains can arise even when the shrinkage primitives remain classical.

Optimisation-based formulations also appear in industrial monitoring and decision-oriented pipelines where the evaluation endpoint is not solely waveform fidelity. For partial-discharge denoising in switchgear insulation, metaheuristic algorithms including Energy Valley Optimisation, Subtraction Average Based Optimisation, and PSO are applied to jointly optimise the wavelet family, decomposition depth, and threshold rule, with RMSE and classification accuracy used as objective measures \cite{Chen2024SwitchgearML}. Similarly, in ultrasonic nondestructive testing, a GWO-VMD framework optimises VMD penalty and mode-number parameters, separates IMF components using correlation coefficients, and applies an improved wavelet-threshold function to noise-dominated modes, with SNR and RMSE used as denoising objectives~\cite{R35}. A comparable metaheuristic formulation has been reported for underwater acoustic denoising, where sand-cat-swarm optimisation adaptively selects the wavelet basis and tunes an improved threshold function, combined with empirical-mode-decomposition preprocessing. The configuration is evaluated by SNR gain, with a reported improvement of 9–13 dB on benchmark and ship-radiated-noise signals~\cite{R21}. A telecommunications instance of this kind of joint optimisation is the tuning of a wavelet-denoising and energy-detection pipeline for cognitive-radio spectrum sensing, where the denoising configuration is optimised against detection-theoretic objectives, namely probability of detection, probability of false alarm, sensing duration, and throughput across SNR, rather than against waveform-fidelity metrics~\cite{R22}.

In pipeline dent assessment, a wavelet-based denoising framework constructs an overcomplete two-dimensional dictionary from Daubechies and Symlet wavelets and then selects atoms via sparse approximation \cite{Lin2024PipelineDents}. In that formulation, the conventional notion of selecting a single mother wavelet is replaced by selecting a structured dictionary, and wavelet selection is effectively achieved through the sparse coding process rather than direct index choice. A practically important advancement is reported in \cite{das2025eegopt}, which formulates an EEG classification pipeline as a hierarchical hyperparameter optimisation problem and applies Bayesian optimisation via Tree-Structured Parzen Estimator (TPE) to choose the denoising method, wavelet packet decomposition or empirical mode decomposition, the wavelet packet mother wavelet, the wavelet packet or empirical mode decomposition threshold, the feature family, principal component analysis variance retention, the classifier type, and classifier hyperparameters. The optimisation objective is cross-validated Matthews correlation coefficient (MCC), and the framework introduces caching and pruning to make repeated signal transformations computationally feasible \cite{das2025eegopt}.

From the perspective of telecommunications and networking, the methodological implication of these works is that wavelet denoising parameters should be treated as coupled receiver design choices. When sampling rate, occupied bandwidth, and disturbance statistics vary across deployments, fixing a single component, such as the decomposition level, while optimising the threshold alone can produce configurations that perform well under one operating point but transfer poorly. This coupling is also reflected in studies where the threshold rule is treated as a categorical variable and optimised jointly with the wavelet and level \cite{Chen2024SwitchgearML,Yang2025PowerDenoising,Brusa2023BearingDenoise}, and in studies where level-dependent thresholds are tuned explicitly by metaheuristics \cite{Xiao2022ThresholdOpt,Hassan2022AcousticAE,Ming2023PDHybridPSO,Zhu2023Chromatography,yang2024suppressing}. These observations motivate future receiver-aligned formulations in which optimisation objectives are defined directly in terms of task endpoints, such as estimation accuracy or detection performance, rather than waveform fidelity alone.

\subsection{TRANSFERABILITY OF CROSS-DOMAIN FINDINGS TO COMMUNICATION SIGNALS}

Much of the evidence synthesized here originates in biomedical, industrial, and geophysical signal processing, since, as Fig.~\ref{fig:heat_map} shows, communication-specific parameter-selection studies remain sparse. The transfer to communication receivers is nonetheless well founded, because the selection criteria, namely similarity, sparsity, entropy, energy concentration, and subband SNR, are statistics of the wavelet representation rather than of domain semantics; their behaviour depends on how well the basis and depth match signal morphology and noise, not on whether the signal is an electrocardiogram, a partial-discharge transient, or an OFDM symbol. What transfers is therefore the selection methodology; what must be re-specified is its inputs, and what must be re-validated is its endpoint.

Three inputs are domain-dependent. The structural prior used by similarity-based selection becomes a pilot, preamble, pulse-shaping waveform, or estimated channel impulse response, in place of a physiological template, as recast in Table~\ref{tab:mw_metrics_comms}. The noise model becomes the AWGN, impulsive (for example Middleton Class A), narrowband, and colored disturbances of communication channels; here the thresholding logic transfers unchanged while the noise statistics that parameterize it differ. The protected spectral support is redefined through the same mechanism in every domain, since the band-alignment heuristic of Section IV-C ties the depth to the occupied bandwidth exactly as it ties to canonical EEG rhythms, the dyadic band edges scaling with the sampling frequency \(f_{\mathrm{s}}\). This shared sampling-bandwidth relation is why the coupling applies identically to receiver design.

The decisive difference is the endpoint. Cross-domain studies optimize waveform-fidelity objectives (SNR, RMSE, PSNR, reference correlation), whereas communication performance is governed by BER, EVM, channel-estimation error, synchronization, and detection probability, and a fidelity-optimal configuration is not necessarily task-optimal. The communication experiments here make this concrete: in Tables~\ref{tab:receiver_level_performance},~\ref{tab:complexity_realtime} and~\ref{tab:friedman_results} the DWT-versus-WPT BER difference is not statistically significant although NMSE and EVM differ, so waveform-fidelity gains need not translate into hard-decision gains. This is why the framework closes on receiver-level validation in Section IV-G rather than on waveform smoothness.

Two features of communication signals limit naive transfer. The baseband is complex valued, whereas most cross-domain methods are real-valued, so applying them to the I/Q components requires per-component processing or a phase-coherent extension such as the DTCWT; the complex-I/Q evaluation in Table~\ref{tab:radioml_fom_efi_wavelet} and the DTCWT channel-estimation instance in~\cite{R8} are included for this reason. Reception is also timing-sensitive, since bursts arrive at arbitrary sample offsets, which favors the shift-invariant SWT. Consequently, the specific wavelet and depth reported as optimal in a non-communication study should be read as evidence for a selection principle, not as a portable recommendation; this review accordingly transfers the methodology, through the receiver-aware framework and the communication-oriented metrics of Tables~\ref{tab:perf_metrics},~\ref{tab:mw_metrics_comms} and~\ref{tab:dl_pros_cons}, while re-deriving the parameter values against communication endpoints.

\subsection{PROPOSED STRATEGIES FOR SELECTION OF INDIVIDUAL WAVELET PARAMETERS IN COMMUNICATION SYSTEMS}

\begin{figure*}[htbp]
\centering
\includegraphics[width=\textwidth,height=0.9\textheight,keepaspectratio]{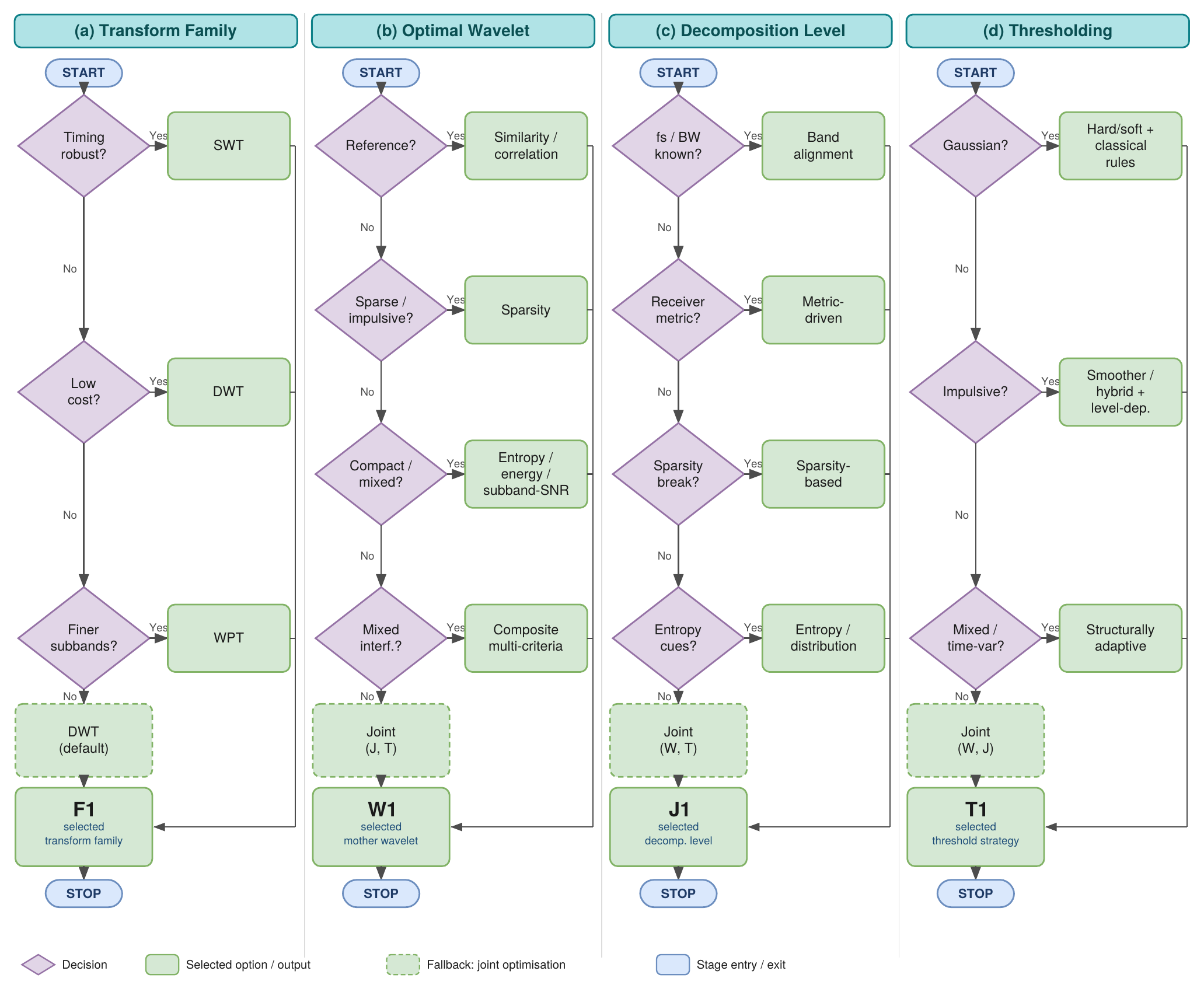}
\caption{Decision Rule for Selection of Wavelet Parameters in Communication Systems}
\label{fig:Decision_Rule}
\end{figure*}

The decision framework of Fig.\ref{fig:unified_decision_framework} is source-grounded, in the sense that every parameter is derived from measurable signal and system attributes rather than assumed by convention. The relevant attributes are the receiver architecture, the signal structure, the sampling frequency \(f_s\), the occupied bandwidth, the sensitivity to temporal alignment, and the disturbance statistics. This grounding allows a single methodology to adapt across heterogeneous environments, from communication receivers and localization systems to preamble-driven synchronization pipelines and bandwidth-constrained embedded platforms, without presuming that one configuration transfers between them. Within this formulation, each parameter assumes a distinct role: the transform family fixes the time-frequency representation and the implementation cost, the mother wavelet controls structural matching to a reference or to sparse signal features, the decomposition depth sets the scale separation relative to the spectral support that must be preserved,
and the thresholding rule governs the bias--variance trade-off under the assumed interference model. The objective in every case is operational, expressed through receiver-level performance and cost rather than through waveform smoothness alone. The detailed decision logic for the four stages is given in the flowcharts of
Fig.~\ref{fig:Decision_Rule}.

The pathway begins with a characterization layer that fixes the operating point before any transform is chosen. It records the receiver scenario, whether OFDM or burst-mode reception, a GNSS or localization pipeline, a channel-estimation or preamble-driven architecture, an underwater acoustic link, or an industrial IoT or power-line environment, together with the sampling frequency \(f_s\), occupied bandwidth, and disturbance model. These descriptors are the common inputs to the four subsequent selection stages and determine which branch of each strategy flowchart is taken. Transform-family selection proceeds on functional and computational grounds. The SWT is preferred where translation invariance is essential, since its undecimated construction stabilizes coefficient alignment under the timing offsets typical of packet and burst reception. The DWT suits real-time, memory-efficient, or low-complexity implementation, reflecting its critically sampled, non-redundant structure. The WPT is reserved for cases requiring finer subband partitioning to match the spectral support of signal and interference, and is useful only when the packet granularity aligns with the actual occupancy structure.

Mother-wavelet selection is keyed to signal morphology and to the prior structure available at the receiver. A reliable reference, such as a pilot, preamble, pulse-shaping waveform, or estimated channel impulse response, favors similarity and correlation criteria that preserve the structure required by the downstream correlator. Sparse, impulsive, or burst-edge-dominated signals favor sparsity criteria, while compact representation under mixed spectral occupancy favors entropy, energy-concentration, or subband-SNR criteria. When neither the mother wavelet nor the decomposition level is known in advance, the framework inserts a coupled wavelet--level resolution step between the wavelet-selection and level-selection blocks. This step supports three operating modes. The first is joint search, in which all admissible \((\psi,J)\) pairs are evaluated using a reference-based or receiver-level objective. The second is profiled search, in which the best level is first estimated for each candidate wavelet and the wavelets are then compared at their corresponding profiled levels. The third is coordinate refinement, in which an initial wavelet and level are chosen from bandwidth or sparsity considerations and then alternately refined until the validation metric stabilizes. This addition prevents circular decision making and makes the framework applicable when no prior wavelet or decomposition depth is available.

Decomposition-level selection links \(J\) directly to sampling and bandwidth. After \(J\) levels, the approximation subband spans approximately \(0\) to \(f_N/2^J\), with \(f_N=f_s/2\), so \(J\) fixes the boundary between the preserved coarse structure and the detail bands subjected to thresholding. When the spectral support is known a priori, band-aligned selection takes the smallest \(J\) that retains the protected band; when a downstream observable is available, \(J\) follows from metric-driven search; when scale transitions are informative, \(J\) is set at the sparsity breakpoint between noise- and signal-dominated scales; and when no clean reference exists, entropy or coefficient-distribution cues are used. Because the dyadic band edges scale with \(f_s\), a fixed integer depth does not transfer across sampling rates or numerologies and must be reassessed at each operating point.

Thresholding selection fixes the shrinkage function and threshold rule together according to the disturbance. Approximately Gaussian noise is handled by hard or soft shrinkage with analytical rules such as the universal (VisuShrink), minimax, or SURE-type thresholds. Impulsive or heavy-tailed interference favors smoother nonlinear, block, or hybrid functions with level-dependent thresholds that resist domination by sporadic large coefficients, while mixed or locally nonstationary disturbances favor structurally adaptive, time-varying schemes that track local noise across time and scale. Where the noise model is uncertain or coupled to the earlier choices, the threshold is optimized jointly with the wavelet and level.

The candidate configuration \(\{F1,W1,J1,T1\}\) is finally submitted to the receiver-level validation and complexity check, which closes the framework on operational rather than visual criteria. It is assessed against task endpoints, including BER, EVM, channel-estimation error (MSE/NMSE), localization error, and complex SNR gain, and weighed against implementation cost through execution time and a FLOP proxy. Only the configuration meeting the chosen performance objective at acceptable cost is consolidated into the final recommendation, yielding a receiver-aware methodology in which transform family, mother wavelet, decomposition level, and thresholding are selected jointly from signal and system context and validated against the endpoints that govern receiver performance.
\color{black}
\subsubsection{Validation of Proposed Decision Framework}

The proposed decision framework, that selects a wavelet-denoising configuration (transform family, mother wavelet, decomposition depth, thresholding policy) is validated at the receiver level. Validation is conducted across two complementary case studies drawn from the telecommunications literature: a synthetic pilot-aided OFDM channel-estimation benchmark and a real-world study using over-the-air measured IEEE 802.11n channels. The framework is evaluated exclusively through receiver‑level performance metrics (BER, EVM, channel‑estimation NMSE and complex SNR gain) thereby deliberately avoiding waveform‑fidelity or generative‑similarity surrogates.

The first case study is a comb-type pilot OFDM link with a frequency-selective Rayleigh channel and QPSK payload, in which each frame is an independent realization with an independent channel and noise draw. The link uses 128 subcarriers, a 32-sample cyclic prefix, comb pilots at spacing four (32 pilots and 96 data subcarriers), and a 12-tap exponential power-delay-profile Rayleigh channel renormalized per frame, at a 30.72 MHz sampling rate and 5 MHz occupied bandwidth. Frames are generated over an SNR grid of 0–20 dB under two noise regimes, AWGN and a bursty-impulsive regime that augments the Gaussian background with sparse high-amplitude impulses. Table~\ref{tab:synthetic_ofdm_holdout} reports the performance of the framework selected wavelet configuration. The performance is compared with fixed‑parameter wavelet baseline and classical baselines (raw least‑squares, DFT(tuned)). The framework attains the lowest channel-estimation NMSE and the highest complex SNR gain in both noise regimes, while its BER is statistically indistinguishable from the strongest classical baseline and below that of the fixed-parameter wavelet. Holm-corrected Wilcoxon gives framework-versus-fixed-wavelet \(p \approx 6.1 \times 10^{-18}\) under AWGN and
\(p \approx 4.8 \times 10^{-12}\) under impulsive noise, and framework-versus-raw-LS
\(p \approx 4.5 \times 10^{-22}\) and \(p \approx 1.5 \times 10^{-16}\), respectively. This indicates that the selected configuration produces the cleanest channel estimate.

To assess the external validity, the decision framework is evaluated on publicly available IEEE 802.11n channel state information (CSI) captured from commodity Atheros 802.11n network interface cards using the Atheros CSI Tool ~\cite{R42},~\cite{Xie2015Splicer}, which patches the card firmware and driver to expose per-subcarrier, per-antenna-pair complex channel coefficients. Because the reported CSI grid and the synthetic link differ in size, each per-frame complex CSI vector is converted to a channel impulse response by an inverse DFT, truncated to the significant delay-spread taps to suppress the noise floor, then zero-padded and transformed back to 128 frequency bins to give the per-frame frequency response. Each realisation is renormalised to unit average channel power so that the operating SNR is set by the added noise rather than by the
measured path loss, and the comb-pilot structure, cyclic prefix, QPSK payload, and noise regimes
of the synthetic benchmark are retained unchanged. This preserves the measured multipath
structure, including its frequency selectivity and its departure from the exponential
power-delay-profile assumption, while keeping every other element of the receiver chain identical
to the synthetic case, so that the difference between Tables~\ref{tab:synthetic_ofdm_holdout} and~\ref{tab:measured_channel_holdout} is attributable to the channel and not to the link configuration.

The performance reported in Table~\ref{tab:measured_channel_holdout} exhibits lowest BER, EVM, and NMSE in both noise regimes. Holm-corrected Wilcoxon gives \(p \approx 4.0 \times 10^{-41}\) versus raw LS and \(p \approx 1.2 \times 10^{-39}\) versus the fixed wavelet.
This demonstrates that the framework generalizes from modeled to genuinely measured propagation, retaining the lowest estimation error and the lowest BER and EVM.These results establish the framework as a technically defensible, use-case-driven validation framework for telecommunications applications.

\begin{table*}[!t]
\centering
\caption{Synthetic OFDM held-out test performance under AWGN and impulsive-noise regimes.}
\label{tab:synthetic_ofdm_holdout}
\footnotesize
\setlength{\tabcolsep}{7pt}
\renewcommand{\arraystretch}{1.08}
\begin{tabular}{llcccc}
\toprule
\textbf{Noise Regime} & \textbf{Method} & \textbf{BER} & \textbf{EVM (\%)} & \textbf{NMSE} & \textbf{SNR Gain (dB)} \\
\midrule
AWGN & Raw LS          & 0.1066 & 60.5 & 0.2063 & 0.00 \\
AWGN & DFT tuned       & \textbf{0.0877} & \textbf{50.0} & 0.0715 & 8.27 \\
AWGN & Fixed wavelet   & 0.0934 & 51.3 & 0.1206 & 7.61 \\
AWGN & Proposed framework & 0.0883 & 50.1 & \textbf{0.0646} & \textbf{8.38} \\
\midrule
Impulsive & Raw LS          & 0.1013 & 60.0 & 0.2208 & 0.00 \\
Impulsive & DFT tuned       & 0.0828 & \textbf{48.8} & 0.0716 & \textbf{8.63} \\
Impulsive & Fixed wavelet   & 0.0905 & 50.4 & 0.1466 & 7.57 \\
Impulsive & Proposed framework & \textbf{0.0826} & \textbf{48.8} & \textbf{0.0693} & 8.61 \\
\bottomrule
\end{tabular}

\vspace{1mm}
\footnotesize
\end{table*}

\begin{table*}[!t]
\centering
\caption{Real measured-channel held-out test results (IEEE 802.11n channels).}
\label{tab:measured_channel_holdout}
\footnotesize
\setlength{\tabcolsep}{7pt}
\renewcommand{\arraystretch}{1.08}
\begin{tabular}{llcccc}
\toprule
\textbf{Noise Regime} & \textbf{Method} & \textbf{BER} & \textbf{EVM (\%)} & \textbf{NMSE} & \textbf{SNR Gain (dB)} \\
\midrule
AWGN & Raw LS          & 0.0713 & 52.4 & 0.2199 & 0.00 \\
AWGN & DFT tuned       & 0.0712 & 52.1 & 0.2209 & -0.60 \\
AWGN & Fixed wavelet   & 0.0706 & 52.0 & 0.1866 & -0.20 \\
AWGN & Proposed framework & \textbf{0.0670} & \textbf{51.3} & \textbf{0.1661} & \textbf{0.20} \\
\midrule
Impulsive & Raw LS          & 0.0541 & 47.9 & 0.2091 & 0.00 \\
Impulsive & DFT tuned       & 0.0546 & 47.8 & 0.2130 & -0.59 \\
Impulsive & Fixed wavelet   & 0.0543 & 47.6 & 0.1872 & -0.26 \\
Impulsive & Proposed framework & \textbf{0.0511} & \textbf{46.9} & \textbf{0.1747} & \textbf{0.12} \\
\bottomrule
\end{tabular}

\vspace{1mm}
\footnotesize
\end{table*}

One limitation of both case studies should be stated explicitly. The classical baselines
used here are raw least squares and a tuned DFT-domain estimator, whereas impulsive-noise
mitigation in deployed receivers is dominated by nonlinear time-domain preprocessing such as
clipping and blanking, as noted in Section~IV-D. A linear minimum mean square error estimator
with known channel and noise statistics would likewise provide an informative upper bound on what
any linear estimator can achieve on this link. Neither comparison is reported here, so the
results should be read as establishing that receiver-aware wavelet-parameter selection outperforms
fixed-parameter wavelet denoising and the estimation-domain baselines considered, and not as
establishing superiority over the nonlinear mitigation schemes that dominate impulsive-noise
practice or as quantifying the residual gap to the linear performance bound. Benchmarking against
clipping, blanking, and an LMMSE bound is identified in Section~IV-I as necessary further work.

\subsection{IMPLEMENTATION CONSTRAINTS ON EMBEDDED DSP AND FPGA PLATFORMS}

Because the proposed framework inserts denoising ahead of synchronization and demodulation (Fig.~\ref{fig:Wavelet-Enhanced Receiver}), its arithmetic, memory, and latency costs add directly to the receiver budget. The framework already treats cost as a first-class axis through the complexity-check stage of Fig~\ref{fig:unified_decision_framework} and the descriptors of Table~\ref{tab:perf_metrics}, namely the per-frame operation count $C_{\mathrm{frame}}$, the per-frame latency
$t_{\mathrm{frame}}$, and the real-time indicator $\mathcal{F}_{\mathrm{RT}}$, using the symbols
defined in that table rather than any separate notation.

The transform families differ in complexity class: the critically sampled DWT admits a Mallat realization~\cite{S1} of cost \(\mathcal{O}(N)\) with in-place storage; the undecimated SWT retains full-length coefficients at every scale, raising arithmetic and memory cost to \(\mathcal{O}(N\log_2 N)\), roughly \(\log_2 N\) times the DWT; and the WPT has a full-tree cost of \(\mathcal{O}(N\log_2 N)\) plus best-basis overhead, matching the SWT in order and in per-level operation count. The tabulated proxies of 62,720, 143,360, and 215,040 operations per frame for DWT, WPT, and SWT in Table~\ref{tab:complexity_realtime} therefore separate the DWT from the other two on structural grounds and separate SWT from WPT on the decomposition depths selected for each, so the shift invariance that benefits impulsive-noise suppression (Table~\ref{tab:receiver_level_performance}) is purchased with a proportional rise in resource demand.

On programmable DSP cores, latency scales with the multiply-accumulate count, hence with filter length and decomposition depth. The longer-support, fidelity-oriented wavelets of Table~\ref{tab:radioml_fom_efi_wavelet} (e.g. db37, coif16) therefore cost more per sample, which is exactly the trade-off penalized by the efficiency index EFI of Section IV-B that favors shorter bases for constrained receivers. Such cores also operate in fixed-point arithmetic, so the filter coefficients, the median-absolute-deviation noise estimate, and the shrinkage nonlinearity are quantized; insufficient wordlength biases the threshold and can erode the BER and EVM gains the framework seeks to certify, making wordlength an implementation-level denoising parameter. The lifting-scheme factorization roughly halves the arithmetic, enables in-place integer-to-integer computation, and is the preferred low-power realization~\cite{R40}.

On FPGA fabrics the binding constraints are spatial. Resource use is set by look-up tables, DSP slices, and block memory, while a polyphase or lifting data-path can be pipelined to one sample per clock~\cite{R40}, so memory and multiplier count, rather than throughput, usually dominate. The undecimated SWT is again the costliest family, since the absence of decimation forces full-length buffers at every level and removes the resource amortization that down sampling affords the DWT; longer filters likewise consume more DSP slices~\cite{R41}, and boundary extension and frame buffering add latency and storage that belong in the budget.

These considerations indicate that the complexity-check stage should be parameterized by the target platform rather than by a single processor-independent cost, expressed through multiply-accumulate operations per sample, memory footprint, latency in clock cycles relative to the receiver timing budget, resource utilization, fixed-point wordlength, and energy per frame, evaluated jointly with the receiver-level endpoints of Section IV-G. The execution-time and FLOP-proxy figures in Tables~\ref{tab:complexity_realtime} and~\ref{tab:perf_metrics} derive from a software benchmark on a general-purpose processor and establish the relative ordering of the transform families rather than absolute embedded performance, since constant factors, memory behaviour, and quantization on a specific DSP or FPGA target can alter the picture. Hardware-in-the-loop validation, including fixed-point sensitivity and resource-versus-fidelity characterization, is therefore a necessary step toward deployment and is noted in Section IV-I.

\subsection{LIMITATIONS AND FUTURE DIRECTIONS}
\label{subsec:limitations_future}

Despite measurable progress toward quantitatively guided parameter selection, current wavelet-denoising studies remain constrained by restricted design spaces and limited robustness validation. In many studies, the mother wavelet search is confined to a narrow subset of orthogonal families and a small range of orders, which can bias claims of optimality and impede transfer across acquisition conditions. Mechanistic interpretation is often underdeveloped, because intrinsic wavelet attributes such as vanishing moments, symmetry and linear phase behaviour, regularity, and compact support are rarely mapped explicitly to the signal structure and noise statistics. Consequently, reported optima may reflect dataset-specific tuning rather than reproducible design principles.

Decomposition depth selection exhibits similar methodological fragility. Although band-alignment heuristics and trial-based scans are widely used, they often neglect boundary artefacts, spectral leakage, and uncertainty in the effective band partitions induced by finite-length signals and implementation details such as padding, signal extension, and filter transients. These effects can distort coefficient statistics and the apparent separation between signal-bearing and noise-dominated subbands, thereby confounding depth-selection conclusions.

Thresholding research, while increasingly framed as an optimisation problem, continues to inherit simplifying assumptions that can be mismatched to real data. Coefficient-wise independence and additive white Gaussian noise (AWGN) like perturbations remain implicit in many pipelines, with classical hard and soft shrinkage and universal or SURE-type rules frequently used as defaults \cite{Baldazzi2020NeuralDenoising,Suhail2020EEGCognitive,Potdar2021PCG,malleswari2021improved,Ghosh2022PCGParametric,daud2022denoising}. Meanwhile, newer parametric or hybrid shrinkage functions are not consistently paired with transparent selection protocols, sensitivity analyses, or fully reproducible disclosure of tuned parameters \cite{Gong2021PowerThreshold,jiang2022wavelet,ouyang2023improved,Qiu2024RadarComplex,Zhu2023Chromatography,wang2020novel,Tian2025BridgeTemp,Ahmed2024ECGThresholding}. Where optimisation is applied, objective functions are often dominated by waveform fidelity criteria such as SNR, MSE, RMSE, or PSNR, even when the operational endpoint is task-driven, such as detection accuracy or classification performance \cite{Chen2024SwitchgearML,das2025eegopt}. This mismatch can yield denoisers that maximise waveform similarity yet degrade downstream decision performance.

Within the screened corpus, learning-based wavelet-parameter selection is already evident, but only outside communication domains. Decision-tree learning has been used to select the optimal decomposition level for partial-discharge denoising ~\cite{soltani2020decision}, while Bayesian optimisation with a Tree-Structured Parzen Estimator has been used to jointly select the transform, mother wavelet, decomposition level, and thresholding configuration in an EEG-classification pipeline ~\cite{das2025eegopt}. Together with the broader optimisation-driven literature reviewed in Section IV-E, these studies show that data-driven wavelet-parameter selection is feasible and effective, but its validated use remains concentrated in biomedical and industrial signal processing.

Related work beyond the screened corpus further indicates that this paradigm is maturing. Examples include AI-based wavelet-parameter optimisation for radar-signal denoising~\cite{A1}, dynamic wavelet-basis selection for inertial-sensor enhancement~\cite{A2}, reinforcement-learning-based mother-wavelet selection for ECG diagnosis~\cite{A3}, and learnable WPT with trainable analysis and synthesis filters for time-series denoising ~\cite{A4}. These studies demonstrate that wavelet parameters can be learned directly, although the evidence still comes from domains adjacent to, rather than within communication links.

In communication systems, learning-based wavelet-domain methods are beginning to appear, but they mainly embed fixed wavelet operations within trainable receivers or detectors instead of explicitly selecting wavelet parameters. The main gap, therefore, lies at the intersection of these two directions: learning-based wavelet-domain processing is emerging in communications, and learning-based wavelet-parameter selection is established elsewhere, but explicit learning-based selection of wavelet parameters for communication links remains largely unexplored. This gap motivates the need to move beyond using learning only as an external selector and toward denoising operators that learn wavelet parameters directly from communication-relevant data.

AI-based selection is currently limited to a small number of studies, such as decision-tree level selection and Bayesian optimisation of denoising within classification pipelines \cite{soltani2020decision,das2025eegopt}, and broader learning-based, generalisable, and interpretable frameworks remain underexplored. A key opportunity is to move beyond using learning only as an outer-loop selector and instead incorporate learning into the denoising operator so that parameters are learned directly from data.
A similar pattern holds for radio-over-fibre and coherent optical links, where impairment mitigation has shifted toward data-driven methods rather than wavelet denoising. Examples include optimised recurrent-neural-network digital predistortion for analog radio-over-fibre~\cite{R23} and real-time machine-learning fibre-nonlinearity compensation in coherent optical networks~\cite{R24}. No dedicated wavelet-denoising parameter-selection study for radio-over-fibre was found within the 2020–2025 search window, and this coverage gap is recorded here for transparency.

Underwater acoustic denoising further shows how domain and method interact across the present corpus. The wavelet studies in this setting, discussed in Sections IV-B, IV-D, and IV-E, mostly target ship-radiated-noise and target-recognition signals rather than the communication link, so they are classified under ocean acoustics rather than telecommunications. By contrast, denoising for the underwater acoustic communication link itself, such as pilot-aided channel estimation and OFDM signal recovery, has over the same period been handled mainly by learning-based denoisers~\cite{R25},~\cite{R26}, including a convolutional-neural-network FBMC receiver that recovers symbols without explicit channel estimation ~\cite{R27}. This contrast between wavelet-dominated ocean-acoustic denoising and learning-dominated communication-link denoising is a clear example of the heuristic-to-data-driven transition emphasised in this review, and it points to where wavelet parameter-selection principles and data-driven optimisation might be combined most productively. Cognitive-radio spectrum sensing has likewise seen wavelet and deep-learning hybrids that embed wavelet denoising inside learned detectors, including a deep wavelet network operating at very low SNR~\cite{R28} and a convolutional model that uses wavelet soft-thresholding as a secondary denoising stage~\cite{R29}. These reflect the same migration of wavelet-domain operations into trainable architectures.A similar trend is visible in OFDM-IM signal detection, where wavelet convolutional neural networks integrate wavelet-domain multiscale feature extraction with CNN-based detectors to improve BER and computational efficiency~\cite{R32}.

\vspace{-6mm}
\subsubsection{Deep unfolding and unrolled wavelet networks}
\label{subsubsec:deep_unfolding}

A promising direction is deep unfolding, also termed unrolling, in which an iterative signal recovery procedure is expressed as a finite-depth neural network whose layers correspond to algorithmic iterations. This paradigm is relevant because a large fraction of wavelet denoising pipelines can be interpreted as repeated application of two operations, namely a linear analysis or synthesis step and a nonlinear shrinkage step. In classical formulations, the shrinkage step is governed by hand-designed choices, including the threshold rule, the threshold values, and the functional form of the shrinkage operator, and these must be tuned either analytically under restrictive assumptions or by empirical search.

In an unfolded formulation, the parameters associated with shrinkage become trainable and can be learned using gradient-based optimisation from representative data. This provides a learning-based replacement for manual parameter tuning and for black-box metaheuristics, while preserving interpretability because the network layers retain a direct correspondence to denoising steps. The link to the optimisation-based evidence in the reviewed literature is direct. Metaheuristic studies already demonstrate that treating scale-dependent thresholds and shrinkage parameters as optimisation variables can yield measurable gains \cite{Xiao2022ThresholdOpt,Hassan2022AcousticAE,wang2020novel,Ming2023PDHybridPSO,Zhu2023Chromatography,yang2024suppressing}. Deep unfolding can be interpreted as replacing these outer-loop searches with a differentiable mechanism in which the thresholds and shrinkage parameters are learned from data, thereby reducing reliance on hand-selected candidate sets and potentially improving cross-condition transfer.

This direction is particularly relevant in telecommunications and networking settings in which channel statistics and disturbance models vary. In such settings, training data can be drawn from channel measurements or from validated simulators, and unfolded architectures can learn threshold schedules and shrinkage shapes that adapt to heavy-tailed and impulsive disturbances without requiring explicit parametric noise modelling. Moreover, the learned parameters remain interpretable as scale-dependent thresholds and shrinkage coefficients, which supports diagnostic analysis and controlled deployment.

A complementary opportunity is to integrate trainable wavelet priors. Several reviewed works restrict the wavelet family to a small candidate set and then select an index through search or optimisation \cite{Ming2023PDHybridPSO,Chen2024SwitchgearML,azzouz2024efficient}. In an unfolded framework, the analysis and synthesis filters associated with a wavelet transform can be parameterised and learned, subject to constraints that preserve perfect reconstruction or approximate orthogonality. This would allow the system to learn wavelet-like filter banks that are matched to signal and interference structures observed in channel data while retaining a multiresolution interpretation. Such trainable wavelet priors can also be combined with structurally adaptive and time-varying thresholding concepts already explored in the literature \cite{Hermawan2024IntervalSWT,Wang2024MAWDUSCA,Yang2025PowerDenoising}, yielding denoisers that adapt across time and scale in a principled and data-driven manner.
\vspace{-6mm}
\subsubsection{Research priorities}
\label{subsubsec:future_priorities}

Future work should prioritize five interlinked directions. The first direction is multi-objective, uncertainty-aware optimization that jointly tunes wavelet family and order, redundancy choice, including DWT, SWT, dual-tree, and wavelet packet variants, decomposition depth, and thresholding under heterogeneous noise models, while explicitly incorporating computational cost and reporting uncertainty through confidence intervals, cross-dataset stability, and sensitivity to candidate-set restrictions. The second direction is principled level selection that moves beyond fixed band heuristics toward data-adaptive criteria validated against leakage and boundary effects, leveraging risk estimators, stability curves, information criteria, and cross-validated task loss. The third direction is dependency-aware shrinkage that replaces coefficient-wise independence assumptions with models that exploit inter-scale and intra-scale dependencies, including structured Bayesian formulations and structured sparsity priors, particularly under nonstationary and impulsive noise. 

The fourth direction is task-aligned benchmarking and reproducibility, which standardizes evaluation with heterogeneous-noise stress tests, cross-domain transfer experiments, and full reporting of parameter coupling, enabling defensible claims of generality and facilitating rigorous comparative analysis. A specific and immediate requirement within this direction is to benchmark receiver-aware wavelet configurations against the nonlinear time-domain schemes that dominate impulsive-noise practice, namely clipping and blanking, and against a linear minimum mean square error estimator with known statistics as an upper bound, so that the residual gap to the linear performance limit is quantified rather than assumed.
The fifth direction is AI-guided parameter selection and end-to-end learning with wavelet priors. Beyond outer-loop selection, learning-based frameworks should be developed that predict wavelet family and order, decomposition depth, and scale-dependent thresholds directly from signal statistics or latent representations and validate cross-domain transfer. Differentiable wavelet transforms and trainable shrinkage functions should be explored so that wavelet, level, and threshold choices can be optimized with gradient-based learning while preserving interpretability through wavelet-domain regularization. The deep unfolding paradigm described in Section~\ref{subsubsec:deep_unfolding} provides a concrete mechanism for achieving this transition because it retains the structure of iterative wavelet denoising while enabling data-driven learning of parameters that otherwise require manual tuning.

\section{CONCLUSION}

This review has synthesised advances in wavelet-based signal denoising reported between 2020 and 2025 through a distinctly parameter-centred perspective. Rather than treating denoising as a generic preprocessing operation, it has examined the mother wavelet, decomposition level, and thresholding rule as coupled design decisions that jointly govern performance. Across biomedical, industrial, geophysical, environmental, and communication domains, the surveyed literature shows a clear movement away from precedent-based defaults toward quantitatively justified, optimisation-oriented, and increasingly data-driven parameter selection. Recent studies make growing use of sparsity, entropy, energy-concentration, and subband-SNR criteria for wavelet choice, more systematic strategies for decomposition-level determination, and adaptive or optimised thresholding rules tailored to signal and noise characteristics. Because many of these criteria are statistics of the wavelet representation rather than domain-specific descriptors, they can support transfer across application domains when the structural prior, disturbance model, protected spectral support, and validation endpoint are appropriately redefined.

A principal outcome of the review is the consolidation of this evidence into a practical Decision Framework for Wavelet-Denoising Parameter Selection in Communication Systems. The framework provides communication-system engineers with a source-grounded and reproducible pathway for selecting the wavelet family, mother wavelet, decomposition depth, and thresholding strategy according to measurable signal and system descriptors, including signal type, sampling conditions, occupied bandwidth, receiver architecture, and noise assumptions. Importantly, the framework is not presented only as a conceptual taxonomy. It is validated at the receiver level using a synthetic pilot-aided OFDM channel-estimation benchmark and a study based on over-the-air measured IEEE 802.11n channels, and is assessed against BER, EVM, channel-estimation NMSE, and complex SNR gain.

The outcome should be stated precisely, because the two case studies do not support an unqualified claim of superiority over every baseline. On the synthetic link the
framework-selected configuration attains the lowest channel-estimation NMSE and significantly
outperforms the fixed-parameter wavelet, but a tuned DFT-domain estimator reaches a marginally
lower BER under AWGN and a marginally higher SNR gain under impulsive noise, and the BER
difference between the framework and that baseline is not statistically significant. The
defensible synthetic-link claim is therefore the lowest estimation error at statistically equivalent bit-error performance, together with a clear and significant advantage over the fixed-parameter wavelet. On the measured channels the framework additionally attains the lowest BER, EVM, and NMSE among all compared methods, so the stronger claim holds there and the advantage persists under genuinely measured propagation. These results reinforce a central methodological point: waveform-fidelity-optimal configurations need not be task-optimal. Consequently, wavelet-parameter selection for communication receivers should close on receiver-level endpoints and implementation cost rather than on visual smoothness or generic signal-fidelity measures alone.

A second qualification delimits where the framework applies at all. The controlled thresholding experiments of Fig.~\ref{fig:Thresholding performance under AWGN and Impulsive Noise}, and the wavelet-family and decomposition-level sensitivity grid of Supplementary Fig.~S1, show that under AWGN alone the no-denoising, soft-thresholding, and Garrote-thresholding BER curves are essentially coincident, and that hard thresholding can incur a high-SNR penalty at larger decomposition depths. Wavelet thresholding
therefore yields no systematic hard-decision benefit against a purely Gaussian background, which is the expected consequence of shrinkage relying on a sparse coefficient-domain signature that AWGN does not produce. The benefit reported throughout this review is contingent on the
disturbance being structured, impulsive, or otherwise sparsely representable in the wavelet
domain, or on the denoising acting on a compressible quantity such as a channel impulse response rather than on the received waveform itself. This null result is reported here rather than only in the supplement because it bounds the scope of the framework as clearly as the positive results establish its value. 

At the same time, the surveyed literature indicates that transferability and robustness remain the dominant unresolved challenges. Many reported optimal settings are obtained from narrow candidate sets, evaluated under idealised noise assumptions, and affected by implementation details such as boundary handling, spectral leakage, and sampling-dependent dyadic scale placement. In addition, optimisation is frequently directed toward waveform fidelity rather than communication-specific task performance, which may limit the practical generalisability of the reported configurations. Although AI-assisted wavelet-parameter selection has begun to emerge, current contributions remain limited in scope, and explicit learning-based selection of wavelet parameters for communication links is still largely unexplored. This leaves a clear need for methods that are interpretable, transferable, and validated under realistic receiver conditions.

Future progress will likely depend on integrated frameworks that combine multi-objective and uncertainty-aware optimisation, principled and leakage-validated level selection, dependency-aware shrinkage models, task-aligned benchmarking, and reproducible reporting. Among these directions, AI-guided parameter selection and differentiable wavelet denoising with learned shrinkage are the most promising routes to jointly optimising wavelet, level, and threshold choices under explicit accuracy, efficiency, and interpretability constraints, and thereby to repositioning wavelet denoising from a largely heuristic preprocessing step into a robust, auditable, and generalisable component of communication-signal pipelines.

\bibliographystyle{IEEEtran}
\bibliography{wavelet_param_review}

\begin{IEEEbiography}[{\includegraphics[width=1in,height=1.25in,clip,keepaspectratio]{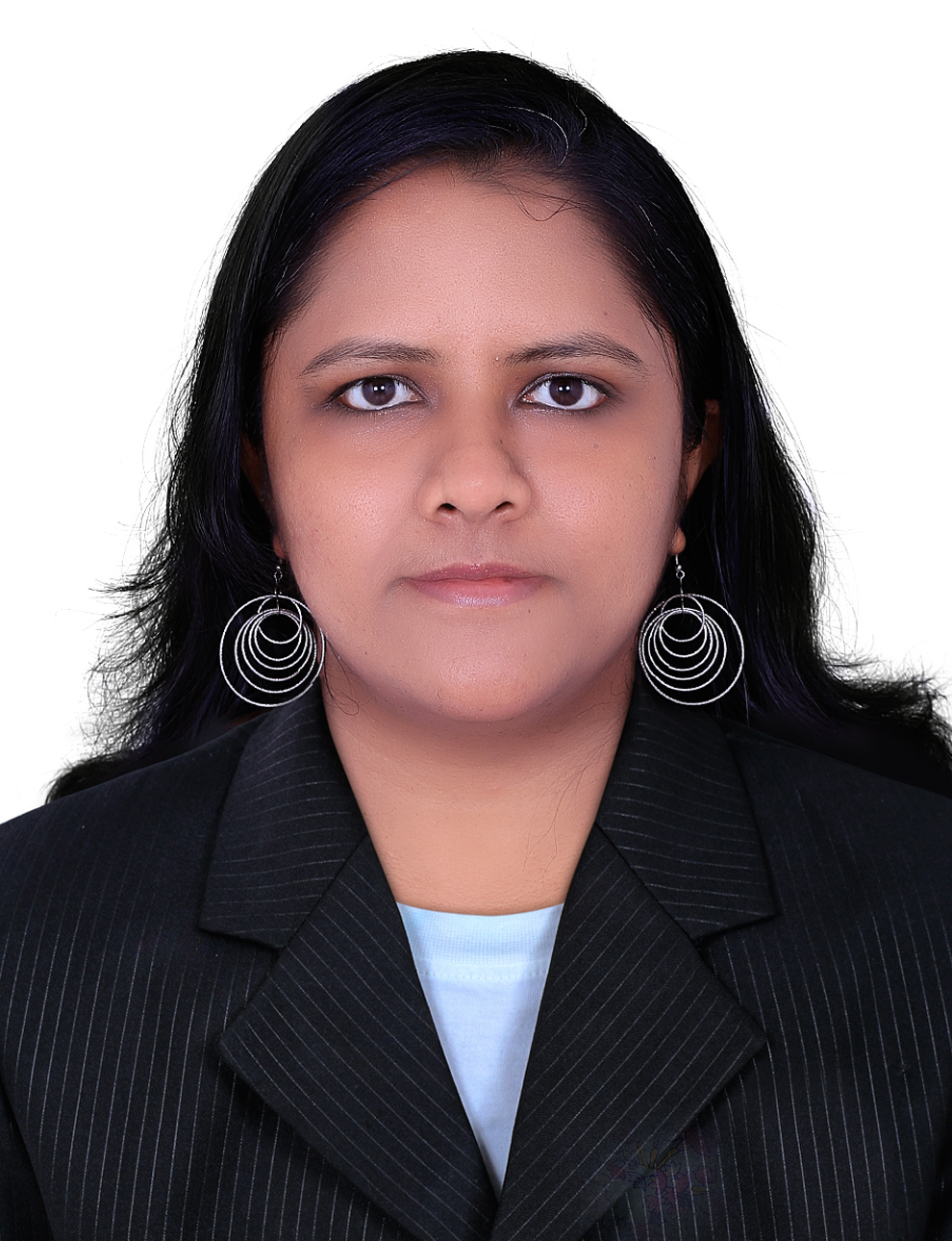}}]{Priyalakshmi Sheela} ~(Member, IEEE) received the B.Tech. degree in Electronics and Communication Engineering from University of Kerala, India, in 2009, and the M.E. degree in Applied Electronics from Anna University of Technology, Tirunelveli, India, in 2011, and thereafter Ph.D. degree in Biosignal Processing with the National Institute of Technology Calicut, India, in 2022. She is currently a Postdoctoral researcher at Walton Institute, South East Technological University, Waterford, Ireland. Her research interests focus on advanced computational frameworks for analysing biosignals and medical images, with applications in neurological disorders, assistive technology, and healthcare innovation. 
\end{IEEEbiography}

\begin{IEEEbiography}
[{\includegraphics[width=1in,height=1.25in,clip,keepaspectratio]
{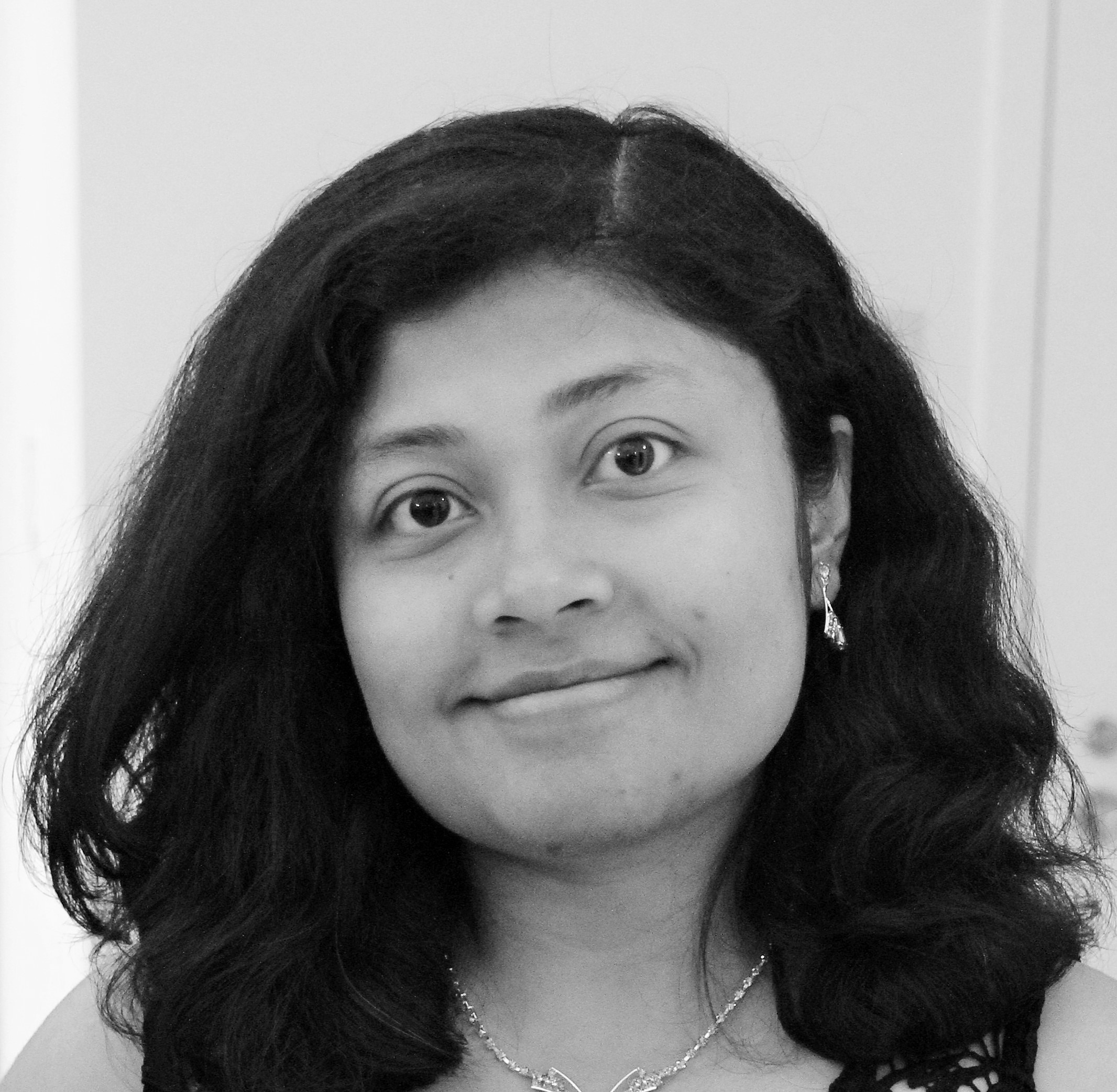}}]
{Indrakshi Dey} (Senior Member, IEEE) received the M.Sc. degree in Wireless
Communications from the University of Southampton, U.K., in 2010,
and the Ph.D. degree in Electrical Engineering from the University
of Calgary, Canada, in 2015. She is currently Lecturer with the Department
of Computing and Mathematics at South East Technological University,
Waterford, Ireland. She is also an Adjunct Assistant Professor at
Trinity College Dublin and a Research Ireland Funded Investigator.
Her research interests include propagation modelling, performance
analysis, signal processing, optimization, AI-aided communication
systems, cyber-physical systems, digital twins, IoT, and terrestrial,
space, and quantum communication networks. Her research sits at the intersection of wireless communications, dynamical systems and number theory. She is also Coordinator and Principle Investigator on multiple national, European and International grants and projects.
\end{IEEEbiography}

\end{document}